\documentclass[%
 aip,
 amsmath,amssymb,
 reprint,%
]{revtex4-1}

\usepackage{graphicx}
\usepackage{dcolumn}
\usepackage{bm}
\usepackage{braket}

\usepackage[utf8]{inputenc}
\usepackage[T1]{fontenc}
\usepackage{mathptmx}
\usepackage{etoolbox}

\makeatletter
\def\@email#1#2{%
 \endgroup
 \patchcmd{\titleblock@produce}
  {\frontmatter@RRAPformat}
  {\frontmatter@RRAPformat{\produce@RRAP{*#1\href{mailto:#2}{#2}}}\frontmatter@RRAPformat}
  {}{}
}%
\makeatother
\begin{document}

\preprint{AIP/123-QED}

\title{Towards direct spatial and intensity characterization of ultra-high intensity laser pulses using ponderomotive scattering of free electrons}
\author{A. Longman}
\affiliation{Lawrence Livermore National Laboratory, Livermore, California 94550, USA}
\email{longman1@llnl.gov}

\author{S. Ravichandran}
\affiliation{Institute for Physical Science and Technology, University of Maryland, College Park, Maryland 20742, USA}
\affiliation{Joint Quantum Institute, University of Maryland, College Park, Maryland 20742, USA}

 \author{L. Manzo}
\affiliation{Lawrence Livermore National Laboratory, Livermore, California 94550, USA}

\author{C. Z. He}
\affiliation{Institute for Physical Science and Technology, University of Maryland, College Park, Maryland 20742, USA}
\affiliation{Joint Quantum Institute, University of Maryland, College Park, Maryland 20742, USA}

\author{R. Lera}
\affiliation{Centro de L\'{a}seres Pulsados (CLPU), 37185 Villamayor, Salamanca, Spain}

\author{N. McLane}
\affiliation{Institute for Physical Science and Technology, University of Maryland, College Park, Maryland 20742, USA}

\author{M. Huault}
\affiliation{Centro de L\'{a}seres Pulsados (CLPU), 37185 Villamayor, Salamanca, Spain}
\affiliation{Departemento de Física Fundamental, Universidad de Salamanca, 37008 Salamanca, Spain }

\author{G. Tiscareno}
\affiliation{The Ohio State University, Columbus, Ohio 43210, USA}

\author{D. Hanggi}
\affiliation{The Ohio State University, Columbus, Ohio 43210, USA}

\author{P. Spingola}
\affiliation{The Ohio State University, Columbus, Ohio 43210, USA}

\author{N. Czapla}
\affiliation{The Ohio State University, Columbus, Ohio 43210, USA}

\author{R. L. Daskalova}
\affiliation{The Ohio State University, Columbus, Ohio 43210, USA}

\author{L. Roso}
\affiliation{Departamento de F\'{i}sica Aplicada, Universidad de Salamanca, 37008 Salamanca, Spain }

\author{R. Fedosejevs}
\affiliation{Department of Electrical and Computer Engineering, University of Alberta, Edmonton T6G1R1, Canada}

\author{W. T. Hill III}
\affiliation{Institute for Physical Science and Technology, University of Maryland, College Park, Maryland 20742, USA}
\affiliation{Joint Quantum Institute, University of Maryland, College Park, Maryland 20742, USA}
\affiliation{Department of Physics, University of Maryland, College Park, Maryland 20742, USA}

\date{\today}

\begin{abstract}
Spatial distributions of electrons ionized and scattered from ultra-low pressure gases are proposed and experimentally demonstrated as a method to directly measure the intensity of an ultra-high intensity laser pulse. Analytic models relating the peak scattered electron energy to the peak laser intensity are derived and compared to paraxial Runge-Kutta simulations highlighting two models suitable for describing electrons scattered from weakly paraxial beams ($f_{\#}>5$) for intensities in  the range of $10^{18}-10^{21}$Wcm$^{-2}$. Scattering energies are shown to be dependant on gas species emphasizing the need for specific gases for given intensity ranges. Direct measurements of the laser intensity at full power of two laser systems is demonstrated both showing a good agreement between indirect methods of intensity measurement and the proposed method. One experiment exhibited the role of spatial aberrations in the scattered electron distribution motivating a qualitative study on the effect. We propose the use of convolutional neural networks as a method for extracting quantitative information of the spatial structure of the laser at full power. We believe the presented technique to be a powerful tool that can be immediately implemented in many high-power laser facilities worldwide.      
\end{abstract}

\maketitle

\section{Introduction}
High-intensity laser-plasma interactions have been a focus of research for several decades driven by their wide range of applications in fields such as fusion energy \cite{Zylstra2022}, high-energy density physics \cite{hoffmann2005}, and table-top particle acceleration \cite{Esarey95,Sarri13,Tajima2020,Robson2007}. In all laser-plasma interactions, the intensity of the laser is a primary parameter that determines the outcome \cite{danson2019,Yoon:21}. Despite its fundamental importance, it remains an open problem to directly measure both the laser intensity, and the spatial mode on shot at full power. 

Direct measurement (independent of other laser parameters) of the intensity of a high-power laser is not trivial as it can surpass values of $10^{18}$ Wcm$^{-2}$, far above the breakdown threshold of any known material. Any solid, liquid, or gaseous material placed within the vicinity of the laser focus is ionized and destroyed. For this reason, the peak intensity of almost every high-intensity laser is inferred from indirect measurements, that is, the intensity is determined from measurements of the pulse duration, pulse energy, and a measured spatial mode at the laser focus. These measurements are often made at reduced energies, or in equivalent planes, and can ignore nonlinear effects in the laser system when increasing the laser energy such as thermal distortion of the gratings and thermal lensing in the amplifiers. Any nonlinear changes in the near-field amplitude or phase induced when running at full power or high repetition rate can lead to significant changes to the far-field focal spot shape, size, pulse duration, and consequently, the peak intensity \cite{Tiwari:19}. 

Several techniques have been suggested to measure the laser intensity directly including nonlinear Thomson scattering \cite{Chen1998,He:19,Gao06,Har-Shemesh:12,Tarbox:15,Harvey18,Mackenroth2019}, ionization state measurements \cite{Perry:89,Chowdhury01,Yamakawa03,Ciappina19,Ciappina20,Ouatu22,Link06,Smeenk:11,Alnaser04}, pair-production \cite{Aleksandrov21}, as well as electron, ion, and proton scattering \cite{Galkin10,kalashnikov_2015,Vais_2017,Ivanov_2018,Vais2018,Kramer2018,Mackenroth_2019,Vais_2020,Vais_2021,Bukharskii22}. These techniques have previously demonstrated successful direct measurements over a range of relativistic laser intensities. However, their implementation can require complicated diagnostics, extensive analyses, and an accurate, simple, and robust experimental demonstration of directly measuring laser intensities above $10^{18}$ Wcm$^{-2}$ has proven difficult. In addition to the peak intensity of the laser pulse, the spatio-temporal mode of the laser is also of great interest and has been shown to play a significant role in laser-plasma interactions \cite{Jeandet:22,Grace22,Wilhelm19}. Several techniques have been deployed to measure the spatio-temporal properties of the pulse but are often conducted either in the laser near-field using pick-off mirrors before final focusing, or collecting light after the focus through re-imaging optics that can also aberrate the image and phase of the focal spot \cite{Grace_2021,Pariente2016}. The development of a diagnostic that can directly measure spatial, temporal, and intensity properties of a beam operating at full power is therefore of great interest. 

In this work, we demonstrate through theory, simulations, and preliminary experiments how the distribution of scattered electrons ionized from ultra-low density gases $(\approx 10-100$nbar$)$ can be used to directly characterize the laser intensity, and possibly spatial profiles at full power. We derive back-of-the-envelope scaling laws that characterize the peak scattered electron energies, their corresponding forward scattering angles, and compare them to single particle simulations in paraxial laser fields. We demonstrate through simulations and experiments that for gases with sufficient ionization potentials, it is possible to make direct measurements of the laser's peak intensity independent of laser energy, beam waist, and pulse duration. The method utilizes a single image plate detector as demonstrated in our complementary paper \cite{Ravichandran23}, and is shown to potentially reduce uncertainties in the measurements from current indirect methods. We extend our analysis to spatially aberrated beams using compositions of higher order laser modes and qualitatively demonstrate spatially modified scattering distributions indicating that it may be possible to directly measure the laser spatial mode at full power. We present preliminary experimental results obtained using two high power laser systems giving a clear demonstration of the models, methods, and a path towards future work.

\section{Theoretical background}
As electrons interact with a laser pulse, they gain energy and momentum proportional to the field amplitude. If the electrons are born infinitely far from the laser pulse and there are no spatial gradients in the field amplitude, then according to the Lawson-Woodward theorem, no net energy can be imparted to the electrons after their interaction \cite{Esarey95,Gibbon,Macchilasers}. To transfer a net energy to the electrons the interaction must violate the Lawson-Woodward theorem either through limiting the interaction volume, utilizing the ponderomotive force of the laser, or using background electric or magnetic fields. 

In the case of a high intensity laser pulse propagating in ultra-low density gas ($\approx10-100$nbar), electrons can be ionized near the region of the peak intensity and then scattered away via the ponderomotive force with an angle relative to the laser $\hat{k}$ vector and proportional to the final energy \cite{Meyerhofer:96,Ivanov_2018,Singh06,Lin2007,Mackenroth_2019,Wang02,He03,Hartemann95,Quesnel98,Yang_2011,Kibble66,Bauer95,Popov08,Hegelich23}. To calculate this scattering angle, we begin by approximating the laser as an infinite plane wave. The relativistic motion of a free electron in an infinite plane wave has been well studied, and its evolution as a function of time and space has known analytic solutions. We refer the reader to the many works on this subject and highlight some of the key results here \cite{Sarachik70,Macchilasers,Gibbon,Hartemann95,Gonoskov22}. 

The motion of an electron in a general electromagnetic field can be given by the normalized Lorentz force, 
\begin{equation} \label{eq1}
    \dot{\vec{p}}=\vec{a}+\vec{p}\times\vec{b}/\gamma,
\end{equation}
where $\vec{p}=\vec{p}/m_ec$ is the normalized momentum, $\vec{a}=e\vec{E}/m_ec\omega$ is the normalized vector potential, $\vec{b}=e\vec{B}/m_e\omega$ is the normalized magnetic field, and the space and time variables are normalized such that $\vec{x}=k\vec{x}$, and $t=\omega t$. $\vec{E}$ and $\vec{B}$ are the oscillatory electric and magnetic fields of the wave, and $\gamma=\sqrt{1+|\vec{p}|^2}$, $m_e$, $e$, $c$, $k$, $\omega$ are the Lorentz factor, electron mass, electron charge, speed of light, and the laser wave-number and angular frequency respectively. Throughout this work we will maintain these normalization's unless otherwise stated. We also neglect radiation reaction as we are working with intensities and initial electron energies well below the point at which these effects become important $(a_0\approx100)$ \cite{Gonoskov22}. 

The electric and magnetic fields of a pulsed infinite plane wave propagating along $\hat{z}$ and polarized along $\hat{x}$ can be written as the real parts of the function \cite{Jackson:100964}, 
\begin{equation}\label{eq2}
    a_x=b_y=a_0f(t)e^{i(z-t)},
\end{equation}
where $f(t)$ is the temporal envelope of the laser, typically a Gaussian, $a_0\approx0.855\lambda_{\mu m}\sqrt{I_0[10^{18}Wcm^{-2}]}$ is the peak normalized vector potential, $I_0$ the laser intensity, and $\lambda_{\mu m}$ is the laser wavelength in microns. Analyzing the motion of the electron in a linearly polarized continuous plane wave according to the Lorentz force shows that the electron will oscillate along the polarization direction due to the real part of the transverse electric field such that $p_x=a_0\cos\left(z(t)-t\right)$, where the longitudinal position of the electron is given by \cite{Macchilasers,Gibbon,Hartemann95}, 
\begin{equation} \label{eq3}
    z(t) = \frac{a_0^2}{4}\left[-z(t)+t+\frac{\sin\left(2\left(z(t)-t\right)\right)}{2}\right].
\end{equation}

If $a_0\gtrsim1$, the electron motion becomes relativistic and the $\vec{p}\times\vec{b}$ force becomes comparable to the electric force causing a significant forward drift along the $\hat{z}$ axis. If the initial electron momentum is much less than the oscillation momentum in the plane wave $(p_0\ll a_0)$, one can derive the relationship \cite{Hartemann95}, 
\begin{equation} \label{eq4}
    p_z=\frac{p_\perp^2}{2}=\frac{p_x^2}{2}.
\end{equation}
Taking the ratio of the electrons forward momentum to its transverse momentum, and the fact that in an infinite plane wave the forward momentum of the laser is conserved into the forward motion of the electron $(p_z=\gamma-1)$, we yield the well known result \cite{Hartemann95,Moore95,Meyerhofer:96}, 
\begin{equation} \label{eq5}
    \tan(\theta)=\frac{p_x}{p_z}=\sqrt{\frac{2}{\gamma-1}}.
\end{equation}
Given that the electron is constrained to the $\hat{x}-\hat{z}$ plane in a linearly polarized plane wave, we can write the Lorentz factor as, 
\begin{equation} \label{eq6}
    \gamma(t)=\sqrt{1+p_x^2+p_z^2}. 
\end{equation}
As the Lorentz factor changes its value periodically, we look to find its average value over one laser cycle to calculate the average scattering energy. Combining (\ref{eq4}) and (\ref{eq6}), we can write the Lorentz factor as, 
\begin{equation} \label{eq7}
    \gamma(t)=\sqrt{1 + p_x^2+\frac{p_x^4}{4}}=1+\frac{p_x^2}{2}.
\end{equation}

\begin{figure}
    \centering
    \includegraphics{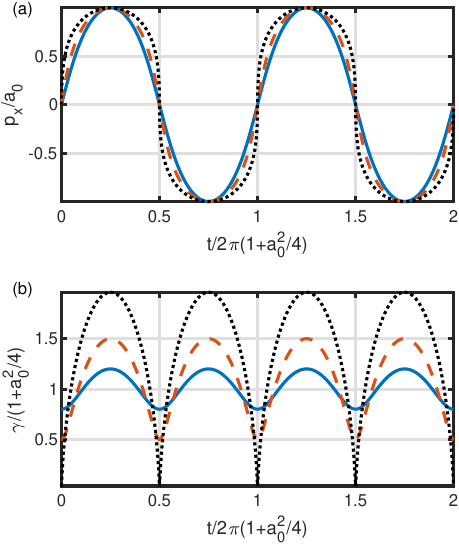}
    \caption{Temporal evolution of the  normalized electron transverse momentum (a), and energy (b), in an infinite plane wave with amplitudes $a_0=1,2$ and $10$, shown by the solid blue, red dashed, and black dotted lines respectively.}
    \label{fig1}
\end{figure}

The motion of an electron in the laboratory frame calculated by numerically integrating (\ref{eq1}) in plane waves of amplitude $a_0=1,2$ and $10$ are plotted in Fig.\ref{fig1} with a blue solid line, red dashed line, and black dotted line respectively. We see in Fig.\ref{fig1}(a) that as the laser intensity increases, the momentum deviates from the sinusoidal trajectory into a more complicated curve given its transcendental dependence on $z(t)$. Similarly when we examine the value of $\gamma$ as a function of time in Fig.\ref{fig1}(b), we find that its functional form changes from a quasi-sinusoid at non-relativistic intensity, into a non-elementary functional form at relativistic intensities. 

To obtain an average energy of the scattered electron in these complicated trajectories, we derive 3 models. The first model assumes that a scattered electron maintains its peak oscillation energy in (\ref{eq7}), 
\begin{equation}\label{eq8}
    \left\langle\gamma\right\rangle_{peak}=1+\frac{a_0^2}{2}.
\end{equation}
This average value of $\gamma$ appears to be what is used in previous works for calculating scattering angles for the electrons \cite{kalashnikov_2015,Ivanov_2018}. A second model calculates the average of (\ref{eq7}) in the co-moving frame, $\xi=z(t)-t$, yielding the result, 
\begin{equation} \label{eq9}
    \left\langle\gamma\right\rangle_{\xi}=1+\frac{a_0^2}{4}.
\end{equation}
Our third model calculates the time average of the electron momentum in the lab frame. This is difficult given the transcendental dependence on the electron position where we have limited analytic tools to perform the integration. We instead rely on numerical techniques to find an approximate formula for the time averaged $\gamma$ in an infinite plane wave, 
\begin{equation} \label{eq10}
    \left\langle\gamma\right\rangle_t\approx\left(1+\frac{a_0^2}{4}\right)\left(1+\frac{1}{2}\left(1+\frac{4}{a_0^2}\right)^{-2}\right).
\end{equation}
For completeness, we also include the expression for $\gamma$ frequently used in relativistic mass expressions \cite{Kibble66,Meyerhofer:96}, and plasma ponderomotive temperature scaling \cite{Macchilasers,Gibbon,Wilks92},
\begin{equation} \label{eq11}
    \gamma_{pond}=\sqrt{1+\frac{a_0^2}{2}}.
\end{equation}
These expressions can then be combined with (\ref{eq5}) to estimate the scattering angle of the electrons relative to the laser $\hat{k}$. With these four plane wave models, we can compare the peak ejected values of $\gamma$ and the corresponding scattering angles to more realistic models including the effects of finite spot sizes, finite pulse durations, and longitudinal field components. 

\section{Scattering from high intensity laser pulses in vacuum}

To model the realistic scattering of the electrons from tightly focused Gaussian laser pulses, we need to include the effects of the spatial profile and the additional field components introduced when focusing a laser pulse. In this work we will restrict ourselves to the paraxial regime $(kw_0\gg 1)$, where $w_0$ is the laser beam waist at best focus. There is an ongoing debate as to when and where the paraxial approximation breaks down after which the use of more sophisticated models of the electric an magnetic fields will be required \cite{Peatross:17,Popov08}. In this work, we treat beams with $5<f_{\#}<10$ as weakly paraxial and do not consider focal geometries with f-numbers smaller than 5. Under the paraxial approximation, we can write our electric and and magnetic fields in the form \cite{Peatross:17,Erikson94,Longman22},
\begin{equation} \label{eq12}
    \vec{a}=a_0w_0\sqrt{\frac{\pi}{2}}\left[\hat{x}+\frac{\hat{y}}{2}\frac{\partial^2}{\partial x\partial y}+i\hat{z}\frac{\partial}{\partial x}\right]\psi^\ell_pf(t)e^{i(z-t)},
\end{equation}
where $\psi_p^\ell$ is the Laguerre-Gaussian basis set describing the laser spatial envelope given by, 
\begin{eqnarray} \label{eq13}
    \psi_p^\ell=&&\sqrt{\frac{2p!}{\pi(p+|\ell|)!}}\frac{1}{w(z)}\left(\frac{r\sqrt{2}}{w(z)}\right)^{|\ell|}L_p^{|\ell|}\left(\frac{2r^2}{w(z)^2}\right)\nonumber\\
    &&\times\exp\left(-\frac{r^2}{w(z)^2}+\frac{ir^2z}{2(z^2+z_0^2)}+i\ell\theta-i\Psi\right).
\end{eqnarray}
Here we have introduced the beam waist at a plane $z$ as $w(z)=w_0\sqrt{1+z^2/z_0^2}$, the Rayleigh range $z_0=w_0^2/2$, the azimuthal and radial mode numbers $\ell\in\mathbb{Z}$ and $p\in|\mathbb{Z}|$ respectively with $\mathbb{Z}$ being the set of integers, the Gouy phase $\Psi=(1+|\ell|+2p)\tan^{-1}(z/z_0)$, and the associated Laguerre polynomials $L_p^\ell(X)$ given by,
\begin{equation} \label{eq14}
    L_p^\ell(X)=\sum_{n=0}^p(-1)^n\frac{(p+\ell)!}{(p-n)!(\ell+n)!n!}X^n.
\end{equation}

The Laguerre-Gaussian functions are an orthogonal set of solutions to the paraxial Helmholtz equation in cylindrical coordinates. Typically, we are concerned with the fundamental $(\ell=0,p=0)$ Gaussian mode as this closely resembles a typical focused laser pulse. As we will see in the following sections, it becomes necessary to include higher order laser modes to model non-Gaussian near-fields, Airy rings, aberrations, and other higher order effects typically found in realistic high-power laser pulses. Full derivations of the field equations from (\ref{eq12}) for the Laguerre-Gaussian basis set is given in another work \cite{Longman22}. Additional corrections to (\ref{eq12}) can be introduced to correct for ultra-short pulse durations when $\tau\leq 2w_0$ \cite{Quesnel98}. In this work we primarily work with laser pulses of duration $\tau=30fs$ and beam waists $w_0>4\mu$m which do not satisfy the inequality and so these corrections are neglected.  


If the electron density is low enough such that the Debye screening length $\lambda_D$ is much larger than the interaction volume, we can treat the electrons as non-interacting single particles and directly integrate the Lorentz force without including collective effects. Assuming an electron temperature of 10 keV, a fully ionized nitrogen gas, a 100 $\mu$m screening length equates to an electron density of roughly $1\times10^{14}cm^{-3}$. Using the ideal gas law at room temperature, we find that we need to operate at pressures roughly below $3\times10^{-4}$mbar. This is easily achievable in current vacuum systems which routinely operate at and below $10^{-6}$mbar. Realistically, the interaction of a tightly focused ultra-high intensity laser and ionized electrons can probably be treated as single particles even in the $10^{-3}$mbar range. 

Direct integration of (\ref{eq1}) using the full 3D field equations given by (\ref{eq12}) is performed using a high-order Runge-Kutta algorithm. On the order of 1 million electrons is found to be sufficient to effectively sample the maximum ejected energies and the corresponding ejection angle of the electrons over a large volume in space. 

The simulation volume is specified in space centered about the focal point, typically $3w_0\times3w_0\times6z_0$ in $x,y,z$ respectively in which gaseous atoms are randomly distributed. A routine is run that identifies when electrons will be ionized based on their ionization potential and the position of each atom, after which it calculates the effective number of electrons scattered per unit pressure of the gas. While this volume is not sufficient to capture all electrons ionized by the laser, it is sufficient to sample the electrons ionized near the peak laser intensity and hence the electrons scattered with the highest energies and smallest angles relative to $\hat{k}$. Electrons ionized in either the temporal foot of the laser, far from the laser axis, or in any pre-pulse are normally ponderomotively swept away before they can interact with the peak laser intensity and for now, are not of interest. 

Given the complexity of the ionization mechanism of the electrons, we opt to use a simple modification of the over-the-barrier ionization model such that the intensity at the ionization will occur is given by \cite{Gibbon}, 
\begin{equation} \label{eq15}
    I_{ion}[Wcm^{-2}]=4\times10^9U_{eV}^4Z^{-2}R,
\end{equation}
where $U_{eV}$ is the ionization potential of the $Z^{th}$ electron shell in electron volts, and $R$ is a random number selected between 1 and 10 for each individual electron. This random number loosely emulates the nature of the measured ionization populations as a function of intensity found in several works \cite{Augst89,Augst:91}. Future work may introduce a more sophisticated model of ionization such as the ADK model but as we will show, the ionization rate is used to determine the intensity range of the gas, rather than the actual intensity measurement, and as such is not the focus of the current work \cite{Tong02}. Example values of electron ionization potentials, and the corresponding ionization intensities for $R=1$ are calculated and given in Table \ref{table1} using the NIST atomic database \cite{NIST_ASD}.

\begin{table}
\caption{\label{table1} Over the barrier ionization potentials for various gases and shells of interest calculated from (\ref{eq15}) and the NIST database \cite{NIST_ASD}.}
\begin{ruledtabular}
\begin{tabular}{ccc}
Element$^Z$   &  $U_{ion} [eV]$    &   $I_{ion}(R=1),[10^{18}Wcm^{-2}]$\\
\hline
N$^{6+}$  & $552.1$   &  $10.3$\\
N$^{7+}$  & $667.1$   &  $16.2$\\
Ar$^{14+}$  & $755.7$   &  $6.7$\\
Ar$^{15+}$  & $854.8$   &  $9.5$\\
Ar$^{16+}$  & $918.0$   &  $11.1$\\
Kr$^{22+}$  & $945$   &  $6.4$\\
Kr$^{23+}$  & $999$   &  $7.5$\\
Kr$^{24+}$  & $1042$   &  $8.5$\\
Kr$^{25+}$  & $1155$   &  $11.2$\\
Kr$^{26+}$  & $1205$   &  $12.5$\\
Kr$^{27+}$  & $2929$   &  $403.3$\\
Kr$^{28+}$  & $3072$   &  $453.2$\\
Kr$^{29+}$  & $3228$   &  $515.8$\\
Kr$^{30+}$  & $3380$   &  $580.8$\\
Kr$^{31+}$  & $3584$   &  $686.8$\\
Kr$^{32+}$  & $3752$   &  $774.1$\\
Kr$^{33+}$  & $3971$   &  $913.3$\\
Kr$^{34+}$  & $4109$   &  $986.4$\\

\end{tabular}
\end{ruledtabular}
\end{table}

\begin{figure}
    \centering
    \includegraphics{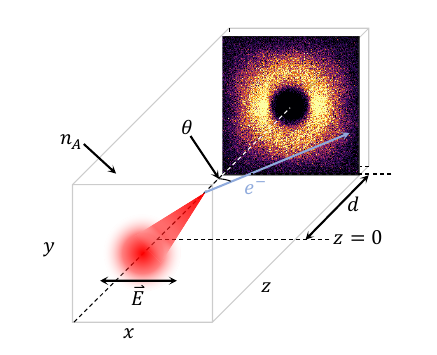}
    \caption{Illustration of the numerical setup. A bivariate histogram detector is placed a distance $d$ from the laser focal plane. The laser is polarized in the horizontal plane along the $\hat{x}$ axis. An atomic number density $n_A$ is specified for the simulation, and is later used to estimate the experimental signal strength.} 
    \label{fig2}
\end{figure}

\begin{figure*}
    \centering
    \includegraphics{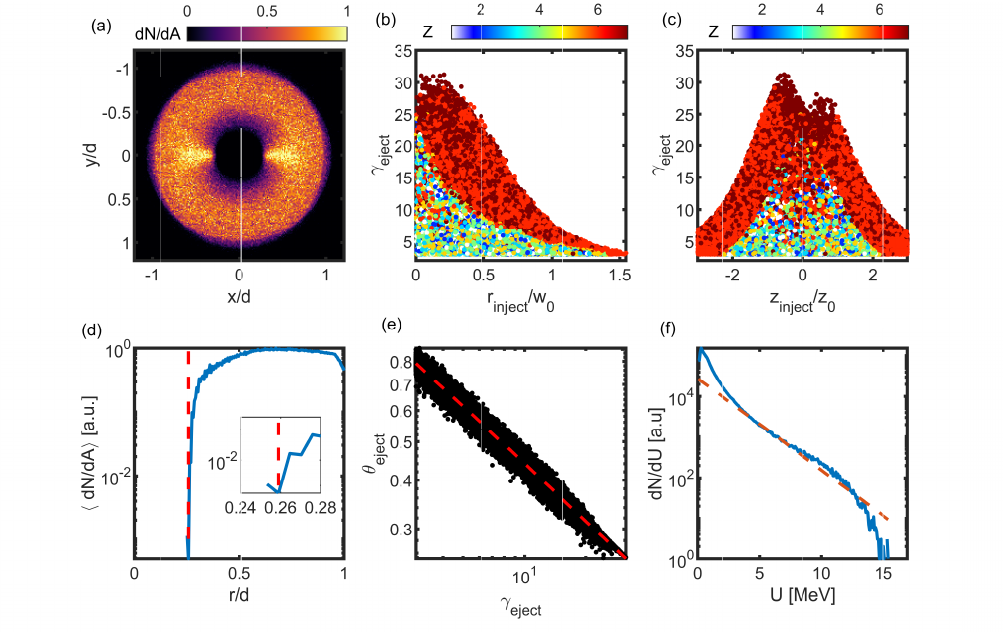}
    \caption{Simulation results from 2 million electrons ionized from nitrogen, being scattered from a laser with parameters $a_0=10$, $w_0=10\mu$m, $\tau=30$fs, $\lambda=800$nm. (a). Flux of the electrons striking the detector removing electrons with energies below 1MeV, the color bar gives the normalized flux. (b) and (c). Ejected Lorentz factor as a function of initial radius and longitudinal position respectively with the colorbar representing the electron ionization number. (d). Average radial lineout of (a) with the inner electron cutoff zoomed in the subplot, the cutoff value is highlighted with the red dashed line. (e). Electron ejected angle as a function of ejected Lorentz factor, and (\ref{eq5}) is shown by the red dashed line. (f). Scattered electron energy distribution fitted with a 1.93MeV Maxwell Boltzmann distribution shown with the red dashed line.} 
    \label{fig3}
\end{figure*}

Using the laser wavelength and our best estimates of the peak vacuum intensity, we can use Table \ref{table1} to select a gas that has an ionization intensity below the estimated peak intensity, but as we will show, should be relatively close to the estimated peak intensity of the laser. Additional gases outside of Table \ref{table1}, for example xenon, neon, and oxygen can also be used but are not listed for brevity. 

An illustration of the simulation setup is given in Fig.\ref{fig2}. A plane detector is placed a distance $d$ from the focal plane of the laser that registers the position and momentum of each electron passing it. Simulations were run for various Gaussian laser pulses to identify a suitable scaling model for the maximum electron energy and the corresponding minimum scattering angle. For the scope of this work, we are interested in values of $a_0$ in the range of 1-30 which roughly corresponds to intensities of $10^{18}$Wcm$^{-2}$ to $10^{21}$Wcm$^{-2}$ at 800nm. Intensities below this range are weakly relativistic and we find that any forward scattering of the electrons becomes negligible, while intensities above this range are possible to measure primarily via the electron energy alone as scattering angles become too small.

Fig.\ref{fig3} shows the scattering of 2 million electrons ionized from nitrogen gas from a laser pulse with $a_0=10$, $w_0=10\mu$m, $\tau=30$fs, and $\lambda=800$nm, where $\tau$ is the temporal intensity full-width half-maximum (FWHM). This corresponds to a peak intensity of $~2.1\times10^{20}$Wcm$^{-2}$.  The normalized flux of electrons striking the detector is given in Fig.\ref{fig3}(a) showing the inner electron cut-off angle and some spatial structure in the electron distribution, notably a higher number of electrons deposited along the polarization axis. We note that unlike the prediction of the plane wave that restricts the electrons to the $\hat{E}-\hat{k}$ plane, the electrons scatter into a 3D ring in agreement with previous work \cite{Quesnel98}. 

Electron energies below 1MeV in Fig.\ref{fig3}(a) are removed to emulate the electron stopping power of an aluminum light blocking shield used in experiments. This gives a lower electron energy cutoff and hence an outer ring as shown in  Fig.\ref{fig3}(a). An average radial lineout of  Fig.\ref{fig3}(a) is plotted in  Fig.\ref{fig3}(d) where the red dashed line indicates the measured cutoff radius of the electrons. For this set of laser parameters, we find the inner electron cutoff angle to be $\tan(\theta)=r/d\approx0.26$. 

  \begin{figure*}
    \centering
    \includegraphics{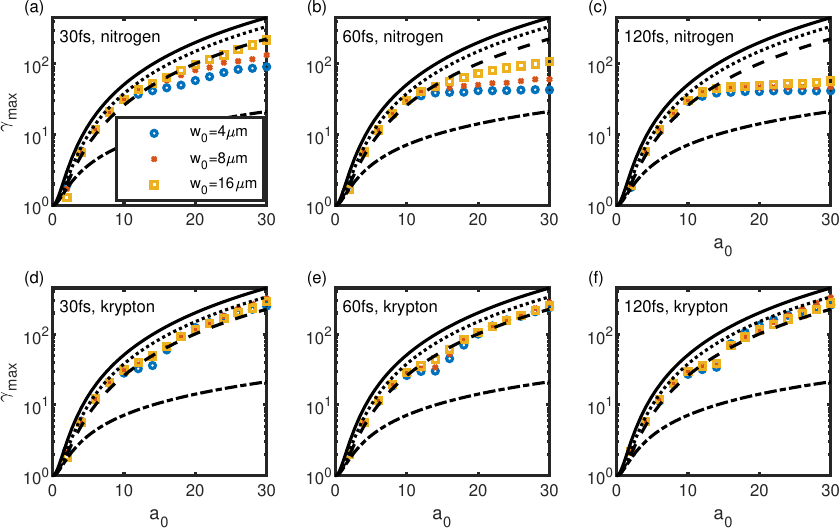}
    \caption{Maximum ejected electron energies averaged over 100 highest energy electrons from nitrogen and krypton gases for laser pulses with $4\mu$m, $8\mu$m, and $16\mu$m beams waists indicated by the blue circles, red crosses, and yellow squares respectively.  (a) 30fs pulse in nitrogen, (b) 60fs pulse in nitrogen, (c) 120fs pulse in nitrogen, (d) 30fs pulse in krypton, (e) 60fs pulse in krypton, and (f) 120fs pulse in krypton. The energy scaling models $\langle\gamma\rangle_{peak}$, $\langle\gamma\rangle_{\xi}$, $\langle\gamma\rangle_{t}$, $\gamma_{pond}$ are given by the solid, dashed, dotted, and dash-dot black lines respectively.} 
    \label{fig4}
\end{figure*}

Fig.\ref{fig3}(b) and (c) give scatter plots of the scattered electron Lorentz factor as a function of the initial radius and longitudinal positions respectively, while the color map corresponds to the ionization shell highlighting that inner shell electrons shown in red are accelerated to the highest energies. We find that electrons born before the focal plane at $z_{inject}\approx-z_0/2$ are accelerated to the highest energies. Fig.\ref{fig3}(e) gives the measured electron ejection angle $\theta_{eject}$ as a function of the ejected energy $\gamma_{eject}$. The red dashed line overlaid on the data points gives the scattered energy-angle relationship derived in (\ref{eq5}). Although there is an excellent agreement between (\ref{eq5}) and the data in Fig.\ref{fig3}(e), we find that there is a broadening of the data points about the red dashed line. We believe this to be a result of the angular spread in the transverse laser $k_{\perp}$ vector components from the shorter f-number. Indeed, as the beam becomes non-paraxial and $w_0\to\lambda$, the spread of electrons about the red dashed line can become significant and even detrimental to our measurement of the peak intensity. Fig.\ref{fig3}(f) gives the electron energy distribution where the red dashed line is a fitted Maxwell-Boltzmann temperature estimated to be 1.93MeV. This temperature further validates the use of the single particle approximation.

We have run simulations for various peak intensities between $a_0=2:30$ for 3 different beam waists, $4\mu$m, $8\mu$m, and $16\mu$m and for 3 different pulse durations, 30fs, 60fs, and 120fs. Simulations were run using nitrogen and krypton gas. Fig.\ref{fig4} summarizes the results of these simulations and plots the average scattered energy of the 100 highest energy electrons ejected from each of the interactions. Fig.\ref{fig4}(a) gives the average final scattering energy for electrons ejected from a 30fs FWHM pulse with $4\mu$m, $8\mu$m, and $16\mu$m beam waists represented by the blue circles, red crosses, and yellow squares respectively. The 4 analytic models described in the previous section are overlaid using the 4 black lines. The $\langle\gamma\rangle_{peak}$ model (\ref{eq8}) is displayed using a solid black line, $\langle\gamma\rangle_{\xi}$ (\ref{eq9}) is displayed using a dashed black line, $\langle\gamma\rangle_{t}$ (\ref{eq10}) is displayed using a dotted black line, and the $\gamma_{pond}$ model (\ref{eq11}) is displayed using a dash-dot black line. We find that for laser intensities of $a_0\leq10$, we have excellent agreement for all three beam waists with the $\langle\gamma\rangle_{\xi}$ and $\langle\gamma\rangle_t$ models. For the 30fs, $w_0=16\mu$m case, we find that the $\langle\gamma\rangle_{\xi}$ model agrees well all the way up to $a_0=30$. For smaller beam waists we find that the electrons tend to scatter with less energy when the laser amplitude is greater than $a_0=10$. 

This is made apparent in the 60fs and 120fs cases with nitrogen in Fig.\ref{fig4}(b) and (c) respectively. For example in the $4\mu$m 60fs case, the peak energy of the electrons essentially clamps at about $\gamma_{eject}\approx30$. The same is true for both the $4\mu$m and $8\mu$m beams when scattered by a 120fs pulse in (c). In these situations, we believe the electrons are not sampling the peak intensity of the laser. Given that the ponderomotive force scales as $F_p\propto w_0^{-2}$, the smaller the beam waist, the faster the electron is expelled from the laser pulse. The inverse appears to be true considering the pulse duration, that is, the longer the pulse duration the less likely the electron will interact with the peak intensity before it is ponderomotively swept away. If the laser pulse is shorter, it requires less time for the electron to slip back to the peak intensity region and therefore is more likely to sample higher intensities before it is swept away. 

 \begin{figure*}
    \centering
    \includegraphics{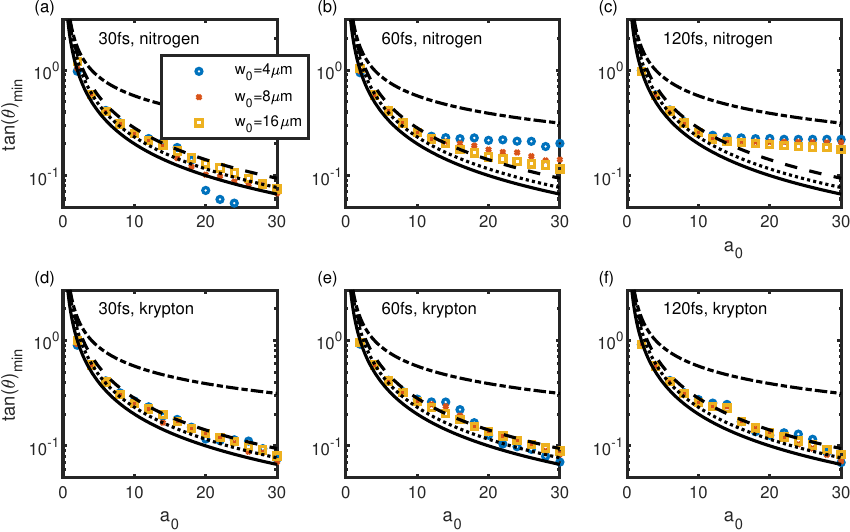}
    \caption{Minimum ejected electron angle averaged over 100 highest energy electrons from nitrogen and krypton gases for laser pulses with $4\mu$m, $8\mu$m, and $16\mu$m beams waists indicated by the blue circles, red crosses, and yellow squares respectively.  (a) 30fs pulse in nitrogen, (b) 60fs pulse in nitrogen, (c) 120fs pulse in nitrogen, (d) 30fs pulse in krypton, (e) 60fs pulse in krypton, and (f) 120fs pulse in krypton. The analytic $\tan(\theta)$ models are derived from \ref{eq5} using the energy scalings $\langle\gamma\rangle_{peak}$, $\langle\gamma\rangle_{\xi}$, $\langle\gamma\rangle_{t}$, $\gamma_{pond}$ and are given by the solid, dashed, dotted, and dash-dot black lines respectively.} 
    \label{fig5}
\end{figure*}

To address this issue, we must select a gas in which the electrons are ionized close to the peak intensity as to increase the interaction probability. Switching to krypton gas in our simulations is shown in Fig.\ref{fig4} (d)-(f). In all pulse durations, we find that the scattered electron energy follows our $\langle\gamma\rangle_{\xi}$ and $\langle\gamma\rangle_t$ scalings much closer for all three beam waists, however, we find that between $a_0=8$ and $a_0=14$ there is a region where the electrons do not follow the scaling laws. This can be explained by the large gap in the ionization potentials between Kr$^{26+}$ and Kr$^{27+}$ as shown in Table \ref{table1}. It is therefore more suitable to use a gas that does not have a large potential gap close to the peak intensity. While not shown, we find that xenon is a suitable candidate for measuring the peak intensity over the whole range of $a_0=1$ to 30 due to its large number of electrons with close ionization spacing in this regime.

This effect highlights the need for future models to introduce more advanced and realistic ionization algorithms as to effectively model the range over which a gas can be used to measure the intensity. Although our model does not capture this effect accurately, we believe that it has emphasized the necessity of an advanced ionization algorithm. We note that while the ionization rate can determine the intensity range over which a particular gas can be used, it does not appear to affect the scattering dynamics of the electrons and therefore does not affect the intensity measurement. 

In Fig.\ref{fig5} we plot the average minimum scattering angle of the 100 highest energy electrons, and as in Fig.\ref{fig4}, as a function of the pulse durations, beam waists and gas species. The scaling laws are computed using (\ref{eq5}) and our 4 models of the various ejected values of $\langle\gamma\rangle$. These plots use the same marker and line styles as in Fig.\ref{fig4}. In Fig.\ref{fig5}(a) we plot the results of the electron scattering from a 30fs laser pulse in nitrogen. We find that all three beam waists follow both the $\langle\gamma\rangle_{\xi}$ and $\langle\gamma\rangle_{t}$ scaling laws well up to $a_0=20$. At $a_0\geq20$ using a $4\mu$m beam waist we see a significant deviation from the scaling laws. This is due to the electrons being accelerated directly forward in this parameters, the exact physics of which is unclear at this point. As the pulse duration is increased to 60fs and 120fs in Fig.\ref{fig5}(b) and (c) respectively, we see the smaller beam waists begin to deviate from the scaling laws similar to in Fig.\ref{fig4}. This is due in part to the clamping of the electron energy, and also due to the increased spectrum of $k_{\perp}$ values in smaller f-number beams. Again, switching to krypton gas in Fig.\ref{fig5}(d) to (f) shows the electrons following both the $\langle\gamma\rangle_{\xi}$ and the $\langle\gamma\rangle_t$ scalings closely for all three beam waists and pulse durations.

To summarize, we have found that the scalings given by $\langle\gamma\rangle_{\xi}$ and $\langle\gamma\rangle_t$ give good upper and lower approximations to the estimated peak scattering energy and scattering angle of the electrons provided they are ionized reasonably close to the peak laser intensity. A suitable choice of gas to ionize can be chosen based on the the over-the-barrier ionization intensities given by (\ref{eq15}), however, more advanced ionization rate algorithms should be used to determine the exact intensity range of any given gas. For amplitudes of $a_0<10$, nitrogen, krypton, and argon look to be suitable gases while for $a_0>10$, krypton and xenon are more suitable. 

While we have demonstrated numerically that the scattered electrons can make a suitable diagnostic for intensities in the range of $a_0=2:30$, and for f-numbers greater than 5, we have not demonstrated for intensities above this range ($I_0>10^{21}$Wcm$^{-2}$) or for non-paraxial beams. We note that while the scattering angle tends to zero at high intensities making precise measurements difficult, the electron energy increases linearly with the intensity. In principle it would be possible to use an electron spectrometer or a stack of image plates separated by various metal plates of various stopping powers to measure the peak intensity, however, as the intensity surpasses $10^{22}$Wcm$^{-2}$, radiation reaction and quantum electrodynamic (QED) effects can begin to play a role and our simple scalings (\ref{eq9}), and (\ref{eq10}) may break down. In addition, the flux of the highest energy electrons will likely reduce as the intensity increases making the measurement more challenging. More advanced models including non-paraxial beams, radiation reaction, and advanced ionization routines are therefore needed to explore the electron scattering beyond the scope of intensities, f-numbers, and pulse durations considered in this work.

\section{Preliminary Experimental Results}

Two experiments have been carried out to measure the scattering distributions of electrons from low pressure gases. In both experiments, image plates were used to record the spatial distributions of the scattered electrons averaged over many shots. The image plates measure the electron dose over a given area, that is, the convolved energy and flux of electrons. Despite only measuring dose, they offer several advantages over other methods given their high signal to noise ratio, dynamic range, aperture size, and overall cost per unit detector area. This allows us to image a large area of the electron distribution and is suitable for both multi-shot and possibly single-shot data acquisition. Fujifilm BAS MS and SR type \cite{Boutoux15,Jackson:100964} image plates were used given their high sensitivity to electrons in the keV to MeV ranges, as well as long signal decay times \cite{Boutoux15}. The general method of using and characterizing image plates to capture electrons has been explored in our complementary paper \cite{Ravichandran23}. 

Given that the image plates are sensitive to the laser light, low energy electrons and laser photons are filtered out using an aluminum foil mounted flush to the image plate. For these experiments, the image plates were mounted approximately $d\approx40$mm  from the laser focus and as such, the laser intensity was still sufficient to ablate the aluminum foil. As a result, a hole was cut in the center of the image plate and aluminum foil to allow the laser to pass through. This configuration restricts us to interactions where the electrons can only be scattered outside of the laser cone, however, one could place the image plate further away such that the aluminum foil is below the ablation threshold to avoid cutting a hole at a cost of signal strength. 

\begin{figure}
    \centering
    \includegraphics{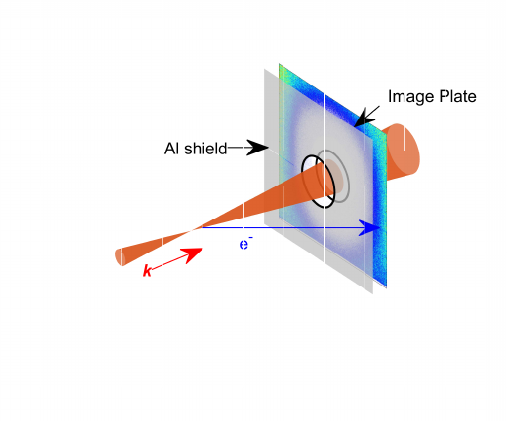}
    \caption{Illustration of the experimental setup. An image plate is placed a distance $d$ from the estimated laser focal plane. An aluminum shield is placed in front of the image plate to prevent laser light from interacting with the image plate, and to reduce general background and x-ray noise.} 
    \label{fig6}
\end{figure}

As the electrons pass through the aluminum foil, they can lose energy, produce bremsstrahlung, and scatter within the foil adding both background and an uncertainty to our measurement. It is therefore desirable to keep the aluminum foil thin to reduce these effects. We can estimate the electron stopping power based on the aluminum thickness and using the NIST stopping power database \cite{NIST_PML}. Given that we expect electrons with energies well above 100keV, we can use foils on the order of 10-100$\mu$m thick to stop low energy electrons and sufficiently block laser photons. At this thickness, we expect the additional uncertainty introduced into the electron scattering angle to be minimal relative to the other experimental uncertainties. The basic experimental arrangement has been illustrated in Fig. \ref{fig6}. 

To estimate the laser intensity using the standard indirect method, images were acquired of the focal spot in a reduced power mode using a long-working distance objective and a charge coupled device (CCD) camera. The magnification of the imaging system is determined using a known calibration allowing us to determine the effective focal spot area and distribution. We normalize the image of the focal spot to acquire an amplitude magnitude function for the laser focal spot profile, 
\begin{equation} \label{eq16}
    \left|\psi_{CCD}\right|^2 = \frac{I_{CCD}}{\sum\sum I_{CCD} \Delta x\Delta y},
\end{equation}
where $\Delta x$ and $\Delta y$ are the magnified CCD pixel width and height respectively, and $I_{CCD}$ is the raw far-field image values.

We note that while we can retrieve the magnitude of the focal spot amplitude, we cannot retrieve its phase with a far-field image alone and require a second measurement to retrieve its phase such as a wavefront sensor, or a near-field amplitude measurement and then using a phase retrieval algorithm such as the Gerchberg-Saxton algorithm \cite{Fienup:82}. 

Normalizing the relative intensity map then allows us to infer the peak laser intensity using the relation \cite{He:19}, 
\begin{equation} \label{eq17}
    I_{infer}\approx\frac{U_{meas}}{\tau_{meas}}\left|\psi_{CCD}\right|^2,
\end{equation}
where $U_{meas}$ is the measured laser energy at full power, and $\tau_{meas}$ is the measured full-width-full-maximum of the temporal intensity envelope, either at full power of a reduced power. The laser energy can be measured at full power upstream from the laser focus using leakage from a mirror and a pre-calibrated ratio of the energy in focus relative to the measurement position. The pulse duration is usually measured using an auto-correlator or a higher order measurement such as a frequency resolved optical gating (FROG) \cite{Trebino}. These measurements often use a small pick-off mirror placed in the beam so as not to disperse the pulse through transmissive optics. Estimating the pulse duration using this method is not trivial as the algorithms for pulse retrieval can have large uncertainties, as well as the pulse duration can change spatially within the beam if spatio-temporally aberrated. 

We emphasize here that knowledge of the pulse energy, pulse duration, and the focal spot profile is needed to infer the peak intensity of the laser focus. These three measurements each have an uncertainty that can increase nonlinearly when operating at full power. 

\subsection{Scarlet Laser, The Ohio State University}
\begin{figure*}
    \centering
    \includegraphics{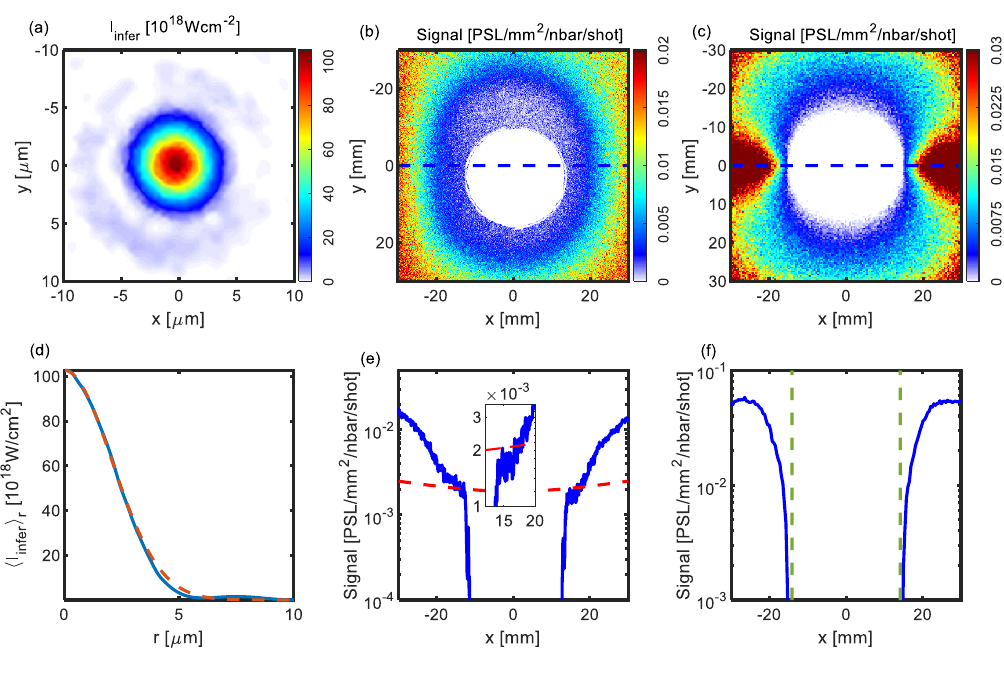}
    \caption{Experimental results from the Scarlet laser operating at $f/6$ at The Ohio State University. (a) Measured experimental focal spot at reduced power and inferred intensity, (b) experimental Fujifilm BAS-MS image plate data normalized to area, vacuum pressure, and number of laser shots, (c) simulated image plate response, (d) average radial lineout of experimental focal spot fitted with a red dashed Gaussian with beam waist $w_0=4.1\mu$m, (e) average radial lineout of experimental image plate data in blue and minimum electron response with red-dashed line, (f) horizontal lineout of simulation data in blue, and the cutoff position plotted in the green dashed line.} 
    \label{fig7}
\end{figure*}

One experiment was carried out at the Scarlet laser facility at The Ohio State University. At the time, this laser was able to deliver 5.5J laser pulses in 35fs at 1 shot/min repetition rate with a central wavelength of 800nm. The laser utilizes an $f/2$ parabola making the focal spot non-paraxial. For this reason, we introduced apodizers in to the beam near-field to give additional control over the laser f-number and consequently the beam waist and peak intensity. For this work, we only consider the weakly paraxial experimental results in which we used an $f/6$ apodizer. The near-field of the laser was apodized to a produce a quasi-uniform illumination across the near-field resulting in a close-to-perfect Airy focal spot absent of any significant aberrations. The indirectly measured focal spot profile and radial lineout is given in Fig.\ref{fig7}(a) and (d) respectively.

The laser energy on target was reduced when apodized to an average value of $1.07\pm0.02$J over 70 shots. An average pulse duration was estimated to be $36$fs before the experiment, however it was not measured on shot due to the invasive nature of the diagnostic. Measured drifts from throughout the campaign found fluctuations in the pulse duration on the order of $10\%$. Using this data and the measured image of the focal spot at a reduced power in Fig.\ref{fig7}(a) in which we estimate an 18$\%$ uncertainty in our image resolution calibration, we infer the peak intensity of the focal spot to be approximately $I_{0,infer}=10.5\pm2.9\times10^{19}$Wcm$^{-2}$. This corresponds to $a_{0,infer}\approx7.0\pm1.0$. The beam waist at best focus can be estimated by curve fitting the average radial lineout of the focal spot as shown in Fig.\ref{fig7}(d) where the blue solid line represents the horizontal lineout of \ref{fig7}(a), and the red dotted line gives a Gaussian fit. We find a best fit beam waist value of $w_0=4.1\mu$m. 

 \begin{figure}
    \centering
    \includegraphics{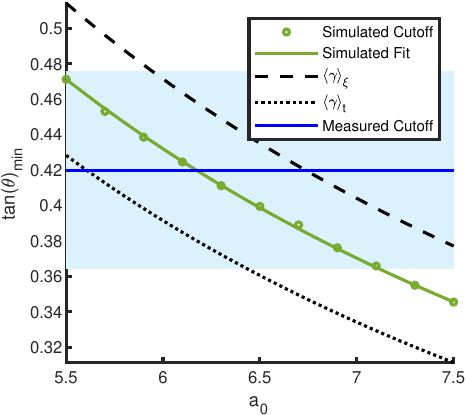}
    \caption{Electron scattering angle - laser intensity relations as measured and estimated for the Scarlet laser system operating at $f/6$. The green circles represent simulation cutoff angles with parameters $w_0=4.1\mu$m, $\tau=36$fs, and $\lambda=800$nm in argon gas for $a_0=5.5:7.5$. The green solid line is the numerical fit to the simulation data with parameters $\tan(\theta)_{min}=2.592/a_0$. The black dashed line and black dotted line give the analytic relations as calculated from (\ref{eq9}) and (\ref{eq10}) respectively. The blue solid line is the measured cutoff angle from Fig.\ref{fig7}(e) and associated error in the shaded blue region. } 
    \label{fig7a}
\end{figure}

The experiment was carried out in a vacuum chamber pumped out to a pressure of approximately $1\times10^{-6}$mbar, and then back filled with argon gas up to a pressure of $~8\times10^{-5}$mbar = 80nbar. To maintain a relatively constant pressure of argon, a slow leak valve was mounted to the chamber wall and adjusted such that the leak rate and the turbo molecular pump rate was approximately equal. 

An 80mm wide, 75mm high Fujifilm BAS-MS image plate was covered with 16$\mu$m of aluminum foil and placed $35\pm2$mm away from the assumed focal plane. A 25mm diameter hole was cut into the image plate and shield to allow the laser to pass unobstructed. This shield is expected to stop electrons below 50keV from reaching the image plate. After being exposed to 70 shots from the laser corresponding to a total of 70 minutes, the chamber was vented and the image plate was scanned 71 minutes after the last laser shot. We estimate that the signal on the image plate will drop to approximately 90$\%$ of its original value in this time \cite{Boutoux15}. The image plate was scanned at a 50$\mu$m resolution, and latitude 5, using a GE Typhoon 7000 image plate scanner. The sensitivity of this reader was not absolutely calibrated and so we estimate our sensitivity value at 10000 based on the photo multiplier tube voltage and published calibrations for identical scanners \cite{Williams14}. The raw image plate $.gel$ files were then converted to photo-stimulated luminescence (PSL) units using the formula \cite{Williams14},
\begin{equation} \label{eq18}
    [PSL]=\left(\frac{G}{2^{16}-1}\right)^2\left(\frac{\mathcal{R} [\mu m]}{100}\right)^2\frac{4000}{S}10^{L/2}. 
\end{equation}
where $G$ is the raw image values, $\mathcal{R}$ is the scanned resolution, $S$ is the sensitivity, and $L$ is the latitude. 

Fig.\ref{fig7}(b) gives the image plate response in PSL normalized to the number of laser shots, the pressure of the chamber, and the image resolution. Despite our best efforts, we found that the hole cut in the image plate did not concentrically align with the cone of the scattered electrons and has resulted in the image plate hole overlapping with some of the cut-off of the electron signal making a precise measurement of the laser intensity difficult. The image has been approximately centered about the scattered electron data given that the simulations indicate the electron cut-off should be symmetric about the laser $\hat{k}$ vector. We note the eccentricity in the electron cut-off radius with the gradient of the signal being strongest in the horizontal (electric field) plane. 

Given the eccentricity of the electron cut-off ring, and the steepest gradient of the electron cut-off occurring in the polarization plane, we opt to take a single horizontal lineout of the image plate response plotted in Fig.\ref{fig7}(e) together with a red dashed line representing the scaled PSL dose of a single electron on the image plate estimated to be $1.9\times10^{-3}$ PSL/electron/nbar/mm$^2$, 1 hour after the final exposure \cite{Boutoux15}. As electrons passing through the image plate at an oblique angle of incidence will deposit more energy, the line is corrected by a factor of $\sec(\theta)=\sqrt{x^2+d^2}/d$. Energy deposited below this line likely corresponds to background x-ray noise. Inset into (e) is a zoomed area of the intersection of the electron signal with the minimum electron dose shown by the red dashed line. Here, we can just see a ledge where the minimum electron signal is expected to occur indicating that this is likely the cut-off of the electrons. The intersection of the lineout and the minimum electron response line occurs at $x=-13.5\pm1.5$mm and $x=16.5\pm1.0$mm. The asymmetry of the two results is likely due to a non-centered choice of center coordinates in the analysis.

Fig.\ref{fig7a} plots the tangent of the minimum scattering angle of the electrons as a function of the laser intensity. Using the relationships derived previously, we plot the $\langle\gamma\rangle_{\xi}$ and $\langle\gamma\rangle_{t}$ scalings using the black dashed and black dotted lines respectively. We have run 10 simulations with parameters $w_0=4.1\mu$m, $\tau=36$fs, and $\lambda=800$nm in argon gas for $a_0=5.5:7.5$ plotted in Fig.\ref{fig7a} using the green circles. A numerical fit given by $\tan(\theta)_{min}=2.592/a_0$ has been overlaid using the solid green line. We remark that the fit lies comfortably between the two analytic models allowing for one to use the models as upper and lower bounds for the intensity via back-of-the-envelope calculations. The measured cutoff radius of the electrons can be related to the cutoff angle via the relation, 
\begin{equation}\label{eq19}
    \tan(\theta)_{min}=r_c/d
\end{equation}
where the distance $d=35\pm2$ and the average value of our cut-off distances $r_c=15.0\pm1.8$mm have been plotted using the blue horizontal line and shaded uncertainty region in Fig.\ref{fig7a}. Finding the intersection of the measured cutoff angle with the numerical model gives a direct measurement of the laser intensity of $a_{0,meas}=6.2\pm0.8$ corresponding to a peak intensity measurement of $I_{0,meas}=8.2\pm2.2\times10^{19}$Wcm$^{-2}$. This agrees well with our inferred measurement and is certainly within the error bounds. We believe that with more accurate positioning of the image plate along $d$, and with a more suitable chosen image plate hole size, we can significantly improve the uncertainty in our direct measurement of the peak intensity. The results have been summarized in Table \ref{table2}. 

An example simulation with $w_0=4.1\mu$m, $\tau=36$fs, and $\lambda=800$nm in argon gas for $a_0=6.2$ is shown in Fig.\ref{fig7}(c) and (f). We have simulated the PSL response of the MS image plate using the known single electron PSL doses as a function of energy and introduced these doses as weights into the bivariate histogram \cite{Boutoux15}. Electrons ejected with energies below 50keV were removed as to simulate the stopping power of the aluminum foil. We reproduce the experimental image plate response well, including the oval shape, but we do not experimentally observe the polarization lobes present in the simulation. The absence of the lobes experimentally could be an indication of deviations from the electric and magnetic field models of the focal spot region even in the weakly paraxial regime. The simulated dose is roughly 50\% times that measured which we attribute to several factors. The image plate signal decay was not modelled in the simulation, background noise from x-rays was not included, and reduced flux and energy from the aluminum foil was not included. Additionally, we had not calibrated the vacuum pressure gauge for use with argon in the experiment and could have been giving an incorrect reading. A horizontal lineout of the simulated response is shown in Fig.\ref{fig7}(f) plotted in blue, and the cutoff radius plotted in the two green vertical dashed lines.

\begin{table}
\caption{\label{table2} Summary of experimental results for the campaigns at the Scarlet laser operating at $f/6$, and the VEGA3 laser at $f/10$.}
\begin{ruledtabular}
\begin{tabular}{lllll}
Laser   &  $U_{meas}$[J]    &   $\tau_{meas}$[fs] & $I_{infer}$[Wcm$^{-2}$] & $I_{meas}$[Wcm$^{-2}$]\\
\hline
Scarlet &  $1.07\pm0.02$  &  $36\pm4$  &   $10.5\pm2.9\times10^{19}$  & $8.2\pm2.2\times10^{19}$\\
VEGA3 &  $23.0\pm0.2$  &  $35.4\pm6.6$  &   $18.4\pm3.5\times10^{19}$  & $9.1\pm0.8\times10^{19}$\\
\end{tabular}
\end{ruledtabular}
\end{table}

\subsection{VEGA3 Laser, Centro de L\'{a}seres Pulsados (CLPU)}
Another experiment was carried out using the VEGA3 PW laser at the Centro de L\'{a}seres Pulsados (CLPU) in Salamanca, Spain \cite{volpe_2019}. This laser can deliver $30$J, 30fs pulses to the focal plane at 1Hz repetition rate. The laser is focused using an $f/10$ parabola which is well within the paraxial regime. 100 shots were delivered on target with an energy of $23.0\pm0.2$J and a pulse duration of approximately $35.4\pm6.6$fs. The temporal pulse width was measured by an autocorrelator outside of the vacuum chamber and so uncertainties in the pulse measurement were introduced due to group velocity dispersion in the vacuum window and other optics in the beam line. Experimental details, methods, and additional results of this experimental campaign are given in more detail in our complementary paper \cite{Ravichandran23}. 

\begin{figure*}
    \centering
    \includegraphics{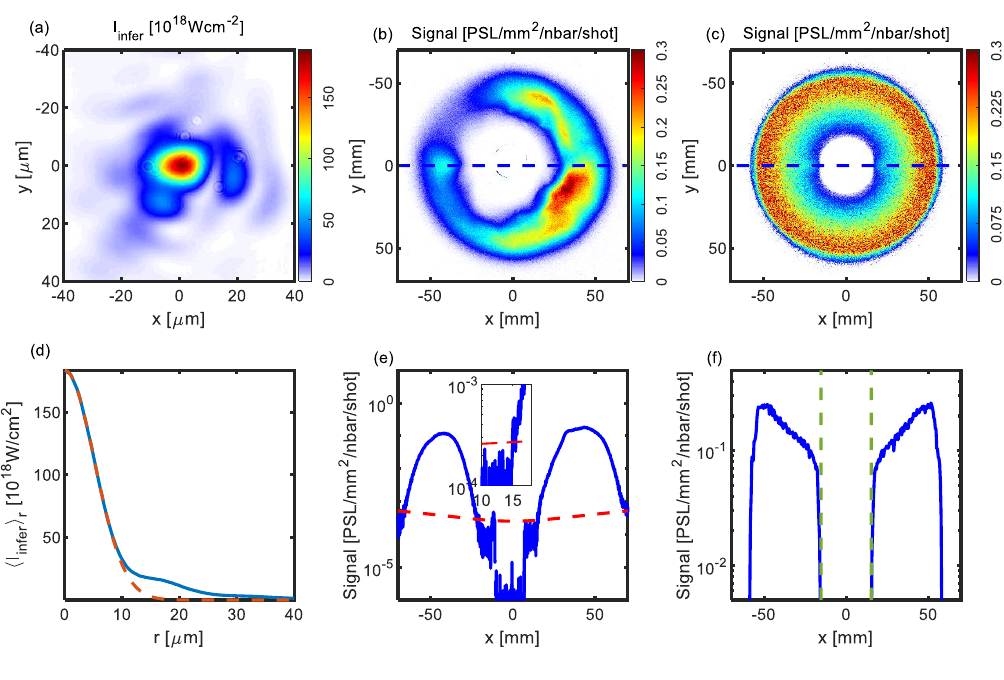}
    \caption{Experimental results from the VEGA3 laser at the Centro de L\'{a}seres Pulsados (CLPU) in Salamanca, Spain. (a) Measured experimental focal spot at reduced power and inferred intensity, (b) experimental Fujifilm BAS-SR image plate data normalized to area, vacuum pressure, and number of laser shots, (c) simulated image plate response, (d) average radial lineout of experimental focal spot fitted with a red-dashed Gaussian with beam waist $w_0=10.8\mu$m, (e) average horizontal lineout of experimental image plate data in blue and minimum electron response with red dashed line, (f) horizontal lineout of simulation data in blue, and the cutoff position plotted in the green dashed line.} 
    \label{fig8}
\end{figure*}

An image of the focal spot is given in Fig.\ref{fig8}(a) highlighting a more aberrated beam in the system. An average radial lineout of the focal spot is given in Fig.\ref{fig8}(d) where we have fitted a Gaussian with a beam waist of approximately $10.8\mu$m. Using these parameters, we infer a peak intensity in the focal plane $I_{0,infer}=18.4\pm3.5\times10^{19}$Wcm$^{-2}$. 

In this experiment a larger 130mm $\times$ 130mm Fujifilm BAS-SR image plate was deployed wrapped with $520\mu$m of aluminum and placed $39\pm1$mm from the assumed focal plane. This shield is expected to stop electrons below $\approx$500keV from reaching the image plate. Nitrogen gas was leaked into the chamber in a similar way to before and maintained at a target chamber pressure of $4\times10^{-5}$mbar = 40nbar. After being exposed to 100 shots from the laser, the chamber was vented and the image plate was scanned approximately 1 hour after the last laser shot. We estimate the signal will drop to roughly 40$\%$ of its original value in this time \cite{Boutoux15}. The image plate was scanned at a $50\mu$m resolution using an Amersham Typhoon 7000 image plate scanning with a sensitivity of 4000, and latitude 5. 

The image plate result is shown in Fig.\ref{fig8}(b). Unlike the Scarlet experiment, we find a circular ring where the outer cut-off radius is governed by the thickness of the aluminum and corresponds to (\ref{eq5}) with 500keV energies. We find that the electron ring is highly structured with a crescent half moon type shape. We believe this asymmetry in the ring is related to the aberrations present in the focal spot which we will discuss in the next section. Similar to the Scarlet data, we find that our scattered electron ring did not concentrically align with the image plate hole as can be seen in Fig.\ref{fig8}(b). Due to the inner electron ring cut-off being aberrated, we opt to use the outer electron cut-off ring to find the center coordinate of the scattering distribution. We will justify this choice in the next section. 

In a similar method to the Scarlet experiment, we restrict our analysis to a single horizontal lineout along the electric field direction of the image plate ring. The lineout is plotted in Fig.\ref{fig8}(e), again scaled with the minimum PSL response for a single electron on a BAS-SR type image plate and the oblique incidence correction factor shown by the red dashed line \cite{Boutoux15}. We find a cleaner background signal below the minimum expected electron signal allowing us to make a more precise measurement. The intercept between the horizontal lineout and the normalized single electron PSL value is found to be $15.0\pm0.5$mm on the right hand side, and $-21\pm1$mm on the left-hand side. Given the ring asymmetry, we will not average this result and instead use only the smaller of the two cut-offs. This reasoning is due in part to our certainty in the center of the ring from the outside cut-off edge, and also due to the role of the aberrations in the scattering as we will investigate in the next section. 

 \begin{figure}
    \centering
    \includegraphics{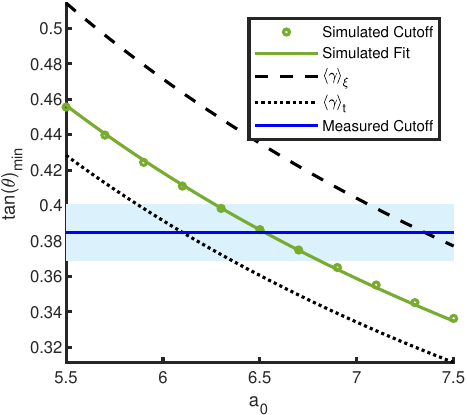}
    \caption{Electron scattering angle - laser intensity relations as measured and estimated for the VEGA3 laser system. The green circles represent simulation cutoff angles with parameters $w_0=10.8\mu$m, $\tau=35.4$fs, and $\lambda=800$nm in nitrogen gas for $a_0=5.5:7.5$. The green solid line is the numerical fit to the simulation data with parameters $\tan(\theta)_{min}=2.511/a_0$. The black dashed line and black dotted line give the analytic relations as calculated from (\ref{eq9}) and (\ref{eq10}) respectively. The blue solid line is the measured cutoff angle from Fig.\ref{fig8}(e) and associated error in the shaded blue region. } 
    \label{fig8a}
\end{figure}

An example simulation with parameters $w_0=10.8\mu$m, $\tau=35.4$fs, and $\lambda=800$nm in nitrogen gas for $a_0=6.5$ is given in Fig.\ref{fig8}(c). We have simulated the PSL response of the SR image plate again using known single electron PSL doses as a function of energy and introduced these doses as weights into the bivariate histogram \cite{Boutoux15}. The simulation was run for 10 million electrons removing ejected electrons below 500keV to simulate the $520\mu$m aluminum foil. We see a similar ring to what is measured experimentally, but without any modulations around the ring azimuthally. The outside cut-off radius looks to agree well with the experimental result, and we find the simulated peak dose to be approximately equal to that measured. A horizontal lineout is plotted in Fig.\ref{fig8}(f) and the electron cutoff radii are shown by the green dashed lines. 

As before, Fig.\ref{fig8a} plots the tangent of the minimum scattering angle of the electrons as a function of the laser intensity. 10 simulations were run with parameters $w_0=10.8\mu$m, $\tau=35.4$fs, and $\lambda=800$nm in nitrogen gas for $a_0=5.5:7.5$ and are plotted in Fig.\ref{fig8a} using the green circles. Numerically fitting the data, we find a simulated angle-intensity relationship of $\tan(\theta)_{min}=2.511/a_0$. Again, we see the simulated data lies between our two models indicating that they can be used to calculate upper and lower bounds of the intensity analytically. Using our experimental values of $d$ and $r_c$ with (\ref{eq19}), we plot the blue line with the associated uncertainty in the shaded blue. The intercept of the measured cutoff and the simulated cutoff occurs at $a_{0,meas}=6.5\pm0.3$. This corresponds to a peak intensity of $I_{0,meas}=9.1\pm0.8\times10^{19}$Wcm$^{-2}$, considerably less than the inferred value of $I_{0,infer}=18.4\pm3.5\times10^{19}$Wcm$^{-2}$. This discrepancy is likely due to the aberrations in the beam worsening as the laser power is increased along with suspected fluctuations in the pulse duration. Additional lineouts at other angles across the image were also taken, but no decrease in cut-off radius was found. Given that these shots were integrated over 100 shots, if the laser was decreasing its intensity due to nonlinear effects in the beam line such as heating in optics from firing continuously at rep rate, the latter data may have dominated the averaging of the shots as early data recorded on the image plate will have decayed in time more.

For both experiments, the image plates integrated electrons over many shots which can lead to a reduction in the accuracy of the measurement. Ideally, we want to minimize the number of shots integrated as to yield the best estimate for the intensity for any given shot. We can estimate the electron dose scaling based on a simple geometry argument. Given that at these ultra-low pressures the electrons are non-interacting, we can assume that the signal strength will scale linearly with background pressure in the chamber. Likewise, increasing the intensity of the laser will increase its effective ionization volume. Increasing the beam waist will increase the areal density as the beam waist squared, and also increase the effective length over which the laser will maintain its intensity (Rayleigh range) $z_0\propto w_0^2$. This allows us to write the dose strength as a function of the base laser parameters, 
\begin{equation} \label{eq19a}
    |PSL|\propto n_eI_0w_0^4 \propto \wp P\lambda^2 f_{\#}^2,
\end{equation}
where $n_e$ is the electron density from the highest energy shell ionized, $P$ is the laser power, and $\wp$ is the background pressure. However, $w_0$ itself is constrained by the requirement to reach the intensity $I_{ion}$ to achieve barrier suppression ionization for a given laser power as given by $w_{ion}\leq \sqrt{2P/\pi I_{ion}}$. Given that the laser power, wavelength, and f-number are typically fixed parameters for a system, the variables to adjust would be the background pressure and the gas species. Gases such as krypton and xenon are useful alternatives given their large electron numbers, and can be introduced into the chamber with few safety concerns. Gaseous mixes could also be explored to provide a wide variety of ionization potentials across a large intensity range. 

We note that increasing the background gas pressure in the vacuum could lead to nonlinear effects if too high. We estimate that at pressures above $10^{-2}$mbar, collective effects could begin to play a role in the interaction and although it could still be possible to make measurements of the intensity, it would complicate the analysis. Previous works have investigated the interaction of high intensity lasers with low density plasmas as a method for determining the peak laser intensity \cite{Vais2018}. 

In summary, we have demonstrated that it is possible to measure the intensity of the laser spot experimentally using image plates: a cost effective, simple, and robust diagnostic. While our experimental measurements have been preliminary, we believe them to be successful and have emphasized that further work is needed to refine the experimental technique, and also highlighted the need for further numerical modelling. 

The two experiments, despite having similar intensities and pulse durations, have extremely different results owing much to the f-number of the system and the aberrations in the VEGA3 beam. We believe this motivates the need to include non-paraxial models of the fields, and also to include aberrations similar to what was observed on VEGA3.

We believe that single shot measurements of the electrons could in principle be possible. Given the pressure used in these experiments and the signal strength on the image plates, one could reduce the number of shots by a factor of 100 and increase the background pressure by a factor of 10 and still have sufficient signal. Utilizing the simulations beforehand, one can predict within a factor of 2 the signal strength, and expected spatial distribution of the electrons allowing for informed decisions of the image plate placement, and expected yields for a given number of shots on a particular laser system.

\section{Scattering from aberrated laser pulses}
As we have seen experimentally, a laser rarely has a perfect Gaussian spatial profile unless heavily corrected or apodized and will likely contain aberrations from the laser near-field profile, misalignment's in the optical system, and any thermal distortions from the optics when operating at full power. We have found numerically and experimentally that these aberrations play a fundamental role in the scattering of electrons. The role of aberrations in various laser-plasma experiments is relatively unexplored. However, there are certain works that claim enhanced effects due to aberrations in acceleration schemes such as wakefield acceleration \cite{Cummings11}, proton acceleration \cite{Grace22}, and vacuum ponderomotive acceleration \cite{Wilhelm19}. 

Spatial aberrations are often described using the Zernike polynomials as they form a complete orthogonal basis set on the unit circle that can be applied to the laser near-field aperture \cite{BornWolf99}. In this work, we omit the contributions of temporal, and spatio-temporal aberrations for brevity but acknowledge that they can be equally important in the electron scattering dynamics and will investigate them in future work \cite{Palastro20}. 

To model spatial aberrations, we write the near-field complex amplitude as, 
\begin{equation} \label{eq20}
    \Psi(\rho,\phi)=U_1(\rho,\phi)\prod_{m,n}\exp\left(ik\alpha_{m,n}Z_n^m(\rho,\phi)\right)
\end{equation}
where $U_1(\rho,\phi)$ is the near-field spatial amplitude (often a Gaussian, super-Gaussian, or a circle), $\alpha_{m,n}$ is the aberration strength in units of $\lambda_0$, and $Z_m^n(\rho,\phi)$ is the Zernike polynomial given in Table \ref{table3}. We note that the unit circle in the near-field can be arbitrarily specified relative to the size of the near-field profile if it is a Gaussian or super-Gaussian. Therefore, we normalize the Zernike unit circle radius to the near-field beam waist $R_0$ such that $U_1(\rho,\phi)\propto\exp\left(-\rho^2/R_0^2\right)^N$, where $N$ is the super-Gaussian parameter.

\begin{figure}
    \centering
    \includegraphics{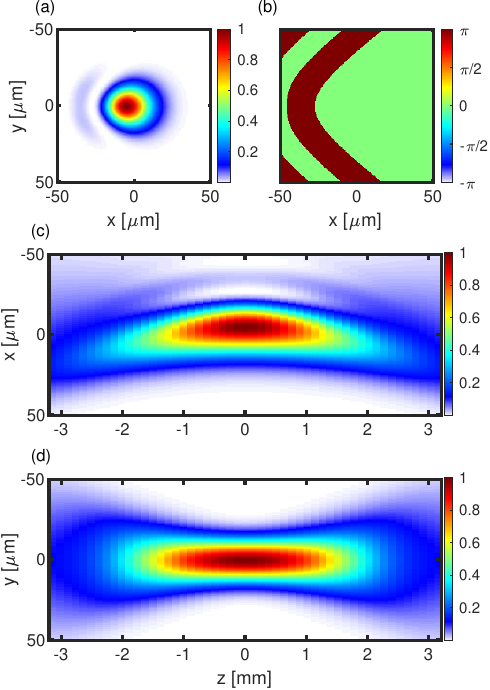}
    \caption{Aberrated focal spot profile with horizontal coma $\alpha_{1,3}=0.05\lambda$, Gaussian near-field, and focused by a $f/10$ optic. (a) Normalized focal spot intensity distribution in the focal plane ($z=0$), (b) focal spot phase profile in the focal plane ($z=0$), (c) normalized longitudinal intensity slice in the $\hat{x}-\hat{z}$ plane, (d) normalized longitudinal intensity slice in the $\hat{y}-\hat{z}$ plane.} 
    \label{fig9}
\end{figure}

\begin{table}
\caption{\label{table3} List of common aberrations found in high power laser systems \cite{BornWolf99}.}
\begin{ruledtabular}
\begin{tabular}{ccc}
$Z_m^n$   &  Name    &  Function  \\
\hline
$Z_2^{-2}$ &  Oblique Astigmatism  &  $\rho^2\sin(2\phi)$  \\
$Z_2^{2}$ &  Vertical Astigmatism  &  $\rho^2\cos(2\phi)$  \\
$Z_3^{-1}$ &  Vertical Coma  &  $(3\rho^3 - 2\rho)\sin(\phi)$  \\
$Z_3^{1}$ &  Horizontal Coma  &  $(3\rho^3 - 2\rho)\cos(\phi)$  \\
$Z_3^{-3}$ &  Vertical Trefoil  &  $\rho^3\sin(3\phi)$  \\
$Z_3^{3}$ &  Oblique Trefoil  &  $\rho^3\cos(3\phi)$  \\

\end{tabular}
\end{ruledtabular}
\end{table}

\begin{figure}
    \centering
    \includegraphics{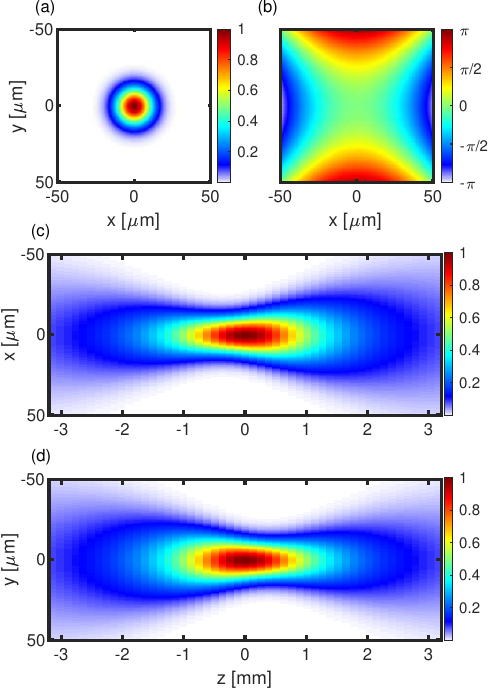}
    \caption{Aberrated focal spot profile with oblique astigmatism $\alpha_{-2,2}=0.05\lambda$, Gaussian near-field, and focused by a $f/10$ optic. (a) Normalized focal spot intensity distribution in the focal plane ($z=0$), (b) focal spot phase profile in the focal plane ($z=0$), (c) normalized longitudinal intensity slice in the $\hat{x}-\hat{z}$ plane, (d) normalized longitudinal intensity slice in the $\hat{y}-\hat{z}$ plane.} 
    \label{fig10}
\end{figure}

Using the angular spectrum method \cite{goodman2005}, we can calculate the three dimensional far-field complex amplitude in the paraxial regime as, 
\begin{equation} \label{eq21}
    \psi_\delta(x,y,z)\approx\mathfrak{F}^{-1}\left\{\Psi(\rho,\phi)\exp\left(\frac{ik_{\perp}^2}{2k_0}z\right)\right\}
\end{equation}
where the subscript $\delta$ denotes the aberrated far-field complex amplitude, and $\mathfrak{F}^{-1}$ is the inverse Fourier transform. Several far-field profiles are plotted in Fig.\ref{fig9} and \ref{fig10} for the case of a horizontal coma of strength $\alpha_{1,3}=0.05\lambda$, and an oblique astigmatism of strength $\alpha_{-2,2}=0.05\lambda$ respectively assuming a purely Gaussian near-field $(N=1)$, and an $f/10$ focusing optic. In Fig.\ref{fig9} and \ref{fig10}, we plot the normalized intensity in the focal plane in (a), and the the corresponding phase in (b). (c) and (d) give longitudinal slices of the normalized intensity profile in the $\hat{x}-\hat{z}$ plane and the $\hat{y}-\hat{z}$ plane respectively. In the case of coma in Fig.\ref{fig9}, we find that the beam essentially bends as it propagates through space. As a result, the ponderomotive force that acts radially outwards is not only asymmetric in the focal plane, but changes as a function of longitudinal position. 

 \begin{figure*}
    \centering
    \includegraphics{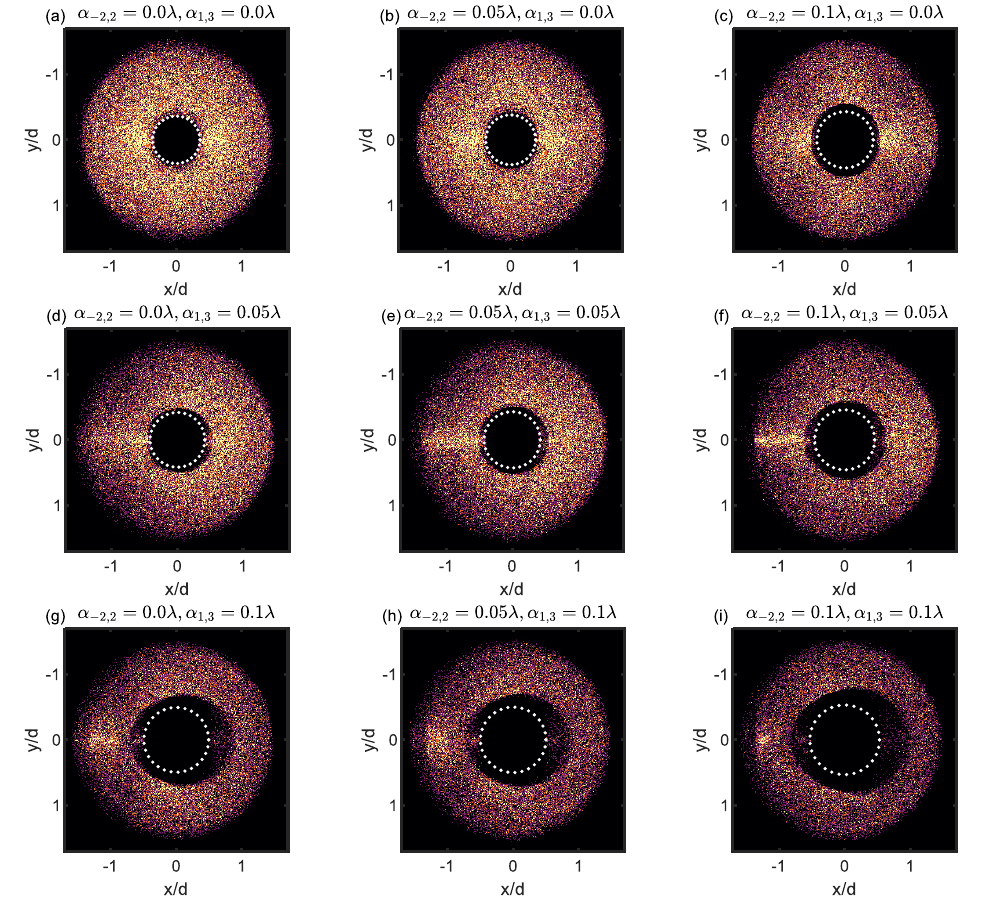}
    \caption{Simulation results of electrons ionized and scattering from aberrated laser focii in nitrogen gas with laser parameters $P=500$TW, $\lambda=800$nm, $f_{\#}\approx20$, and $\tau=30$fs. Electron energies below 500keV are removed to emulate an aluminum shield. The color map of each simulation result is relative to one another. Aberration information of each simulation is listed as subplot titles where $\alpha_{-2,2}$ gives the astigmatism strength, and $\alpha_{1,3}$ gives the coma strength in units of $\lambda$. The white-dotted circles in each tile represents the electron cutoff radius calculated from (\ref{eq9}) for each of aberrated beams.}  
    \label{fig11}
\end{figure*}

In the case of astigmatism in Fig.\ref{fig10}, the beam propagates in a straight line, but in this case the focal spot becomes elliptical outside of the focal plane as well as rotating 90 degrees before and after focus. It is not clear at this stage how these aberrations will affect the scattering of electrons but given that the ponderomotive force becomes asymmetric in both cases, we can predict that the aberrations may be imprinted into the scattered electron distribution. 

To integrate the Lorentz force in (\ref{eq1}) efficiently, we need an analytic description of the field equations in and around the laser focus. The result of (\ref{eq21}) only gives the field amplitude and phase, not the electric and magnetic field equations. We could use (\ref{eq12}) to numerically calculate the fields from our aberrated near-field and interpolate the fields at each time step within the Runge-Kutta, however we have found this to add a large numerical overhead. 

An alternative approach would be to decompose the aberrated amplitude and phase into the previously given Laguerre-Gaussian (LG) functions (\ref{eq13}). We use a modal decomposition into the LG basis set given by \cite{Longman:17},
\begin{equation} \label{eq22}
    \eta_{\ell,p}=\braket{\psi_p^\ell|\psi_\delta}.
\end{equation}
This requires the aberrated wavefunction is normalized such that $\braket{\psi_\delta|\psi_\delta}=1$. The aberrated focal spot can then be reconstructed via summing over all LG modes weighted by the $\eta_{\ell,p}$ parameter. For completeness, this is written as, 
\begin{equation}
    \psi_{total}(r,\theta)=\sum_{\ell,p}\eta_{\ell,p}\psi_p^\ell. 
\end{equation}
The electric and magnetic fields are then computed as a linear superposition of the electric and magnetic fields of each mode. Using this technique, we find that the number of modes required for the decomposition can become large if the aberration is strong, so we restrict ourselves to weakly aberrated beams where $\alpha_{m,n}\leq0.1\lambda_0$.

Simulations using aberrated focal spots were run for various strengths of coma and astigmatism, the two most common types of beam aberration found in the laboratory. The scattered distributions of 500000 electrons scattered into the screen placed at $z=d$ is shown in Fig.\ref{fig11}. The figure is broken into 9 simulations where oblique astigmatism is increased horizontally as you move left to right, and horizontal coma is increased as you move vertically from top to bottom. All images utilize the same color scale. The laser parameters used in all simulations is $\lambda=800nm$, $P=500$TW, $f_{\#}\approx20$, $\tau=30$fs, nitrogen gas, and cutting off electrons below 500keV, where $P\approx U/\tau$ is the laser power. For the aberrated beam simulations we opt to keep the laser power constant as the amplitude of each mode in the composition varies from aberration to aberration.  

The plot in Fig.\ref{fig11} shows two unique characteristics of the scattering distributions of the electrons with aberrated beams. As the astigmatism is increased in Fig.\ref{fig11}(a)-(c), we find the scattering distribution develops lobes of brighter flux in the horizontal and vertical planes. This contrasts the scattering from a laser pulse with coma as shown in Fig.\ref{fig11}(d) and (g) in which we see an asymmetry form in the scattered distribution. On the left side of the distribution we see a bright lobe, whereas on the right side we see a shape that resembles a half moon, similar to what was observed experimentally in Fig.\ref{fig8}(b). Combining the two aberrations produces distinct spatial distributions of the electrons as shown in Fig.\ref{fig11}(e), (f), (h) and (i).

We note that the outside cutoff radius of the low energy electrons remains concentric with the laser $\hat{k}$, independent of the aberration in the beam, and we utilize this observation as a justification for centering the electron ring from VEGA3 using the outside cut-off rather than the highly aberrated inside cut-off angle. Referring to Fig.\ref{fig8}(b), we notice that there is a high level of structure within the electron ring. To simulate this will likely require many laser modes and the introduction of higher order aberrations outside of just the coma and astigmatism as well as spatio-temporal aberrations. To achieve this we are developing new models that can solve for non-paraxial fields with temporal and spatial aberrations similar to other works \cite{Bukharskii22}.  

We note in Fig.\ref{fig11} that as the aberration strength in the beam increases, the apparent inside cut-off angle of the two horizontal lobes increases. This is due to the aberration increasing the effective focal spot size area, and as a result decreases the focal spot intensity. Overlaid in the images are white dotted circles which indicate the theoretical electron cutoff from (\ref{eq9}) and the known peak intensity of the aberrated beam. We note that despite the presence of coma or astigmatism, the ideal position for the cut-off measurement remains in the horizontal polarization plane. 

At this stage we are still some distance away from solving the inverse problem of measuring the aberration given the scattered distribution of electrons. The scattered distribution will likely contain a wealth of information about the focal spot including the intensity, beam waist, spatial aberrations, and possibly temporal information about the laser pulse. For a complete reconstruction, it may be necessary to retrieve the full distribution function of the electrons, that is, their complete position-momentum phase space information \cite{Bukharskii22}. This is in principle possible using spatially resolved spectrometers \cite{Meyerhofer:96}, but to build a device that can measure the spatial and spectral distributions with comparable resolution to the image plate will be challenging. 

We believe that it may be possible to make quantitative estimates on the spatial characteristics of the beam based on the scattered distribution of the electron dose alone. By running a large number of simulations for various laser intensities, beam waists, and aberrations one could construct a suitable training set for a convolutional neural network (CNN). Once suitably trained, the CNN could solve the inverse problem and give quantitative information about the laser spatial profile at full power given the electron distribution on an image plate obtained experimentally. 

In summary, we have qualitatively demonstrated that beam aberrations such as coma and astigmatism imprint distinct angular distributions into the scattered electrons and are likely the cause of the aberrated ring structure found in the VEGA3 data. The mapping between the aberrated focal spot and the scattered electron distribution will likely be highly nonlinear making analytic relations between the two unlikely. We propose the use of machine learning tools to solve the inverse problem through training a convolutional neural network on both simulation and experimental data sets. With a suitably trained network, it could be in practice possible to make quantitative estimates on both the spatial, and temporal aberrations in the beam when operating at full power. Experiments with user specified phase aberrations using spatial light modulators are currently being planned and should provide a verification of our current aberration models, and as training sets for neural networks.  

\section{Conclusions}
With the increasing number of high-power, high-intensity laser facilities worldwide there is an increasing need to develop new methods for accurately and precisely characterizing the laser spatially, spectrally, and temporally both in the near-field, and the far-field. In this work, we have demonstrated the use of electrons ionized and scattered from high-intensity laser pulses propagating through ultra-low density gases as a way to directly measure the laser intensity independently from the laser pulse duration, energy and spatial mode, as well as possibly measuring the laser spatial mode at full-power. 

Extracting information about the laser pulse intensity from the scattered distribution was achieved through an elementary analysis of single electrons scattering from plane waves relating the maximum ejected scattering energy and corresponding scattering angle to the peak intensity of the laser. Three models were developed and verified using numerical modeling of single electrons scattering from paraxial laser pulses highlighting two of the models as candidates to relate the laser intensity to the scattered electron energy and corresponding angle. The models were shown to work well for intensities in the range of $a_0=1:30$, pulse durations $\tau>30$fs at 800nm, and for f-numbers greater than 5. For shorter f-numbers, higher intensities, or shorter pulse durations, higher order corrections and more sophisticated models are likely necessary to measure the intensity. 

The ionization rate was shown to play an important role in the scattered energy of electrons indicating that if the electrons are ionized at intensities much less than the peak laser intensity, they are ponderomotively scattered without effectively sampling the peak intensity. It is therefore necessary to select a gas with ionization levels reasonably close to the peak intensity. We showed through a simplified model that gases such as argon and nitrogen are suitable for measuring intensities of $a_0<10$, while krypton or xenon are more suitable for intensities above this range. More sophisticated models of the ionization rates and injection points are necessary to obtain more precise intensity bounds for a specific gas.  

Two experimental campaigns utilized image plates to capture the convolved electron flux and energy (dose) in a plane approximately 40mm from the focus giving a two dimensional image of the scattered electron distribution. The analytic back-of-the-envelope models and simulations were then applied to the experimental results, the first of which demonstrated a good agreement between the inferred experimental intensity made from indirect measurements to our measured intensity using the direct method. The second experiment exhibited an aberrated focal spot which produced an asymmetric scattered electron distribution. This experiment yielded a measured intensity approximately $50\%$ of that inferred through indirect measurements. We believe this discrepancy is related to the fact that aberrations present in the focal spot observed at low power measurements may have worsened at high power leading to a reduction in the on shot peak intensity.

The measured doses on the image plates integrated from 70-100 shots indicate that single shot measurements for some laser systems may be possible if background pressures are on the order of $10^{-4}$mbar. It was shown through scaling arguments that the signal scales proportionally to the gaseous pressure and the $f_{\#}^2$ indicating that tightly focused beams may require more shots to get suitable signals for measurement. 

Motivated by the experimental results of aberrated beams, we demonstrated qualitative numerical relationships between the focal spot aberrations and the scattered electron distributions in approximate agreement with the experimental results. The mapping from the aberration in the focal plane to the electron distribution at the measurement plane appears to be highly nonlinear and an analytic transform between the two may not exist. We propose the use of machine learning as a possible tool to solve the inverse mapping of the electron distribution to the focal plane. A convolutional neural network trained on both simulation and experimental data sets could be suitable way to make quantitative estimates on the spatial beam aberrations, and hence the laser phase on shot at full power. 

We believe the use of image plates, coupled with our back-of-the-envelope scaling law, is a simple, cost-effective, and robust method to make direct measurements of the focal spot intensity at full power. Many high-power laser facilities are already equipped with image plate scanners allowing for the immediate implementation of the technique. Future work will push towards non-paraxial corrections to our scaling law, radiation reaction effects for intensities approaching and surpassing $a_0=100$, advanced ionization algorithms, and corrections for temporal effects. We believe that as we push towards single-shot, ultra-short (few cycle) laser pulses, additional effects such as the carrier-envelope phase may also begin to play a role, but for now they have been omitted for brevity.

\begin{acknowledgments}
This work was performed under the auspices of the U.S. Department of Energy by Lawrence Livermore National Laboratory under Contract No. DE-AC52-07NA27344, and the DOE Office of Science, Fusion Energy Sciences under Contract No. DE-SC0021231: the LaserNetUS initiative at The Scarlet Laser Facility. G.T, D.H, P.S, N.C, and R.D acknowledge support under Contract Nos. DE-SC0022092 and DE-SC0022094. R.F acknowledges support by the Natural Sciences and Engineering Research Council of Canada (Grant No. RGPIN-2019-05013). S.R, C.Z.H, N.M, and W.T.H acknowledge support by the National Science Foundation (Grant No. PHY2010392). R.L, M.H, and L.R acknowledge support by Laserlab Europe V (Grant No. 871124), and Junta de Castilla y L\'{e}on (Grant No. CLP087U16).  The authors would like to thank and acknowledge D. Schumacher, C. Mendez, E. Chowdhury, and A. Raymond for helpful discussions.  
\end{acknowledgments}

\section*{Conflict of Interest Statement }
The authors have no conflicts to disclose.

\section*{Data Availability Statement}
The data that support the findings of this study are available from the corresponding author upon reasonable request.

\nocite{*}
\bibliography{references}

\providecommand{\noopsort}[1]{}\providecommand{\singleletter}[1]{#1}%
\begin{thebibliography}{79}%
\makeatletter
\providecommand \@ifxundefined [1]{%
 \@ifx{#1\undefined}
}%
\providecommand \@ifnum [1]{%
 \ifnum #1\expandafter \@firstoftwo
 \else \expandafter \@secondoftwo
 \fi
}%
\providecommand \@ifx [1]{%
 \ifx #1\expandafter \@firstoftwo
 \else \expandafter \@secondoftwo
 \fi
}%
\providecommand \natexlab [1]{#1}%
\providecommand \enquote  [1]{``#1''}%
\providecommand \bibnamefont  [1]{#1}%
\providecommand \bibfnamefont [1]{#1}%
\providecommand \citenamefont [1]{#1}%
\providecommand \href@noop [0]{\@secondoftwo}%
\providecommand \href [0]{\begingroup \@sanitize@url \@href}%
\providecommand \@href[1]{\@@startlink{#1}\@@href}%
\providecommand \@@href[1]{\endgroup#1\@@endlink}%
\providecommand \@sanitize@url [0]{\catcode `\\12\catcode `\$12\catcode
  `\&12\catcode `\#12\catcode `\^12\catcode `\_12\catcode `\%12\relax}%
\providecommand \@@startlink[1]{}%
\providecommand \@@endlink[0]{}%
\providecommand \url  [0]{\begingroup\@sanitize@url \@url }%
\providecommand \@url [1]{\endgroup\@href {#1}{\urlprefix }}%
\providecommand \urlprefix  [0]{URL }%
\providecommand \Eprint [0]{\href }%
\providecommand \doibase [0]{http://dx.doi.org/}%
\providecommand \selectlanguage [0]{\@gobble}%
\providecommand \bibinfo  [0]{\@secondoftwo}%
\providecommand \bibfield  [0]{\@secondoftwo}%
\providecommand \translation [1]{[#1]}%
\providecommand \BibitemOpen [0]{}%
\providecommand \bibitemStop [0]{}%
\providecommand \bibitemNoStop [0]{.\EOS\space}%
\providecommand \EOS [0]{\spacefactor3000\relax}%
\providecommand \BibitemShut  [1]{\csname bibitem#1\endcsname}%
\let\auto@bib@innerbib\@empty
\bibitem [{\citenamefont {Zylstra}\ \emph {et~al.}(2022)\citenamefont
  {Zylstra}, \citenamefont {Hurricane}, \citenamefont {Callahan}, \citenamefont
  {Kritcher}, \citenamefont {Ralph}, \citenamefont {Robey}, \citenamefont
  {Ross}, \citenamefont {Young}, \citenamefont {Baker}, \citenamefont {Casey},
  \citenamefont {D{\"o}ppner}, \citenamefont {Divol}, \citenamefont
  {Hohenberger}, \citenamefont {Le~Pape}, \citenamefont {Pak}, \citenamefont
  {Patel}, \citenamefont {Tommasini}, \citenamefont {Ali}, \citenamefont
  {Amendt}, \citenamefont {Atherton}, \citenamefont {Bachmann}, \citenamefont
  {Bailey}, \citenamefont {Benedetti}, \citenamefont {Berzak~Hopkins},
  \citenamefont {Betti}, \citenamefont {Bhandarkar}, \citenamefont {Biener},
  \citenamefont {Bionta}, \citenamefont {Birge}, \citenamefont {Bond},
  \citenamefont {Bradley}, \citenamefont {Braun}, \citenamefont {Briggs},
  \citenamefont {Bruhn}, \citenamefont {Celliers}, \citenamefont {Chang},
  \citenamefont {Chapman}, \citenamefont {Chen}, \citenamefont {Choate},
  \citenamefont {Christopherson}, \citenamefont {Clark}, \citenamefont
  {Crippen}, \citenamefont {Dewald}, \citenamefont {Dittrich}, \citenamefont
  {Edwards}, \citenamefont {Farmer}, \citenamefont {Field}, \citenamefont
  {Fittinghoff}, \citenamefont {Frenje}, \citenamefont {Gaffney}, \citenamefont
  {Gatu~Johnson}, \citenamefont {Glenzer}, \citenamefont {Grim}, \citenamefont
  {Haan}, \citenamefont {Hahn}, \citenamefont {Hall}, \citenamefont {Hammel},
  \citenamefont {Harte}, \citenamefont {Hartouni}, \citenamefont {Heebner},
  \citenamefont {Hernandez}, \citenamefont {Herrmann}, \citenamefont
  {Herrmann}, \citenamefont {Hinkel}, \citenamefont {Ho}, \citenamefont
  {Holder}, \citenamefont {Hsing}, \citenamefont {Huang}, \citenamefont
  {Humbird}, \citenamefont {Izumi}, \citenamefont {Jarrott}, \citenamefont
  {Jeet}, \citenamefont {Jones}, \citenamefont {Kerbel}, \citenamefont {Kerr},
  \citenamefont {Khan}, \citenamefont {Kilkenny}, \citenamefont {Kim},
  \citenamefont {Geppert~Kleinrath}, \citenamefont {Geppert~Kleinrath},
  \citenamefont {Kong}, \citenamefont {Koning}, \citenamefont {Kroll},
  \citenamefont {Kruse}, \citenamefont {Kustowski}, \citenamefont {Landen},
  \citenamefont {Langer}, \citenamefont {Larson}, \citenamefont {Lemos},
  \citenamefont {Lindl}, \citenamefont {Ma}, \citenamefont {MacDonald},
  \citenamefont {MacGowan}, \citenamefont {Mackinnon}, \citenamefont
  {MacLaren}, \citenamefont {MacPhee}, \citenamefont {Marinak}, \citenamefont
  {Mariscal}, \citenamefont {Marley}, \citenamefont {Masse}, \citenamefont
  {Meaney}, \citenamefont {Meezan}, \citenamefont {Michel}, \citenamefont
  {Millot}, \citenamefont {Milovich}, \citenamefont {Moody}, \citenamefont
  {Moore}, \citenamefont {Morton}, \citenamefont {Murphy}, \citenamefont
  {Newman}, \citenamefont {Di~Nicola}, \citenamefont {Nikroo}, \citenamefont
  {Nora}, \citenamefont {Patel}, \citenamefont {Pelz}, \citenamefont
  {Peterson}, \citenamefont {Ping}, \citenamefont {Pollock}, \citenamefont
  {Ratledge}, \citenamefont {Rice}, \citenamefont {Rinderknecht}, \citenamefont
  {Rosen}, \citenamefont {Rubery}, \citenamefont {Salmonson}, \citenamefont
  {Sater}, \citenamefont {Schiaffino}, \citenamefont {Schlossberg},
  \citenamefont {Schneider}, \citenamefont {Schroeder}, \citenamefont {Scott},
  \citenamefont {Sepke}, \citenamefont {Sequoia}, \citenamefont {Sherlock},
  \citenamefont {Shin}, \citenamefont {Smalyuk}, \citenamefont {Spears},
  \citenamefont {Springer}, \citenamefont {Stadermann}, \citenamefont
  {Stoupin}, \citenamefont {Strozzi}, \citenamefont {Suter}, \citenamefont
  {Thomas}, \citenamefont {Town}, \citenamefont {Tubman}, \citenamefont
  {Trosseille}, \citenamefont {Volegov}, \citenamefont {Weber}, \citenamefont
  {Widmann}, \citenamefont {Wild}, \citenamefont {Wilde}, \citenamefont
  {Van~Wonterghem}, \citenamefont {Woods}, \citenamefont {Woodworth},
  \citenamefont {Yamaguchi}, \citenamefont {Yang},\ and\ \citenamefont
  {Zimmerman}}]{Zylstra2022}%
  \BibitemOpen
  \bibfield  {author} {\bibinfo {author} {\bibfnamefont {A.~B.}\ \bibnamefont
  {Zylstra}}, \bibinfo {author} {\bibfnamefont {O.~A.}\ \bibnamefont
  {Hurricane}}, \bibinfo {author} {\bibfnamefont {D.~A.}\ \bibnamefont
  {Callahan}}, \bibinfo {author} {\bibfnamefont {A.~L.}\ \bibnamefont
  {Kritcher}}, \bibinfo {author} {\bibfnamefont {J.~E.}\ \bibnamefont {Ralph}},
  \bibinfo {author} {\bibfnamefont {H.~F.}\ \bibnamefont {Robey}}, \bibinfo
  {author} {\bibfnamefont {J.~S.}\ \bibnamefont {Ross}}, \bibinfo {author}
  {\bibfnamefont {C.~V.}\ \bibnamefont {Young}}, \bibinfo {author}
  {\bibfnamefont {K.~L.}\ \bibnamefont {Baker}}, \bibinfo {author}
  {\bibfnamefont {D.~T.}\ \bibnamefont {Casey}}, \bibinfo {author}
  {\bibfnamefont {T.}~\bibnamefont {D{\"o}ppner}}, \bibinfo {author}
  {\bibfnamefont {L.}~\bibnamefont {Divol}}, \bibinfo {author} {\bibfnamefont
  {M.}~\bibnamefont {Hohenberger}}, \bibinfo {author} {\bibfnamefont
  {S.}~\bibnamefont {Le~Pape}}, \bibinfo {author} {\bibfnamefont
  {A.}~\bibnamefont {Pak}}, \bibinfo {author} {\bibfnamefont {P.~K.}\
  \bibnamefont {Patel}}, \bibinfo {author} {\bibfnamefont {R.}~\bibnamefont
  {Tommasini}}, \bibinfo {author} {\bibfnamefont {S.~J.}\ \bibnamefont {Ali}},
  \bibinfo {author} {\bibfnamefont {P.~A.}\ \bibnamefont {Amendt}}, \bibinfo
  {author} {\bibfnamefont {L.~J.}\ \bibnamefont {Atherton}}, \bibinfo {author}
  {\bibfnamefont {B.}~\bibnamefont {Bachmann}}, \bibinfo {author}
  {\bibfnamefont {D.}~\bibnamefont {Bailey}}, \bibinfo {author} {\bibfnamefont
  {L.~R.}\ \bibnamefont {Benedetti}}, \bibinfo {author} {\bibfnamefont
  {L.}~\bibnamefont {Berzak~Hopkins}}, \bibinfo {author} {\bibfnamefont
  {R.}~\bibnamefont {Betti}}, \bibinfo {author} {\bibfnamefont {S.~D.}\
  \bibnamefont {Bhandarkar}}, \bibinfo {author} {\bibfnamefont
  {J.}~\bibnamefont {Biener}}, \bibinfo {author} {\bibfnamefont {R.~M.}\
  \bibnamefont {Bionta}}, \bibinfo {author} {\bibfnamefont {N.~W.}\
  \bibnamefont {Birge}}, \bibinfo {author} {\bibfnamefont {E.~J.}\ \bibnamefont
  {Bond}}, \bibinfo {author} {\bibfnamefont {D.~K.}\ \bibnamefont {Bradley}},
  \bibinfo {author} {\bibfnamefont {T.}~\bibnamefont {Braun}}, \bibinfo
  {author} {\bibfnamefont {T.~M.}\ \bibnamefont {Briggs}}, \bibinfo {author}
  {\bibfnamefont {M.~W.}\ \bibnamefont {Bruhn}}, \bibinfo {author}
  {\bibfnamefont {P.~M.}\ \bibnamefont {Celliers}}, \bibinfo {author}
  {\bibfnamefont {B.}~\bibnamefont {Chang}}, \bibinfo {author} {\bibfnamefont
  {T.}~\bibnamefont {Chapman}}, \bibinfo {author} {\bibfnamefont
  {H.}~\bibnamefont {Chen}}, \bibinfo {author} {\bibfnamefont {C.}~\bibnamefont
  {Choate}}, \bibinfo {author} {\bibfnamefont {A.~R.}\ \bibnamefont
  {Christopherson}}, \bibinfo {author} {\bibfnamefont {D.~S.}\ \bibnamefont
  {Clark}}, \bibinfo {author} {\bibfnamefont {J.~W.}\ \bibnamefont {Crippen}},
  \bibinfo {author} {\bibfnamefont {E.~L.}\ \bibnamefont {Dewald}}, \bibinfo
  {author} {\bibfnamefont {T.~R.}\ \bibnamefont {Dittrich}}, \bibinfo {author}
  {\bibfnamefont {M.~J.}\ \bibnamefont {Edwards}}, \bibinfo {author}
  {\bibfnamefont {W.~A.}\ \bibnamefont {Farmer}}, \bibinfo {author}
  {\bibfnamefont {J.~E.}\ \bibnamefont {Field}}, \bibinfo {author}
  {\bibfnamefont {D.}~\bibnamefont {Fittinghoff}}, \bibinfo {author}
  {\bibfnamefont {J.}~\bibnamefont {Frenje}}, \bibinfo {author} {\bibfnamefont
  {J.}~\bibnamefont {Gaffney}}, \bibinfo {author} {\bibfnamefont
  {M.}~\bibnamefont {Gatu~Johnson}}, \bibinfo {author} {\bibfnamefont {S.~H.}\
  \bibnamefont {Glenzer}}, \bibinfo {author} {\bibfnamefont {G.~P.}\
  \bibnamefont {Grim}}, \bibinfo {author} {\bibfnamefont {S.}~\bibnamefont
  {Haan}}, \bibinfo {author} {\bibfnamefont {K.~D.}\ \bibnamefont {Hahn}},
  \bibinfo {author} {\bibfnamefont {G.~N.}\ \bibnamefont {Hall}}, \bibinfo
  {author} {\bibfnamefont {B.~A.}\ \bibnamefont {Hammel}}, \bibinfo {author}
  {\bibfnamefont {J.}~\bibnamefont {Harte}}, \bibinfo {author} {\bibfnamefont
  {E.}~\bibnamefont {Hartouni}}, \bibinfo {author} {\bibfnamefont {J.~E.}\
  \bibnamefont {Heebner}}, \bibinfo {author} {\bibfnamefont {V.~J.}\
  \bibnamefont {Hernandez}}, \bibinfo {author} {\bibfnamefont {H.}~\bibnamefont
  {Herrmann}}, \bibinfo {author} {\bibfnamefont {M.~C.}\ \bibnamefont
  {Herrmann}}, \bibinfo {author} {\bibfnamefont {D.~E.}\ \bibnamefont
  {Hinkel}}, \bibinfo {author} {\bibfnamefont {D.~D.}\ \bibnamefont {Ho}},
  \bibinfo {author} {\bibfnamefont {J.~P.}\ \bibnamefont {Holder}}, \bibinfo
  {author} {\bibfnamefont {W.~W.}\ \bibnamefont {Hsing}}, \bibinfo {author}
  {\bibfnamefont {H.}~\bibnamefont {Huang}}, \bibinfo {author} {\bibfnamefont
  {K.~D.}\ \bibnamefont {Humbird}}, \bibinfo {author} {\bibfnamefont
  {N.}~\bibnamefont {Izumi}}, \bibinfo {author} {\bibfnamefont {L.~C.}\
  \bibnamefont {Jarrott}}, \bibinfo {author} {\bibfnamefont {J.}~\bibnamefont
  {Jeet}}, \bibinfo {author} {\bibfnamefont {O.}~\bibnamefont {Jones}},
  \bibinfo {author} {\bibfnamefont {G.~D.}\ \bibnamefont {Kerbel}}, \bibinfo
  {author} {\bibfnamefont {S.~M.}\ \bibnamefont {Kerr}}, \bibinfo {author}
  {\bibfnamefont {S.~F.}\ \bibnamefont {Khan}}, \bibinfo {author}
  {\bibfnamefont {J.}~\bibnamefont {Kilkenny}}, \bibinfo {author}
  {\bibfnamefont {Y.}~\bibnamefont {Kim}}, \bibinfo {author} {\bibfnamefont
  {H.}~\bibnamefont {Geppert~Kleinrath}}, \bibinfo {author} {\bibfnamefont
  {V.}~\bibnamefont {Geppert~Kleinrath}}, \bibinfo {author} {\bibfnamefont
  {C.}~\bibnamefont {Kong}}, \bibinfo {author} {\bibfnamefont {J.~M.}\
  \bibnamefont {Koning}}, \bibinfo {author} {\bibfnamefont {J.~J.}\
  \bibnamefont {Kroll}}, \bibinfo {author} {\bibfnamefont {M.~K.~G.}\
  \bibnamefont {Kruse}}, \bibinfo {author} {\bibfnamefont {B.}~\bibnamefont
  {Kustowski}}, \bibinfo {author} {\bibfnamefont {O.~L.}\ \bibnamefont
  {Landen}}, \bibinfo {author} {\bibfnamefont {S.}~\bibnamefont {Langer}},
  \bibinfo {author} {\bibfnamefont {D.}~\bibnamefont {Larson}}, \bibinfo
  {author} {\bibfnamefont {N.~C.}\ \bibnamefont {Lemos}}, \bibinfo {author}
  {\bibfnamefont {J.~D.}\ \bibnamefont {Lindl}}, \bibinfo {author}
  {\bibfnamefont {T.}~\bibnamefont {Ma}}, \bibinfo {author} {\bibfnamefont
  {M.~J.}\ \bibnamefont {MacDonald}}, \bibinfo {author} {\bibfnamefont {B.~J.}\
  \bibnamefont {MacGowan}}, \bibinfo {author} {\bibfnamefont {A.~J.}\
  \bibnamefont {Mackinnon}}, \bibinfo {author} {\bibfnamefont {S.~A.}\
  \bibnamefont {MacLaren}}, \bibinfo {author} {\bibfnamefont {A.~G.}\
  \bibnamefont {MacPhee}}, \bibinfo {author} {\bibfnamefont {M.~M.}\
  \bibnamefont {Marinak}}, \bibinfo {author} {\bibfnamefont {D.~A.}\
  \bibnamefont {Mariscal}}, \bibinfo {author} {\bibfnamefont {E.~V.}\
  \bibnamefont {Marley}}, \bibinfo {author} {\bibfnamefont {L.}~\bibnamefont
  {Masse}}, \bibinfo {author} {\bibfnamefont {K.}~\bibnamefont {Meaney}},
  \bibinfo {author} {\bibfnamefont {N.~B.}\ \bibnamefont {Meezan}}, \bibinfo
  {author} {\bibfnamefont {P.~A.}\ \bibnamefont {Michel}}, \bibinfo {author}
  {\bibfnamefont {M.}~\bibnamefont {Millot}}, \bibinfo {author} {\bibfnamefont
  {J.~L.}\ \bibnamefont {Milovich}}, \bibinfo {author} {\bibfnamefont {J.~D.}\
  \bibnamefont {Moody}}, \bibinfo {author} {\bibfnamefont {A.~S.}\ \bibnamefont
  {Moore}}, \bibinfo {author} {\bibfnamefont {J.~W.}\ \bibnamefont {Morton}},
  \bibinfo {author} {\bibfnamefont {T.}~\bibnamefont {Murphy}}, \bibinfo
  {author} {\bibfnamefont {K.}~\bibnamefont {Newman}}, \bibinfo {author}
  {\bibfnamefont {J.-M.~G.}\ \bibnamefont {Di~Nicola}}, \bibinfo {author}
  {\bibfnamefont {A.}~\bibnamefont {Nikroo}}, \bibinfo {author} {\bibfnamefont
  {R.}~\bibnamefont {Nora}}, \bibinfo {author} {\bibfnamefont {M.~V.}\
  \bibnamefont {Patel}}, \bibinfo {author} {\bibfnamefont {L.~J.}\ \bibnamefont
  {Pelz}}, \bibinfo {author} {\bibfnamefont {J.~L.}\ \bibnamefont {Peterson}},
  \bibinfo {author} {\bibfnamefont {Y.}~\bibnamefont {Ping}}, \bibinfo {author}
  {\bibfnamefont {B.~B.}\ \bibnamefont {Pollock}}, \bibinfo {author}
  {\bibfnamefont {M.}~\bibnamefont {Ratledge}}, \bibinfo {author}
  {\bibfnamefont {N.~G.}\ \bibnamefont {Rice}}, \bibinfo {author}
  {\bibfnamefont {H.}~\bibnamefont {Rinderknecht}}, \bibinfo {author}
  {\bibfnamefont {M.}~\bibnamefont {Rosen}}, \bibinfo {author} {\bibfnamefont
  {M.~S.}\ \bibnamefont {Rubery}}, \bibinfo {author} {\bibfnamefont {J.~D.}\
  \bibnamefont {Salmonson}}, \bibinfo {author} {\bibfnamefont {J.}~\bibnamefont
  {Sater}}, \bibinfo {author} {\bibfnamefont {S.}~\bibnamefont {Schiaffino}},
  \bibinfo {author} {\bibfnamefont {D.~J.}\ \bibnamefont {Schlossberg}},
  \bibinfo {author} {\bibfnamefont {M.~B.}\ \bibnamefont {Schneider}}, \bibinfo
  {author} {\bibfnamefont {C.~R.}\ \bibnamefont {Schroeder}}, \bibinfo {author}
  {\bibfnamefont {H.~A.}\ \bibnamefont {Scott}}, \bibinfo {author}
  {\bibfnamefont {S.~M.}\ \bibnamefont {Sepke}}, \bibinfo {author}
  {\bibfnamefont {K.}~\bibnamefont {Sequoia}}, \bibinfo {author} {\bibfnamefont
  {M.~W.}\ \bibnamefont {Sherlock}}, \bibinfo {author} {\bibfnamefont
  {S.}~\bibnamefont {Shin}}, \bibinfo {author} {\bibfnamefont {V.~A.}\
  \bibnamefont {Smalyuk}}, \bibinfo {author} {\bibfnamefont {B.~K.}\
  \bibnamefont {Spears}}, \bibinfo {author} {\bibfnamefont {P.~T.}\
  \bibnamefont {Springer}}, \bibinfo {author} {\bibfnamefont {M.}~\bibnamefont
  {Stadermann}}, \bibinfo {author} {\bibfnamefont {S.}~\bibnamefont {Stoupin}},
  \bibinfo {author} {\bibfnamefont {D.~J.}\ \bibnamefont {Strozzi}}, \bibinfo
  {author} {\bibfnamefont {L.~J.}\ \bibnamefont {Suter}}, \bibinfo {author}
  {\bibfnamefont {C.~A.}\ \bibnamefont {Thomas}}, \bibinfo {author}
  {\bibfnamefont {R.~P.~J.}\ \bibnamefont {Town}}, \bibinfo {author}
  {\bibfnamefont {E.~R.}\ \bibnamefont {Tubman}}, \bibinfo {author}
  {\bibfnamefont {C.}~\bibnamefont {Trosseille}}, \bibinfo {author}
  {\bibfnamefont {P.~L.}\ \bibnamefont {Volegov}}, \bibinfo {author}
  {\bibfnamefont {C.~R.}\ \bibnamefont {Weber}}, \bibinfo {author}
  {\bibfnamefont {K.}~\bibnamefont {Widmann}}, \bibinfo {author} {\bibfnamefont
  {C.}~\bibnamefont {Wild}}, \bibinfo {author} {\bibfnamefont {C.~H.}\
  \bibnamefont {Wilde}}, \bibinfo {author} {\bibfnamefont {B.~M.}\ \bibnamefont
  {Van~Wonterghem}}, \bibinfo {author} {\bibfnamefont {D.~T.}\ \bibnamefont
  {Woods}}, \bibinfo {author} {\bibfnamefont {B.~N.}\ \bibnamefont
  {Woodworth}}, \bibinfo {author} {\bibfnamefont {M.}~\bibnamefont
  {Yamaguchi}}, \bibinfo {author} {\bibfnamefont {S.~T.}\ \bibnamefont {Yang}},
  \ and\ \bibinfo {author} {\bibfnamefont {G.~B.}\ \bibnamefont {Zimmerman}},\
  }\bibfield  {title} {\enquote {\bibinfo {title} {Burning plasma achieved in
  inertial fusion},}\ }\href {\doibase 10.1038/s41586-021-04281-w} {\bibfield
  {journal} {\bibinfo  {journal} {Nature}\ }\textbf {\bibinfo {volume} {601}},\
  \bibinfo {pages} {542--548} (\bibinfo {year} {2022})}\BibitemShut {NoStop}%
\bibitem [{\citenamefont {Hoffmann}\ \emph {et~al.}(2005)\citenamefont
  {Hoffmann}, \citenamefont {Blazevic}, \citenamefont {Ni}, \citenamefont
  {Rosmej}, \citenamefont {Roth}, \citenamefont {Tahir}, \citenamefont
  {Tauschwitz}, \citenamefont {Udrea}, \citenamefont {Varentsov}, \citenamefont
  {Weyrich},\ and\ \citenamefont {et~al.}}]{hoffmann2005}%
  \BibitemOpen
  \bibfield  {author} {\bibinfo {author} {\bibfnamefont {D.}~\bibnamefont
  {Hoffmann}}, \bibinfo {author} {\bibfnamefont {A.}~\bibnamefont {Blazevic}},
  \bibinfo {author} {\bibfnamefont {P.}~\bibnamefont {Ni}}, \bibinfo {author}
  {\bibfnamefont {O.}~\bibnamefont {Rosmej}}, \bibinfo {author} {\bibfnamefont
  {M.}~\bibnamefont {Roth}}, \bibinfo {author} {\bibfnamefont {N.}~\bibnamefont
  {Tahir}}, \bibinfo {author} {\bibfnamefont {A.}~\bibnamefont {Tauschwitz}},
  \bibinfo {author} {\bibfnamefont {S.}~\bibnamefont {Udrea}}, \bibinfo
  {author} {\bibfnamefont {D.}~\bibnamefont {Varentsov}}, \bibinfo {author}
  {\bibfnamefont {K.}~\bibnamefont {Weyrich}}, \ and\ \bibinfo {author}
  {\bibnamefont {et~al.}},\ }\bibfield  {title} {\enquote {\bibinfo {title}
  {Present and future perspectives for high energy density physics with intense
  heavy ion and laser beams},}\ }\href {\doibase 10.1017/S026303460505010X}
  {\bibfield  {journal} {\bibinfo  {journal} {Laser and Particle Beams}\
  }\textbf {\bibinfo {volume} {23}},\ \bibinfo {pages} {47–53} (\bibinfo
  {year} {2005})}\BibitemShut {NoStop}%
\bibitem [{\citenamefont {Esarey}, \citenamefont {Sprangle},\ and\
  \citenamefont {Krall}(1995)}]{Esarey95}%
  \BibitemOpen
  \bibfield  {author} {\bibinfo {author} {\bibfnamefont {E.}~\bibnamefont
  {Esarey}}, \bibinfo {author} {\bibfnamefont {P.}~\bibnamefont {Sprangle}}, \
  and\ \bibinfo {author} {\bibfnamefont {J.}~\bibnamefont {Krall}},\ }\bibfield
   {title} {\enquote {\bibinfo {title} {Laser acceleration of electrons in
  vacuum},}\ }\href@noop {} {\bibfield  {journal} {\bibinfo  {journal} {Phys.
  Rev. E}\ }\textbf {\bibinfo {volume} {52}},\ \bibinfo {pages} {5443--5453}
  (\bibinfo {year} {1995})}\BibitemShut {NoStop}%
\bibitem [{\citenamefont {Sarri}\ \emph {et~al.}(2013)\citenamefont {Sarri},
  \citenamefont {Schumaker}, \citenamefont {Di~Piazza}, \citenamefont {Vargas},
  \citenamefont {Dromey}, \citenamefont {Dieckmann}, \citenamefont {Chvykov},
  \citenamefont {Maksimchuk}, \citenamefont {Yanovsky}, \citenamefont {He},
  \citenamefont {Hou}, \citenamefont {Nees}, \citenamefont {Thomas},
  \citenamefont {Keitel}, \citenamefont {Zepf},\ and\ \citenamefont
  {Krushelnick}}]{Sarri13}%
  \BibitemOpen
  \bibfield  {author} {\bibinfo {author} {\bibfnamefont {G.}~\bibnamefont
  {Sarri}}, \bibinfo {author} {\bibfnamefont {W.}~\bibnamefont {Schumaker}},
  \bibinfo {author} {\bibfnamefont {A.}~\bibnamefont {Di~Piazza}}, \bibinfo
  {author} {\bibfnamefont {M.}~\bibnamefont {Vargas}}, \bibinfo {author}
  {\bibfnamefont {B.}~\bibnamefont {Dromey}}, \bibinfo {author} {\bibfnamefont
  {M.~E.}\ \bibnamefont {Dieckmann}}, \bibinfo {author} {\bibfnamefont
  {V.}~\bibnamefont {Chvykov}}, \bibinfo {author} {\bibfnamefont
  {A.}~\bibnamefont {Maksimchuk}}, \bibinfo {author} {\bibfnamefont
  {V.}~\bibnamefont {Yanovsky}}, \bibinfo {author} {\bibfnamefont {Z.~H.}\
  \bibnamefont {He}}, \bibinfo {author} {\bibfnamefont {B.~X.}\ \bibnamefont
  {Hou}}, \bibinfo {author} {\bibfnamefont {J.~A.}\ \bibnamefont {Nees}},
  \bibinfo {author} {\bibfnamefont {A.~G.~R.}\ \bibnamefont {Thomas}}, \bibinfo
  {author} {\bibfnamefont {C.~H.}\ \bibnamefont {Keitel}}, \bibinfo {author}
  {\bibfnamefont {M.}~\bibnamefont {Zepf}}, \ and\ \bibinfo {author}
  {\bibfnamefont {K.}~\bibnamefont {Krushelnick}},\ }\bibfield  {title}
  {\enquote {\bibinfo {title} {Table-top laser-based source of femtosecond,
  collimated, ultrarelativistic positron beams},}\ }\href {\doibase
  10.1103/PhysRevLett.110.255002} {\bibfield  {journal} {\bibinfo  {journal}
  {Phys. Rev. Lett.}\ }\textbf {\bibinfo {volume} {110}},\ \bibinfo {pages}
  {255002} (\bibinfo {year} {2013})}\BibitemShut {NoStop}%
\bibitem [{\citenamefont {Tajima}, \citenamefont {Yan},\ and\ \citenamefont
  {Ebisuzaki}(2020)}]{Tajima2020}%
  \BibitemOpen
  \bibfield  {author} {\bibinfo {author} {\bibfnamefont {T.}~\bibnamefont
  {Tajima}}, \bibinfo {author} {\bibfnamefont {X.~Q.}\ \bibnamefont {Yan}}, \
  and\ \bibinfo {author} {\bibfnamefont {T.}~\bibnamefont {Ebisuzaki}},\
  }\bibfield  {title} {\enquote {\bibinfo {title} {Wakefield acceleration},}\
  }\href {\doibase 10.1007/s41614-020-0043-z} {\bibfield  {journal} {\bibinfo
  {journal} {Reviews of Modern Plasma Physics}\ }\textbf {\bibinfo {volume}
  {4}},\ \bibinfo {pages} {7} (\bibinfo {year} {2020})}\BibitemShut {NoStop}%
\bibitem [{\citenamefont {Robson}\ \emph {et~al.}(2007)\citenamefont {Robson},
  \citenamefont {Simpson}, \citenamefont {Clarke}, \citenamefont {Ledingham},
  \citenamefont {Lindau}, \citenamefont {Lundh}, \citenamefont {McCanny},
  \citenamefont {Mora}, \citenamefont {Neely}, \citenamefont {Wahlstr{\"o}m},
  \citenamefont {Zepf},\ and\ \citenamefont {McKenna}}]{Robson2007}%
  \BibitemOpen
  \bibfield  {author} {\bibinfo {author} {\bibfnamefont {L.}~\bibnamefont
  {Robson}}, \bibinfo {author} {\bibfnamefont {P.~T.}\ \bibnamefont {Simpson}},
  \bibinfo {author} {\bibfnamefont {R.~J.}\ \bibnamefont {Clarke}}, \bibinfo
  {author} {\bibfnamefont {K.~W.~D.}\ \bibnamefont {Ledingham}}, \bibinfo
  {author} {\bibfnamefont {F.}~\bibnamefont {Lindau}}, \bibinfo {author}
  {\bibfnamefont {O.}~\bibnamefont {Lundh}}, \bibinfo {author} {\bibfnamefont
  {T.}~\bibnamefont {McCanny}}, \bibinfo {author} {\bibfnamefont
  {P.}~\bibnamefont {Mora}}, \bibinfo {author} {\bibfnamefont {D.}~\bibnamefont
  {Neely}}, \bibinfo {author} {\bibfnamefont {C.-G.}\ \bibnamefont
  {Wahlstr{\"o}m}}, \bibinfo {author} {\bibfnamefont {M.}~\bibnamefont {Zepf}},
  \ and\ \bibinfo {author} {\bibfnamefont {P.}~\bibnamefont {McKenna}},\
  }\bibfield  {title} {\enquote {\bibinfo {title} {Scaling of proton
  acceleration driven by petawatt-laser--plasma interactions},}\ }\href
  {\doibase 10.1038/nphys476} {\bibfield  {journal} {\bibinfo  {journal}
  {Nature Physics}\ }\textbf {\bibinfo {volume} {3}},\ \bibinfo {pages}
  {58--62} (\bibinfo {year} {2007})}\BibitemShut {NoStop}%
\bibitem [{\citenamefont {Danson}\ \emph {et~al.}(2019)\citenamefont {Danson},
  \citenamefont {Haefner}, \citenamefont {Bromage}, \citenamefont {Butcher},
  \citenamefont {Chanteloup}, \citenamefont {Chowdhury}, \citenamefont
  {Galvanauskas}, \citenamefont {Gizzi}, \citenamefont {Hein}, \citenamefont
  {Hillier},\ and\ \citenamefont {et~al.}}]{danson2019}%
  \BibitemOpen
  \bibfield  {author} {\bibinfo {author} {\bibfnamefont {C.~N.}\ \bibnamefont
  {Danson}}, \bibinfo {author} {\bibfnamefont {C.}~\bibnamefont {Haefner}},
  \bibinfo {author} {\bibfnamefont {J.}~\bibnamefont {Bromage}}, \bibinfo
  {author} {\bibfnamefont {T.}~\bibnamefont {Butcher}}, \bibinfo {author}
  {\bibfnamefont {J.-C.~F.}\ \bibnamefont {Chanteloup}}, \bibinfo {author}
  {\bibfnamefont {E.~A.}\ \bibnamefont {Chowdhury}}, \bibinfo {author}
  {\bibfnamefont {A.}~\bibnamefont {Galvanauskas}}, \bibinfo {author}
  {\bibfnamefont {L.~A.}\ \bibnamefont {Gizzi}}, \bibinfo {author}
  {\bibfnamefont {J.}~\bibnamefont {Hein}}, \bibinfo {author} {\bibfnamefont
  {D.~I.}\ \bibnamefont {Hillier}}, \ and\ \bibinfo {author} {\bibnamefont
  {et~al.}},\ }\bibfield  {title} {\enquote {\bibinfo {title} {Petawatt and
  exawatt class lasers worldwide},}\ }\href@noop {} {\bibfield  {journal}
  {\bibinfo  {journal} {High Power Laser Science and Engineering}\ }\textbf
  {\bibinfo {volume} {7}} (\bibinfo {year} {2019})}\BibitemShut {NoStop}%
\bibitem [{\citenamefont {Yoon}\ \emph {et~al.}(2021)\citenamefont {Yoon},
  \citenamefont {Kim}, \citenamefont {Choi}, \citenamefont {Sung},
  \citenamefont {Lee}, \citenamefont {Lee},\ and\ \citenamefont
  {Nam}}]{Yoon:21}%
  \BibitemOpen
  \bibfield  {author} {\bibinfo {author} {\bibfnamefont {J.~W.}\ \bibnamefont
  {Yoon}}, \bibinfo {author} {\bibfnamefont {Y.~G.}\ \bibnamefont {Kim}},
  \bibinfo {author} {\bibfnamefont {I.~W.}\ \bibnamefont {Choi}}, \bibinfo
  {author} {\bibfnamefont {J.~H.}\ \bibnamefont {Sung}}, \bibinfo {author}
  {\bibfnamefont {H.~W.}\ \bibnamefont {Lee}}, \bibinfo {author} {\bibfnamefont
  {S.~K.}\ \bibnamefont {Lee}}, \ and\ \bibinfo {author} {\bibfnamefont
  {C.~H.}\ \bibnamefont {Nam}},\ }\bibfield  {title} {\enquote {\bibinfo
  {title} {Realization of laser intensity over 1e23w/cm2},}\ }\href@noop {}
  {\bibfield  {journal} {\bibinfo  {journal} {Optica}\ }\textbf {\bibinfo
  {volume} {8}},\ \bibinfo {pages} {630--635} (\bibinfo {year}
  {2021})}\BibitemShut {NoStop}%
\bibitem [{\citenamefont {Tiwari}\ \emph {et~al.}(2019)\citenamefont {Tiwari},
  \citenamefont {Gaul}, \citenamefont {Martinez}, \citenamefont {Dyer},
  \citenamefont {Gordon}, \citenamefont {Spinks}, \citenamefont {Toncian},
  \citenamefont {Bowers}, \citenamefont {Jiao}, \citenamefont {Kupfer},
  \citenamefont {Lisi}, \citenamefont {McCary}, \citenamefont {Roycroft},
  \citenamefont {Yandow}, \citenamefont {Glenn}, \citenamefont {Donovan},
  \citenamefont {Ditmire},\ and\ \citenamefont {Hegelich}}]{Tiwari:19}%
  \BibitemOpen
  \bibfield  {author} {\bibinfo {author} {\bibfnamefont {G.}~\bibnamefont
  {Tiwari}}, \bibinfo {author} {\bibfnamefont {E.}~\bibnamefont {Gaul}},
  \bibinfo {author} {\bibfnamefont {M.}~\bibnamefont {Martinez}}, \bibinfo
  {author} {\bibfnamefont {G.}~\bibnamefont {Dyer}}, \bibinfo {author}
  {\bibfnamefont {J.}~\bibnamefont {Gordon}}, \bibinfo {author} {\bibfnamefont
  {M.}~\bibnamefont {Spinks}}, \bibinfo {author} {\bibfnamefont
  {T.}~\bibnamefont {Toncian}}, \bibinfo {author} {\bibfnamefont
  {B.}~\bibnamefont {Bowers}}, \bibinfo {author} {\bibfnamefont
  {X.}~\bibnamefont {Jiao}}, \bibinfo {author} {\bibfnamefont {R.}~\bibnamefont
  {Kupfer}}, \bibinfo {author} {\bibfnamefont {L.}~\bibnamefont {Lisi}},
  \bibinfo {author} {\bibfnamefont {E.}~\bibnamefont {McCary}}, \bibinfo
  {author} {\bibfnamefont {R.}~\bibnamefont {Roycroft}}, \bibinfo {author}
  {\bibfnamefont {A.}~\bibnamefont {Yandow}}, \bibinfo {author} {\bibfnamefont
  {G.~D.}\ \bibnamefont {Glenn}}, \bibinfo {author} {\bibfnamefont
  {M.}~\bibnamefont {Donovan}}, \bibinfo {author} {\bibfnamefont
  {T.}~\bibnamefont {Ditmire}}, \ and\ \bibinfo {author} {\bibfnamefont
  {B.~M.}\ \bibnamefont {Hegelich}},\ }\bibfield  {title} {\enquote {\bibinfo
  {title} {Beam distortion effects upon focusing an ultrashort petawatt laser
  pulse to greater than $10^{22}$w/cm2},}\ }\href {\doibase
  10.1364/OL.44.002764} {\bibfield  {journal} {\bibinfo  {journal} {Opt.
  Lett.}\ }\textbf {\bibinfo {volume} {44}},\ \bibinfo {pages} {2764--2767}
  (\bibinfo {year} {2019})}\BibitemShut {NoStop}%
\bibitem [{\citenamefont {Chen}, \citenamefont {Maksimchuk},\ and\
  \citenamefont {Umstadter}(1998)}]{Chen1998}%
  \BibitemOpen
  \bibfield  {author} {\bibinfo {author} {\bibfnamefont {S.-y.}\ \bibnamefont
  {Chen}}, \bibinfo {author} {\bibfnamefont {A.}~\bibnamefont {Maksimchuk}}, \
  and\ \bibinfo {author} {\bibfnamefont {D.}~\bibnamefont {Umstadter}},\
  }\bibfield  {title} {\enquote {\bibinfo {title} {Experimental observation of
  relativistic nonlinear thomson scattering},}\ }\href {\doibase 10.1038/25303}
  {\bibfield  {journal} {\bibinfo  {journal} {Nature}\ }\textbf {\bibinfo
  {volume} {396}},\ \bibinfo {pages} {653--655} (\bibinfo {year}
  {1998})}\BibitemShut {NoStop}%
\bibitem [{\citenamefont {He}\ \emph {et~al.}(2019)\citenamefont {He},
  \citenamefont {Longman}, \citenamefont {P\'{e}rez-Hern\'{a}ndez},
  \citenamefont {de~Marco}, \citenamefont {Salgado}, \citenamefont {Zeraouli},
  \citenamefont {Gatti}, \citenamefont {Roso}, \citenamefont {Fedosejevs},\
  and\ \citenamefont {Hill}}]{He:19}%
  \BibitemOpen
  \bibfield  {author} {\bibinfo {author} {\bibfnamefont {C.~Z.}\ \bibnamefont
  {He}}, \bibinfo {author} {\bibfnamefont {A.}~\bibnamefont {Longman}},
  \bibinfo {author} {\bibfnamefont {J.~A.}\ \bibnamefont
  {P\'{e}rez-Hern\'{a}ndez}}, \bibinfo {author} {\bibfnamefont
  {M.}~\bibnamefont {de~Marco}}, \bibinfo {author} {\bibfnamefont
  {C.}~\bibnamefont {Salgado}}, \bibinfo {author} {\bibfnamefont
  {G.}~\bibnamefont {Zeraouli}}, \bibinfo {author} {\bibfnamefont
  {G.}~\bibnamefont {Gatti}}, \bibinfo {author} {\bibfnamefont
  {L.}~\bibnamefont {Roso}}, \bibinfo {author} {\bibfnamefont {R.}~\bibnamefont
  {Fedosejevs}}, \ and\ \bibinfo {author} {\bibfnamefont {W.~T.}\ \bibnamefont
  {Hill}},\ }\bibfield  {title} {\enquote {\bibinfo {title} {Towards an in
  situ, full-power gauge of the focal-volume intensity of petawatt-class
  lasers},}\ }\href {\doibase 10.1364/OE.27.030020} {\bibfield  {journal}
  {\bibinfo  {journal} {Opt. Express}\ }\textbf {\bibinfo {volume} {27}},\
  \bibinfo {pages} {30020--30030} (\bibinfo {year} {2019})}\BibitemShut
  {NoStop}%
\bibitem [{\citenamefont {Gao}(2006)}]{Gao06}%
  \BibitemOpen
  \bibfield  {author} {\bibinfo {author} {\bibfnamefont {J.}~\bibnamefont
  {Gao}},\ }\bibfield  {title} {\enquote {\bibinfo {title} {{Laser intensity
  measurement by Thomson scattering}},}\ }\href {\doibase 10.1063/1.2180869}
  {\bibfield  {journal} {\bibinfo  {journal} {Applied Physics Letters}\
  }\textbf {\bibinfo {volume} {88}} (\bibinfo {year} {2006}),\
  10.1063/1.2180869},\ \bibinfo {note} {091105}\BibitemShut {NoStop}%
\bibitem [{\citenamefont {Har-Shemesh}\ and\ \citenamefont
  {Piazza}(2012)}]{Har-Shemesh:12}%
  \BibitemOpen
  \bibfield  {author} {\bibinfo {author} {\bibfnamefont {O.}~\bibnamefont
  {Har-Shemesh}}\ and\ \bibinfo {author} {\bibfnamefont {A.~D.}\ \bibnamefont
  {Piazza}},\ }\bibfield  {title} {\enquote {\bibinfo {title} {Peak intensity
  measurement of relativistic lasers via nonlinear thomson scattering},}\
  }\href {\doibase 10.1364/OL.37.001352} {\bibfield  {journal} {\bibinfo
  {journal} {Opt. Lett.}\ }\textbf {\bibinfo {volume} {37}},\ \bibinfo {pages}
  {1352--1354} (\bibinfo {year} {2012})}\BibitemShut {NoStop}%
\bibitem [{\citenamefont {Tarbox}\ \emph {et~al.}(2015)\citenamefont {Tarbox},
  \citenamefont {Cunningham}, \citenamefont {Sandberg}, \citenamefont
  {Peatross},\ and\ \citenamefont {Ware}}]{Tarbox:15}%
  \BibitemOpen
  \bibfield  {author} {\bibinfo {author} {\bibfnamefont {G.}~\bibnamefont
  {Tarbox}}, \bibinfo {author} {\bibfnamefont {E.}~\bibnamefont {Cunningham}},
  \bibinfo {author} {\bibfnamefont {R.}~\bibnamefont {Sandberg}}, \bibinfo
  {author} {\bibfnamefont {J.}~\bibnamefont {Peatross}}, \ and\ \bibinfo
  {author} {\bibfnamefont {M.}~\bibnamefont {Ware}},\ }\bibfield  {title}
  {\enquote {\bibinfo {title} {Radiation from free electrons in a laser focus
  at 1018 w/cm2: modeling of photon yields and required focal conditions},}\
  }\href {\doibase 10.1364/JOSAB.32.000743} {\bibfield  {journal} {\bibinfo
  {journal} {J. Opt. Soc. Am. B}\ }\textbf {\bibinfo {volume} {32}},\ \bibinfo
  {pages} {743--750} (\bibinfo {year} {2015})}\BibitemShut {NoStop}%
\bibitem [{\citenamefont {Harvey}(2018)}]{Harvey18}%
  \BibitemOpen
  \bibfield  {author} {\bibinfo {author} {\bibfnamefont {C.~N.}\ \bibnamefont
  {Harvey}},\ }\bibfield  {title} {\enquote {\bibinfo {title} {In situ
  characterization of ultraintense laser pulses},}\ }\href {\doibase
  10.1103/PhysRevAccelBeams.21.114001} {\bibfield  {journal} {\bibinfo
  {journal} {Phys. Rev. Accel. Beams}\ }\textbf {\bibinfo {volume} {21}},\
  \bibinfo {pages} {114001} (\bibinfo {year} {2018})}\BibitemShut {NoStop}%
\bibitem [{\citenamefont {Mackenroth}\ and\ \citenamefont
  {Holkundkar}(2019)}]{Mackenroth2019}%
  \BibitemOpen
  \bibfield  {author} {\bibinfo {author} {\bibfnamefont {F.}~\bibnamefont
  {Mackenroth}}\ and\ \bibinfo {author} {\bibfnamefont {A.~R.}\ \bibnamefont
  {Holkundkar}},\ }\bibfield  {title} {\enquote {\bibinfo {title} {Determining
  the duration of an ultra-intense laser pulse directly in its focus},}\ }\href
  {\doibase 10.1038/s41598-019-55949-3} {\bibfield  {journal} {\bibinfo
  {journal} {Scientific Reports}\ }\textbf {\bibinfo {volume} {9}},\ \bibinfo
  {pages} {19607} (\bibinfo {year} {2019})}\BibitemShut {NoStop}%
\bibitem [{\citenamefont {Perry}, \citenamefont {Landen},\ and\ \citenamefont
  {Sz\"{o}ke}(1989)}]{Perry:89}%
  \BibitemOpen
  \bibfield  {author} {\bibinfo {author} {\bibfnamefont {M.~D.}\ \bibnamefont
  {Perry}}, \bibinfo {author} {\bibfnamefont {O.~L.}\ \bibnamefont {Landen}}, \
  and\ \bibinfo {author} {\bibfnamefont {A.}~\bibnamefont {Sz\"{o}ke}},\
  }\bibfield  {title} {\enquote {\bibinfo {title} {Measurement of the local
  laser intensity by photoelectron energy shifts in multiphoton ionization},}\
  }\href {\doibase 10.1364/JOSAB.6.000344} {\bibfield  {journal} {\bibinfo
  {journal} {J. Opt. Soc. Am. B}\ }\textbf {\bibinfo {volume} {6}},\ \bibinfo
  {pages} {344--349} (\bibinfo {year} {1989})}\BibitemShut {NoStop}%
\bibitem [{\citenamefont {Chowdhury}, \citenamefont {Barty},\ and\
  \citenamefont {Walker}(2001)}]{Chowdhury01}%
  \BibitemOpen
  \bibfield  {author} {\bibinfo {author} {\bibfnamefont {E.~A.}\ \bibnamefont
  {Chowdhury}}, \bibinfo {author} {\bibfnamefont {C.~P.~J.}\ \bibnamefont
  {Barty}}, \ and\ \bibinfo {author} {\bibfnamefont {B.~C.}\ \bibnamefont
  {Walker}},\ }\bibfield  {title} {\enquote {\bibinfo {title}
  {``nonrelativistic'' ionization of the l-shell states in argon by a
  ``relativistic'' ${10}^{19}\mathrm{W}/{\mathrm{cm}}^{2}$ laser field},}\
  }\href {\doibase 10.1103/PhysRevA.63.042712} {\bibfield  {journal} {\bibinfo
  {journal} {Phys. Rev. A}\ }\textbf {\bibinfo {volume} {63}},\ \bibinfo
  {pages} {042712} (\bibinfo {year} {2001})}\BibitemShut {NoStop}%
\bibitem [{\citenamefont {Yamakawa}\ \emph {et~al.}(2003)\citenamefont
  {Yamakawa}, \citenamefont {Akahane}, \citenamefont {Fukuda}, \citenamefont
  {Aoyama}, \citenamefont {Inoue},\ and\ \citenamefont {Ueda}}]{Yamakawa03}%
  \BibitemOpen
  \bibfield  {author} {\bibinfo {author} {\bibfnamefont {K.}~\bibnamefont
  {Yamakawa}}, \bibinfo {author} {\bibfnamefont {Y.}~\bibnamefont {Akahane}},
  \bibinfo {author} {\bibfnamefont {Y.}~\bibnamefont {Fukuda}}, \bibinfo
  {author} {\bibfnamefont {M.}~\bibnamefont {Aoyama}}, \bibinfo {author}
  {\bibfnamefont {N.}~\bibnamefont {Inoue}}, \ and\ \bibinfo {author}
  {\bibfnamefont {H.}~\bibnamefont {Ueda}},\ }\bibfield  {title} {\enquote
  {\bibinfo {title} {Ionization of many-electron atoms by ultrafast laser
  pulses with peak intensities greater than
  ${10}^{19}{\mathrm{w}/\mathrm{c}\mathrm{m}}^{2}$},}\ }\href {\doibase
  10.1103/PhysRevA.68.065403} {\bibfield  {journal} {\bibinfo  {journal} {Phys.
  Rev. A}\ }\textbf {\bibinfo {volume} {68}},\ \bibinfo {pages} {065403}
  (\bibinfo {year} {2003})}\BibitemShut {NoStop}%
\bibitem [{\citenamefont {Ciappina}\ \emph {et~al.}(2019)\citenamefont
  {Ciappina}, \citenamefont {Popruzhenko}, \citenamefont {Bulanov},
  \citenamefont {Ditmire}, \citenamefont {Korn},\ and\ \citenamefont
  {Weber}}]{Ciappina19}%
  \BibitemOpen
  \bibfield  {author} {\bibinfo {author} {\bibfnamefont {M.~F.}\ \bibnamefont
  {Ciappina}}, \bibinfo {author} {\bibfnamefont {S.~V.}\ \bibnamefont
  {Popruzhenko}}, \bibinfo {author} {\bibfnamefont {S.~V.}\ \bibnamefont
  {Bulanov}}, \bibinfo {author} {\bibfnamefont {T.}~\bibnamefont {Ditmire}},
  \bibinfo {author} {\bibfnamefont {G.}~\bibnamefont {Korn}}, \ and\ \bibinfo
  {author} {\bibfnamefont {S.}~\bibnamefont {Weber}},\ }\bibfield  {title}
  {\enquote {\bibinfo {title} {Progress toward atomic diagnostics of ultrahigh
  laser intensities},}\ }\href {\doibase 10.1103/PhysRevA.99.043405} {\bibfield
   {journal} {\bibinfo  {journal} {Phys. Rev. A}\ }\textbf {\bibinfo {volume}
  {99}},\ \bibinfo {pages} {043405} (\bibinfo {year} {2019})}\BibitemShut
  {NoStop}%
\bibitem [{\citenamefont {Ciappina}, \citenamefont {Peganov},\ and\
  \citenamefont {Popruzhenko}(2020)}]{Ciappina20}%
  \BibitemOpen
  \bibfield  {author} {\bibinfo {author} {\bibfnamefont {M.~F.}\ \bibnamefont
  {Ciappina}}, \bibinfo {author} {\bibfnamefont {E.~E.}\ \bibnamefont
  {Peganov}}, \ and\ \bibinfo {author} {\bibfnamefont {S.~V.}\ \bibnamefont
  {Popruzhenko}},\ }\bibfield  {title} {\enquote {\bibinfo {title}
  {{Focal-shape effects on the efficiency of the tunnel-ionization probe for
  extreme laser intensities}},}\ }\href {\doibase 10.1063/5.0005380} {\bibfield
   {journal} {\bibinfo  {journal} {Matter and Radiation at Extremes}\ }\textbf
  {\bibinfo {volume} {5}} (\bibinfo {year} {2020}),\ 10.1063/5.0005380},\
  \bibinfo {note} {044401}\BibitemShut {NoStop}%
\bibitem [{\citenamefont {Ouatu}\ \emph {et~al.}(2022)\citenamefont {Ouatu},
  \citenamefont {Spiers}, \citenamefont {Aboushelbaya}, \citenamefont {Feng},
  \citenamefont {von~der Leyen}, \citenamefont {Paddock}, \citenamefont
  {Timmis}, \citenamefont {Ticos}, \citenamefont {Krushelnick},\ and\
  \citenamefont {Norreys}}]{Ouatu22}%
  \BibitemOpen
  \bibfield  {author} {\bibinfo {author} {\bibfnamefont {I.}~\bibnamefont
  {Ouatu}}, \bibinfo {author} {\bibfnamefont {B.~T.}\ \bibnamefont {Spiers}},
  \bibinfo {author} {\bibfnamefont {R.}~\bibnamefont {Aboushelbaya}}, \bibinfo
  {author} {\bibfnamefont {Q.}~\bibnamefont {Feng}}, \bibinfo {author}
  {\bibfnamefont {M.~W.}\ \bibnamefont {von~der Leyen}}, \bibinfo {author}
  {\bibfnamefont {R.~W.}\ \bibnamefont {Paddock}}, \bibinfo {author}
  {\bibfnamefont {R.}~\bibnamefont {Timmis}}, \bibinfo {author} {\bibfnamefont
  {C.}~\bibnamefont {Ticos}}, \bibinfo {author} {\bibfnamefont {K.~M.}\
  \bibnamefont {Krushelnick}}, \ and\ \bibinfo {author} {\bibfnamefont {P.~A.}\
  \bibnamefont {Norreys}},\ }\bibfield  {title} {\enquote {\bibinfo {title}
  {Ionization states for the multipetawatt laser-qed regime},}\ }\href
  {\doibase 10.1103/PhysRevE.106.015205} {\bibfield  {journal} {\bibinfo
  {journal} {Phys. Rev. E}\ }\textbf {\bibinfo {volume} {106}},\ \bibinfo
  {pages} {015205} (\bibinfo {year} {2022})}\BibitemShut {NoStop}%
\bibitem [{\citenamefont {Link}\ \emph {et~al.}(2006)\citenamefont {Link},
  \citenamefont {Chowdhury}, \citenamefont {Morrison}, \citenamefont
  {Ovchinnikov}, \citenamefont {Offermann}, \citenamefont {Van~Woerkom},
  \citenamefont {Freeman}, \citenamefont {Pasley}, \citenamefont {Shipton},
  \citenamefont {Beg}, \citenamefont {Rambo}, \citenamefont {Schwarz},
  \citenamefont {Geissel}, \citenamefont {Edens},\ and\ \citenamefont
  {Porter}}]{Link06}%
  \BibitemOpen
  \bibfield  {author} {\bibinfo {author} {\bibfnamefont {A.}~\bibnamefont
  {Link}}, \bibinfo {author} {\bibfnamefont {E.~A.}\ \bibnamefont {Chowdhury}},
  \bibinfo {author} {\bibfnamefont {J.~T.}\ \bibnamefont {Morrison}}, \bibinfo
  {author} {\bibfnamefont {V.~M.}\ \bibnamefont {Ovchinnikov}}, \bibinfo
  {author} {\bibfnamefont {D.}~\bibnamefont {Offermann}}, \bibinfo {author}
  {\bibfnamefont {L.}~\bibnamefont {Van~Woerkom}}, \bibinfo {author}
  {\bibfnamefont {R.~R.}\ \bibnamefont {Freeman}}, \bibinfo {author}
  {\bibfnamefont {J.}~\bibnamefont {Pasley}}, \bibinfo {author} {\bibfnamefont
  {E.}~\bibnamefont {Shipton}}, \bibinfo {author} {\bibfnamefont
  {F.}~\bibnamefont {Beg}}, \bibinfo {author} {\bibfnamefont {P.}~\bibnamefont
  {Rambo}}, \bibinfo {author} {\bibfnamefont {J.}~\bibnamefont {Schwarz}},
  \bibinfo {author} {\bibfnamefont {M.}~\bibnamefont {Geissel}}, \bibinfo
  {author} {\bibfnamefont {A.}~\bibnamefont {Edens}}, \ and\ \bibinfo {author}
  {\bibfnamefont {J.~L.}\ \bibnamefont {Porter}},\ }\bibfield  {title}
  {\enquote {\bibinfo {title} {{Development of an in situ peak intensity
  measurement method for ultraintense single shot laser-plasma experiments at
  the Sandia Z petawatt facility}},}\ }\href {\doibase 10.1063/1.2336469}
  {\bibfield  {journal} {\bibinfo  {journal} {Review of Scientific
  Instruments}\ }\textbf {\bibinfo {volume} {77}} (\bibinfo {year} {2006}),\
  10.1063/1.2336469}\BibitemShut {NoStop}%
\bibitem [{\citenamefont {Smeenk}\ \emph {et~al.}(2011)\citenamefont {Smeenk},
  \citenamefont {Salvail}, \citenamefont {Arissian}, \citenamefont {Corkum},
  \citenamefont {Hebeisen},\ and\ \citenamefont {Staudte}}]{Smeenk:11}%
  \BibitemOpen
  \bibfield  {author} {\bibinfo {author} {\bibfnamefont {C.}~\bibnamefont
  {Smeenk}}, \bibinfo {author} {\bibfnamefont {J.~Z.}\ \bibnamefont {Salvail}},
  \bibinfo {author} {\bibfnamefont {L.}~\bibnamefont {Arissian}}, \bibinfo
  {author} {\bibfnamefont {P.~B.}\ \bibnamefont {Corkum}}, \bibinfo {author}
  {\bibfnamefont {C.~T.}\ \bibnamefont {Hebeisen}}, \ and\ \bibinfo {author}
  {\bibfnamefont {A.}~\bibnamefont {Staudte}},\ }\bibfield  {title} {\enquote
  {\bibinfo {title} {Precise in-situ measurement of laser pulse intensity using
  strong field ionization},}\ }\href {\doibase 10.1364/OE.19.009336} {\bibfield
   {journal} {\bibinfo  {journal} {Opt. Express}\ }\textbf {\bibinfo {volume}
  {19}},\ \bibinfo {pages} {9336--9344} (\bibinfo {year} {2011})}\BibitemShut
  {NoStop}%
\bibitem [{\citenamefont {Alnaser}\ \emph {et~al.}(2004)\citenamefont
  {Alnaser}, \citenamefont {Tong}, \citenamefont {Osipov}, \citenamefont
  {Voss}, \citenamefont {Maharjan}, \citenamefont {Shan}, \citenamefont
  {Chang},\ and\ \citenamefont {Cocke}}]{Alnaser04}%
  \BibitemOpen
  \bibfield  {author} {\bibinfo {author} {\bibfnamefont {A.~S.}\ \bibnamefont
  {Alnaser}}, \bibinfo {author} {\bibfnamefont {X.~M.}\ \bibnamefont {Tong}},
  \bibinfo {author} {\bibfnamefont {T.}~\bibnamefont {Osipov}}, \bibinfo
  {author} {\bibfnamefont {S.}~\bibnamefont {Voss}}, \bibinfo {author}
  {\bibfnamefont {C.~M.}\ \bibnamefont {Maharjan}}, \bibinfo {author}
  {\bibfnamefont {B.}~\bibnamefont {Shan}}, \bibinfo {author} {\bibfnamefont
  {Z.}~\bibnamefont {Chang}}, \ and\ \bibinfo {author} {\bibfnamefont {C.~L.}\
  \bibnamefont {Cocke}},\ }\bibfield  {title} {\enquote {\bibinfo {title}
  {Laser-peak-intensity calibration using recoil-ion momentum imaging},}\
  }\href {\doibase 10.1103/PhysRevA.70.023413} {\bibfield  {journal} {\bibinfo
  {journal} {Phys. Rev. A}\ }\textbf {\bibinfo {volume} {70}},\ \bibinfo
  {pages} {023413} (\bibinfo {year} {2004})}\BibitemShut {NoStop}%
\bibitem [{\citenamefont {Aleksandrov}\ and\ \citenamefont
  {Andreev}(2021)}]{Aleksandrov21}%
  \BibitemOpen
  \bibfield  {author} {\bibinfo {author} {\bibfnamefont {I.~A.}\ \bibnamefont
  {Aleksandrov}}\ and\ \bibinfo {author} {\bibfnamefont {A.~A.}\ \bibnamefont
  {Andreev}},\ }\bibfield  {title} {\enquote {\bibinfo {title} {Pair production
  seeded by electrons in noble gases as a method for laser intensity
  diagnostics},}\ }\href {\doibase 10.1103/PhysRevA.104.052801} {\bibfield
  {journal} {\bibinfo  {journal} {Phys. Rev. A}\ }\textbf {\bibinfo {volume}
  {104}},\ \bibinfo {pages} {052801} (\bibinfo {year} {2021})}\BibitemShut
  {NoStop}%
\bibitem [{\citenamefont {Galkin}\ \emph {et~al.}(2010)\citenamefont {Galkin},
  \citenamefont {Kalashnikov}, \citenamefont {Klinkov}, \citenamefont
  {Korobkin}, \citenamefont {Romanovsky},\ and\ \citenamefont
  {Shiryaev}}]{Galkin10}%
  \BibitemOpen
  \bibfield  {author} {\bibinfo {author} {\bibfnamefont {A.~L.}\ \bibnamefont
  {Galkin}}, \bibinfo {author} {\bibfnamefont {M.~P.}\ \bibnamefont
  {Kalashnikov}}, \bibinfo {author} {\bibfnamefont {V.~K.}\ \bibnamefont
  {Klinkov}}, \bibinfo {author} {\bibfnamefont {V.~V.}\ \bibnamefont
  {Korobkin}}, \bibinfo {author} {\bibfnamefont {M.~Y.}\ \bibnamefont
  {Romanovsky}}, \ and\ \bibinfo {author} {\bibfnamefont {O.~B.}\ \bibnamefont
  {Shiryaev}},\ }\bibfield  {title} {\enquote {\bibinfo {title}
  {{Electrodynamics of electron in a superintense laser field: New principles
  of diagnostics of relativistic laser intensity}},}\ }\href {\doibase
  10.1063/1.3425864} {\bibfield  {journal} {\bibinfo  {journal} {Physics of
  Plasmas}\ }\textbf {\bibinfo {volume} {17}} (\bibinfo {year} {2010}),\
  10.1063/1.3425864},\ \bibinfo {note} {053105}\BibitemShut {NoStop}%
\bibitem [{\citenamefont {Kalashnikov}\ \emph {et~al.}(2015)\citenamefont
  {Kalashnikov}, \citenamefont {Andreev}, \citenamefont {Ivanov}, \citenamefont
  {Galkin}, \citenamefont {Korobkin}, \citenamefont {Romanovsky}, \citenamefont
  {Shiryaev}, \citenamefont {Schnuerer}, \citenamefont {Braenzel},
  \citenamefont {Trofimov},\ and\ \citenamefont {et~al.}}]{kalashnikov_2015}%
  \BibitemOpen
  \bibfield  {author} {\bibinfo {author} {\bibfnamefont {M.}~\bibnamefont
  {Kalashnikov}}, \bibinfo {author} {\bibfnamefont {A.}~\bibnamefont
  {Andreev}}, \bibinfo {author} {\bibfnamefont {K.}~\bibnamefont {Ivanov}},
  \bibinfo {author} {\bibfnamefont {A.}~\bibnamefont {Galkin}}, \bibinfo
  {author} {\bibfnamefont {V.}~\bibnamefont {Korobkin}}, \bibinfo {author}
  {\bibfnamefont {M.}~\bibnamefont {Romanovsky}}, \bibinfo {author}
  {\bibfnamefont {O.}~\bibnamefont {Shiryaev}}, \bibinfo {author}
  {\bibfnamefont {M.}~\bibnamefont {Schnuerer}}, \bibinfo {author}
  {\bibfnamefont {J.}~\bibnamefont {Braenzel}}, \bibinfo {author}
  {\bibfnamefont {V.}~\bibnamefont {Trofimov}}, \ and\ \bibinfo {author}
  {\bibnamefont {et~al.}},\ }\bibfield  {title} {\enquote {\bibinfo {title}
  {Diagnostics of peak laser intensity based on the measurement of energy of
  electrons emitted from laser focal region},}\ }\href {\doibase
  10.1017/S0263034615000403} {\bibfield  {journal} {\bibinfo  {journal} {Laser
  and Particle Beams}\ }\textbf {\bibinfo {volume} {33}} (\bibinfo {year}
  {2015}),\ 10.1017/S0263034615000403}\BibitemShut {NoStop}%
\bibitem [{\citenamefont {Vais}\ \emph {et~al.}(2017)\citenamefont {Vais},
  \citenamefont {Bochkarev}, \citenamefont {Ter-Avetisyan},\ and\ \citenamefont
  {Bychenkov}}]{Vais_2017}%
  \BibitemOpen
  \bibfield  {author} {\bibinfo {author} {\bibfnamefont {O.}~\bibnamefont
  {Vais}}, \bibinfo {author} {\bibfnamefont {S.}~\bibnamefont {Bochkarev}},
  \bibinfo {author} {\bibfnamefont {S.}~\bibnamefont {Ter-Avetisyan}}, \ and\
  \bibinfo {author} {\bibfnamefont {V.}~\bibnamefont {Bychenkov}},\ }\bibfield
  {title} {\enquote {\bibinfo {title} {Angular distribution of electrons
  directly accelerated by an intense tightly focused laser pulse*},}\ }\href
  {\doibase 10.1070/QEL16259} {\bibfield  {journal} {\bibinfo  {journal}
  {Quantum Electronics}\ }\textbf {\bibinfo {volume} {47}},\ \bibinfo {pages}
  {38} (\bibinfo {year} {2017})}\BibitemShut {NoStop}%
\bibitem [{\citenamefont {Ivanov}\ \emph {et~al.}(2018)\citenamefont {Ivanov},
  \citenamefont {Tsymbalov}, \citenamefont {Vais}, \citenamefont {Bochkarev},
  \citenamefont {Volkov}, \citenamefont {Bychenkov},\ and\ \citenamefont
  {Savel’ev}}]{Ivanov_2018}%
  \BibitemOpen
  \bibfield  {author} {\bibinfo {author} {\bibfnamefont {K.~A.}\ \bibnamefont
  {Ivanov}}, \bibinfo {author} {\bibfnamefont {I.~N.}\ \bibnamefont
  {Tsymbalov}}, \bibinfo {author} {\bibfnamefont {O.~E.}\ \bibnamefont {Vais}},
  \bibinfo {author} {\bibfnamefont {S.~G.}\ \bibnamefont {Bochkarev}}, \bibinfo
  {author} {\bibfnamefont {R.~V.}\ \bibnamefont {Volkov}}, \bibinfo {author}
  {\bibfnamefont {V.~Y.}\ \bibnamefont {Bychenkov}}, \ and\ \bibinfo {author}
  {\bibfnamefont {A.~B.}\ \bibnamefont {Savel’ev}},\ }\bibfield  {title}
  {\enquote {\bibinfo {title} {Accelerated electrons for in situ peak intensity
  monitoring of tightly focused femtosecond laser radiation at high
  intensities},}\ }\href {\doibase 10.1088/1361-6587/aada60} {\bibfield
  {journal} {\bibinfo  {journal} {Plasma Physics and Controlled Fusion}\
  }\textbf {\bibinfo {volume} {60}},\ \bibinfo {pages} {105011} (\bibinfo
  {year} {2018})}\BibitemShut {NoStop}%
\bibitem [{\citenamefont {Vais}\ and\ \citenamefont
  {Bychenkov}(2018)}]{Vais2018}%
  \BibitemOpen
  \bibfield  {author} {\bibinfo {author} {\bibfnamefont {O.~E.}\ \bibnamefont
  {Vais}}\ and\ \bibinfo {author} {\bibfnamefont {V.~Y.}\ \bibnamefont
  {Bychenkov}},\ }\bibfield  {title} {\enquote {\bibinfo {title} {Direct
  electron acceleration for diagnostics of a laser pulse focused by an off-axis
  parabolic mirror},}\ }\href {\doibase 10.1007/s00340-018-7084-9} {\bibfield
  {journal} {\bibinfo  {journal} {Applied Physics B}\ }\textbf {\bibinfo
  {volume} {124}},\ \bibinfo {pages} {211} (\bibinfo {year}
  {2018})}\BibitemShut {NoStop}%
\bibitem [{\citenamefont {Kr{\"a}mer}\ \emph {et~al.}(2018)\citenamefont
  {Kr{\"a}mer}, \citenamefont {Jochmann}, \citenamefont {Budde}, \citenamefont
  {Bussmann}, \citenamefont {Couperus}, \citenamefont {Cowan}, \citenamefont
  {Debus}, \citenamefont {K{\"o}hler}, \citenamefont {Kuntzsch}, \citenamefont
  {Laso~Garc{\'i}a}, \citenamefont {Lehnert}, \citenamefont {Michel},
  \citenamefont {Pausch}, \citenamefont {Zarini}, \citenamefont {Schramm},\
  and\ \citenamefont {Irman}}]{Kramer2018}%
  \BibitemOpen
  \bibfield  {author} {\bibinfo {author} {\bibfnamefont {J.~M.}\ \bibnamefont
  {Kr{\"a}mer}}, \bibinfo {author} {\bibfnamefont {A.}~\bibnamefont
  {Jochmann}}, \bibinfo {author} {\bibfnamefont {M.}~\bibnamefont {Budde}},
  \bibinfo {author} {\bibfnamefont {M.}~\bibnamefont {Bussmann}}, \bibinfo
  {author} {\bibfnamefont {J.~P.}\ \bibnamefont {Couperus}}, \bibinfo {author}
  {\bibfnamefont {T.~E.}\ \bibnamefont {Cowan}}, \bibinfo {author}
  {\bibfnamefont {A.}~\bibnamefont {Debus}}, \bibinfo {author} {\bibfnamefont
  {A.}~\bibnamefont {K{\"o}hler}}, \bibinfo {author} {\bibfnamefont
  {M.}~\bibnamefont {Kuntzsch}}, \bibinfo {author} {\bibfnamefont
  {A.}~\bibnamefont {Laso~Garc{\'i}a}}, \bibinfo {author} {\bibfnamefont
  {U.}~\bibnamefont {Lehnert}}, \bibinfo {author} {\bibfnamefont
  {P.}~\bibnamefont {Michel}}, \bibinfo {author} {\bibfnamefont
  {R.}~\bibnamefont {Pausch}}, \bibinfo {author} {\bibfnamefont
  {O.}~\bibnamefont {Zarini}}, \bibinfo {author} {\bibfnamefont
  {U.}~\bibnamefont {Schramm}}, \ and\ \bibinfo {author} {\bibfnamefont
  {A.}~\bibnamefont {Irman}},\ }\bibfield  {title} {\enquote {\bibinfo {title}
  {Making spectral shape measurements in inverse compton scattering a tool for
  advanced diagnostic applications},}\ }\href {\doibase
  10.1038/s41598-018-19546-0} {\bibfield  {journal} {\bibinfo  {journal}
  {Scientific Reports}\ }\textbf {\bibinfo {volume} {8}},\ \bibinfo {pages}
  {1398} (\bibinfo {year} {2018})}\BibitemShut {NoStop}%
\bibitem [{\citenamefont {Mackenroth}, \citenamefont {Holkundkar},\ and\
  \citenamefont {Schlenvoigt}(2019)}]{Mackenroth_2019}%
  \BibitemOpen
  \bibfield  {author} {\bibinfo {author} {\bibfnamefont {F.}~\bibnamefont
  {Mackenroth}}, \bibinfo {author} {\bibfnamefont {A.~R.}\ \bibnamefont
  {Holkundkar}}, \ and\ \bibinfo {author} {\bibfnamefont {H.-P.}\ \bibnamefont
  {Schlenvoigt}},\ }\bibfield  {title} {\enquote {\bibinfo {title}
  {Ultra-intense laser pulse characterization using ponderomotive electron
  scattering},}\ }\href {\doibase 10.1088/1367-2630/ab5c4d} {\bibfield
  {journal} {\bibinfo  {journal} {New Journal of Physics}\ }\textbf {\bibinfo
  {volume} {21}},\ \bibinfo {pages} {123028} (\bibinfo {year}
  {2019})}\BibitemShut {NoStop}%
\bibitem [{\citenamefont {Vais}\ \emph {et~al.}(2020)\citenamefont {Vais},
  \citenamefont {Thomas}, \citenamefont {Maksimchuk}, \citenamefont
  {Krushelnick},\ and\ \citenamefont {Bychenkov}}]{Vais_2020}%
  \BibitemOpen
  \bibfield  {author} {\bibinfo {author} {\bibfnamefont {O.~E.}\ \bibnamefont
  {Vais}}, \bibinfo {author} {\bibfnamefont {A.~G.~R.}\ \bibnamefont {Thomas}},
  \bibinfo {author} {\bibfnamefont {A.~M.}\ \bibnamefont {Maksimchuk}},
  \bibinfo {author} {\bibfnamefont {K.}~\bibnamefont {Krushelnick}}, \ and\
  \bibinfo {author} {\bibfnamefont {V.~Y.}\ \bibnamefont {Bychenkov}},\
  }\bibfield  {title} {\enquote {\bibinfo {title} {Characterizing extreme laser
  intensities by ponderomotive acceleration of protons from rarified gas},}\
  }\href {\doibase 10.1088/1367-2630/ab6eac} {\bibfield  {journal} {\bibinfo
  {journal} {New Journal of Physics}\ }\textbf {\bibinfo {volume} {22}},\
  \bibinfo {pages} {023003} (\bibinfo {year} {2020})}\BibitemShut {NoStop}%
\bibitem [{\citenamefont {Vais}\ and\ \citenamefont
  {Bychenkov}(2020)}]{Vais_2021}%
  \BibitemOpen
  \bibfield  {author} {\bibinfo {author} {\bibfnamefont {O.~E.}\ \bibnamefont
  {Vais}}\ and\ \bibinfo {author} {\bibfnamefont {V.~Y.}\ \bibnamefont
  {Bychenkov}},\ }\bibfield  {title} {\enquote {\bibinfo {title} {Complementary
  diagnostics of high-intensity femtosecond laser pulses via vacuum
  acceleration of protons and electrons},}\ }\href {\doibase
  10.1088/1361-6587/abc92a} {\bibfield  {journal} {\bibinfo  {journal} {Plasma
  Physics and Controlled Fusion}\ }\textbf {\bibinfo {volume} {63}},\ \bibinfo
  {pages} {014002} (\bibinfo {year} {2020})}\BibitemShut {NoStop}%
\bibitem [{\citenamefont {Bukharskii}\ \emph {et~al.}(2022)\citenamefont
  {Bukharskii}, \citenamefont {Vais}, \citenamefont {Korneev},\ and\
  \citenamefont {Bychenkov}}]{Bukharskii22}%
  \BibitemOpen
  \bibfield  {author} {\bibinfo {author} {\bibfnamefont {N.~D.}\ \bibnamefont
  {Bukharskii}}, \bibinfo {author} {\bibfnamefont {O.~E.}\ \bibnamefont
  {Vais}}, \bibinfo {author} {\bibfnamefont {P.~A.}\ \bibnamefont {Korneev}}, \
  and\ \bibinfo {author} {\bibfnamefont {V.~Y.}\ \bibnamefont {Bychenkov}},\
  }\bibfield  {title} {\enquote {\bibinfo {title} {{Restoration of the focal
  parameters for an extreme-power laser pulse with ponderomotively scattered
  proton spectra by using a neural network algorithm}},}\ }\href {\doibase
  10.1063/5.0126571} {\bibfield  {journal} {\bibinfo  {journal} {Matter and
  Radiation at Extremes}\ }\textbf {\bibinfo {volume} {8}} (\bibinfo {year}
  {2022}),\ 10.1063/5.0126571}\BibitemShut {NoStop}%
\bibitem [{\citenamefont {Jeandet}\ \emph {et~al.}(2022)\citenamefont
  {Jeandet}, \citenamefont {Jolly}, \citenamefont {Borot}, \citenamefont
  {Bussi\`{e}re}, \citenamefont {Dumont}, \citenamefont {Gautier},
  \citenamefont {Gobert}, \citenamefont {Goddet}, \citenamefont {Gonsalves},
  \citenamefont {Irman}, \citenamefont {Leemans}, \citenamefont
  {Lopez-Martens}, \citenamefont {Mennerat}, \citenamefont {Nakamura},
  \citenamefont {Ouill\'{e}}, \citenamefont {Pariente}, \citenamefont
  {Pittman}, \citenamefont {P\"{u}schel}, \citenamefont {Sanson}, \citenamefont
  {Sylla}, \citenamefont {Thaury}, \citenamefont {Zeil},\ and\ \citenamefont
  {Qu\'{e}r\'{e}}}]{Jeandet:22}%
  \BibitemOpen
  \bibfield  {author} {\bibinfo {author} {\bibfnamefont {A.}~\bibnamefont
  {Jeandet}}, \bibinfo {author} {\bibfnamefont {S.~W.}\ \bibnamefont {Jolly}},
  \bibinfo {author} {\bibfnamefont {A.}~\bibnamefont {Borot}}, \bibinfo
  {author} {\bibfnamefont {B.}~\bibnamefont {Bussi\`{e}re}}, \bibinfo {author}
  {\bibfnamefont {P.}~\bibnamefont {Dumont}}, \bibinfo {author} {\bibfnamefont
  {J.}~\bibnamefont {Gautier}}, \bibinfo {author} {\bibfnamefont
  {O.}~\bibnamefont {Gobert}}, \bibinfo {author} {\bibfnamefont {J.-P.}\
  \bibnamefont {Goddet}}, \bibinfo {author} {\bibfnamefont {A.}~\bibnamefont
  {Gonsalves}}, \bibinfo {author} {\bibfnamefont {A.}~\bibnamefont {Irman}},
  \bibinfo {author} {\bibfnamefont {W.~P.}\ \bibnamefont {Leemans}}, \bibinfo
  {author} {\bibfnamefont {R.}~\bibnamefont {Lopez-Martens}}, \bibinfo {author}
  {\bibfnamefont {G.}~\bibnamefont {Mennerat}}, \bibinfo {author}
  {\bibfnamefont {K.}~\bibnamefont {Nakamura}}, \bibinfo {author}
  {\bibfnamefont {M.}~\bibnamefont {Ouill\'{e}}}, \bibinfo {author}
  {\bibfnamefont {G.}~\bibnamefont {Pariente}}, \bibinfo {author}
  {\bibfnamefont {M.}~\bibnamefont {Pittman}}, \bibinfo {author} {\bibfnamefont
  {T.}~\bibnamefont {P\"{u}schel}}, \bibinfo {author} {\bibfnamefont
  {F.}~\bibnamefont {Sanson}}, \bibinfo {author} {\bibfnamefont
  {F.}~\bibnamefont {Sylla}}, \bibinfo {author} {\bibfnamefont
  {C.}~\bibnamefont {Thaury}}, \bibinfo {author} {\bibfnamefont
  {K.}~\bibnamefont {Zeil}}, \ and\ \bibinfo {author} {\bibfnamefont
  {F.}~\bibnamefont {Qu\'{e}r\'{e}}},\ }\bibfield  {title} {\enquote {\bibinfo
  {title} {Survey of spatio-temporal couplings throughout high-power ultrashort
  lasers},}\ }\href {\doibase 10.1364/OE.444564} {\bibfield  {journal}
  {\bibinfo  {journal} {Opt. Express}\ }\textbf {\bibinfo {volume} {30}},\
  \bibinfo {pages} {3262--3288} (\bibinfo {year} {2022})}\BibitemShut {NoStop}%
\bibitem [{\citenamefont {Grace}\ \emph {et~al.}(2022)\citenamefont {Grace},
  \citenamefont {Djordjevic}, \citenamefont {Guang}, \citenamefont {Mariscal},
  \citenamefont {Scott}, \citenamefont {Simpson}, \citenamefont {Swanson},
  \citenamefont {Zeraouli}, \citenamefont {Stuart}, \citenamefont {Trebino},\
  and\ \citenamefont {Ma}}]{Grace22}%
  \BibitemOpen
  \bibfield  {author} {\bibinfo {author} {\bibfnamefont {E.~S.}\ \bibnamefont
  {Grace}}, \bibinfo {author} {\bibfnamefont {B.~Z.}\ \bibnamefont
  {Djordjevic}}, \bibinfo {author} {\bibfnamefont {Z.}~\bibnamefont {Guang}},
  \bibinfo {author} {\bibfnamefont {D.}~\bibnamefont {Mariscal}}, \bibinfo
  {author} {\bibfnamefont {G.~G.}\ \bibnamefont {Scott}}, \bibinfo {author}
  {\bibfnamefont {R.~A.}\ \bibnamefont {Simpson}}, \bibinfo {author}
  {\bibfnamefont {K.~K.}\ \bibnamefont {Swanson}}, \bibinfo {author}
  {\bibfnamefont {G.}~\bibnamefont {Zeraouli}}, \bibinfo {author}
  {\bibfnamefont {B.}~\bibnamefont {Stuart}}, \bibinfo {author} {\bibfnamefont
  {R.}~\bibnamefont {Trebino}}, \ and\ \bibinfo {author} {\bibfnamefont
  {T.}~\bibnamefont {Ma}},\ }\bibfield  {title} {\enquote {\bibinfo {title}
  {{Single-shot measurements of pulse-front tilt in intense ps laser pulses and
  its effect on accelerated electron and ion beam characteristics
  (invited)}},}\ }\href {\doibase 10.1063/5.0101803} {\bibfield  {journal}
  {\bibinfo  {journal} {Review of Scientific Instruments}\ }\textbf {\bibinfo
  {volume} {93}} (\bibinfo {year} {2022}),\ 10.1063/5.0101803},\ \bibinfo
  {note} {123508}\BibitemShut {NoStop}%
\bibitem [{\citenamefont {Wilhelm}\ and\ \citenamefont
  {Durfee}(2019)}]{Wilhelm19}%
  \BibitemOpen
  \bibfield  {author} {\bibinfo {author} {\bibfnamefont {A.~M.}\ \bibnamefont
  {Wilhelm}}\ and\ \bibinfo {author} {\bibfnamefont {C.~G.}\ \bibnamefont
  {Durfee}},\ }\bibfield  {title} {\enquote {\bibinfo {title} {Tilted snowplow
  ponderomotive electron acceleration with spatio-temporally shaped ultrafast
  laser pulses},}\ }\href {\doibase 10.3389/fphy.2019.00066} {\bibfield
  {journal} {\bibinfo  {journal} {Frontiers in Physics}\ }\textbf {\bibinfo
  {volume} {7}} (\bibinfo {year} {2019}),\ 10.3389/fphy.2019.00066}\BibitemShut
  {NoStop}%
\bibitem [{\citenamefont {Grace}\ \emph {et~al.}(2021)\citenamefont {Grace},
  \citenamefont {Ma}, \citenamefont {Guang}, \citenamefont {Jafari},
  \citenamefont {Park}, \citenamefont {Clark}, \citenamefont {Kemp},
  \citenamefont {Moody}, \citenamefont {Rhodes}, \citenamefont {Ping},
  \citenamefont {Shepherd}, \citenamefont {Stuart},\ and\ \citenamefont
  {Trebino}}]{Grace_2021}%
  \BibitemOpen
  \bibfield  {author} {\bibinfo {author} {\bibfnamefont {E.}~\bibnamefont
  {Grace}}, \bibinfo {author} {\bibfnamefont {T.}~\bibnamefont {Ma}}, \bibinfo
  {author} {\bibfnamefont {Z.}~\bibnamefont {Guang}}, \bibinfo {author}
  {\bibfnamefont {R.}~\bibnamefont {Jafari}}, \bibinfo {author} {\bibfnamefont
  {J.}~\bibnamefont {Park}}, \bibinfo {author} {\bibfnamefont {J.}~\bibnamefont
  {Clark}}, \bibinfo {author} {\bibfnamefont {G.}~\bibnamefont {Kemp}},
  \bibinfo {author} {\bibfnamefont {J.}~\bibnamefont {Moody}}, \bibinfo
  {author} {\bibfnamefont {M.}~\bibnamefont {Rhodes}}, \bibinfo {author}
  {\bibfnamefont {Y.}~\bibnamefont {Ping}}, \bibinfo {author} {\bibfnamefont
  {R.}~\bibnamefont {Shepherd}}, \bibinfo {author} {\bibfnamefont
  {B.}~\bibnamefont {Stuart}}, \ and\ \bibinfo {author} {\bibfnamefont
  {R.}~\bibnamefont {Trebino}},\ }\bibfield  {title} {\enquote {\bibinfo
  {title} {Single-shot complete spatiotemporal measurement of terawatt laser
  pulses},}\ }\href {\doibase 10.1088/2040-8986/ac0e1b} {\bibfield  {journal}
  {\bibinfo  {journal} {Journal of Optics}\ }\textbf {\bibinfo {volume} {23}},\
  \bibinfo {pages} {075505} (\bibinfo {year} {2021})}\BibitemShut {NoStop}%
\bibitem [{\citenamefont {Pariente}\ \emph {et~al.}(2016)\citenamefont
  {Pariente}, \citenamefont {Gallet}, \citenamefont {Borot}, \citenamefont
  {Gobert},\ and\ \citenamefont {Qu{\'e}r{\'e}}}]{Pariente2016}%
  \BibitemOpen
  \bibfield  {author} {\bibinfo {author} {\bibfnamefont {G.}~\bibnamefont
  {Pariente}}, \bibinfo {author} {\bibfnamefont {V.}~\bibnamefont {Gallet}},
  \bibinfo {author} {\bibfnamefont {A.}~\bibnamefont {Borot}}, \bibinfo
  {author} {\bibfnamefont {O.}~\bibnamefont {Gobert}}, \ and\ \bibinfo {author}
  {\bibfnamefont {F.}~\bibnamefont {Qu{\'e}r{\'e}}},\ }\bibfield  {title}
  {\enquote {\bibinfo {title} {Space--time characterization of ultra-intense
  femtosecond laser beams},}\ }\href {\doibase 10.1038/nphoton.2016.140}
  {\bibfield  {journal} {\bibinfo  {journal} {Nature Photonics}\ }\textbf
  {\bibinfo {volume} {10}},\ \bibinfo {pages} {547--553} (\bibinfo {year}
  {2016})}\BibitemShut {NoStop}%
\bibitem [{\citenamefont {Ravichandran}\ \emph {et~al.}(2023)\citenamefont
  {Ravichandran}, \citenamefont {Longman}, \citenamefont {Huault},
  \citenamefont {Lera}, \citenamefont {He}, \citenamefont {Fedosejevs},
  \citenamefont {Roso},\ and\ \citenamefont {III}}]{Ravichandran23}%
  \BibitemOpen
  \bibfield  {author} {\bibinfo {author} {\bibfnamefont {S.}~\bibnamefont
  {Ravichandran}}, \bibinfo {author} {\bibfnamefont {A.}~\bibnamefont
  {Longman}}, \bibinfo {author} {\bibfnamefont {M.}~\bibnamefont {Huault}},
  \bibinfo {author} {\bibfnamefont {R.}~\bibnamefont {Lera}}, \bibinfo {author}
  {\bibfnamefont {C.~Z.}\ \bibnamefont {He}}, \bibinfo {author} {\bibfnamefont
  {R.}~\bibnamefont {Fedosejevs}}, \bibinfo {author} {\bibfnamefont
  {L.}~\bibnamefont {Roso}}, \ and\ \bibinfo {author} {\bibfnamefont
  {W.~T.~H.}\ \bibnamefont {III}},\ }\href@noop {} {\enquote {\bibinfo {title}
  {Imaging electron angular distributions to assess a full-power petawatt-class
  laser focus},}\ } (\bibinfo {year} {2023}),\ \Eprint
  {http://arxiv.org/abs/arXiv:2305.14581} {arXiv:2305.14581} \BibitemShut
  {NoStop}%
\bibitem [{\citenamefont {Gibbon}(2005)}]{Gibbon}%
  \BibitemOpen
  \bibfield  {author} {\bibinfo {author} {\bibfnamefont {P.}~\bibnamefont
  {Gibbon}},\ }\href@noop {} {\emph {\bibinfo {title} {Short Pulse Laser
  Interactions with Matter}}},\ \bibinfo {edition} {1st}\ ed.\ (\bibinfo
  {publisher} {Imperial College Press},\ \bibinfo {year} {2005})\BibitemShut
  {NoStop}%
\bibitem [{\citenamefont {Macchi}(2013)}]{Macchilasers}%
  \BibitemOpen
  \bibfield  {author} {\bibinfo {author} {\bibfnamefont {A.}~\bibnamefont
  {Macchi}},\ }\href@noop {} {\emph {\bibinfo {title} {{A Superintense
  Laser-Plasma Interaction Theory Primer}}}},\ \bibinfo {edition} {1st}\ ed.\
  (\bibinfo  {publisher} {Springer Netherlands},\ \bibinfo {year}
  {2013})\BibitemShut {NoStop}%
\bibitem [{\citenamefont {Meyerhofer}\ \emph {et~al.}(1996)\citenamefont
  {Meyerhofer}, \citenamefont {Knauer}, \citenamefont {McNaught},\ and\
  \citenamefont {Moore}}]{Meyerhofer:96}%
  \BibitemOpen
  \bibfield  {author} {\bibinfo {author} {\bibfnamefont {D.~D.}\ \bibnamefont
  {Meyerhofer}}, \bibinfo {author} {\bibfnamefont {J.~P.}\ \bibnamefont
  {Knauer}}, \bibinfo {author} {\bibfnamefont {S.~J.}\ \bibnamefont
  {McNaught}}, \ and\ \bibinfo {author} {\bibfnamefont {C.~I.}\ \bibnamefont
  {Moore}},\ }\bibfield  {title} {\enquote {\bibinfo {title} {Observation of
  relativistic mass shift effects during high-intensity laser--electron
  interactions},}\ }\href {\doibase 10.1364/JOSAB.13.000113} {\bibfield
  {journal} {\bibinfo  {journal} {J. Opt. Soc. Am. B}\ }\textbf {\bibinfo
  {volume} {13}},\ \bibinfo {pages} {113--117} (\bibinfo {year}
  {1996})}\BibitemShut {NoStop}%
\bibitem [{\citenamefont {Singh}(2006)}]{Singh06}%
  \BibitemOpen
  \bibfield  {author} {\bibinfo {author} {\bibfnamefont {K.~P.}\ \bibnamefont
  {Singh}},\ }\bibfield  {title} {\enquote {\bibinfo {title} {{Self-injection
  and acceleration of electrons during ionization of gas atoms by a short laser
  pulse}},}\ }\href {\doibase 10.1063/1.2187449} {\bibfield  {journal}
  {\bibinfo  {journal} {Physics of Plasmas}\ }\textbf {\bibinfo {volume} {13}}
  (\bibinfo {year} {2006}),\ 10.1063/1.2187449},\ \bibinfo {note}
  {043101}\BibitemShut {NoStop}%
\bibitem [{\citenamefont {Lin}\ \emph {et~al.}(2007)\citenamefont {Lin},
  \citenamefont {Kong}, \citenamefont {Chen}, \citenamefont {Wang},
  \citenamefont {Xu},\ and\ \citenamefont {Ho}}]{Lin2007}%
  \BibitemOpen
  \bibfield  {author} {\bibinfo {author} {\bibfnamefont {D.}~\bibnamefont
  {Lin}}, \bibinfo {author} {\bibfnamefont {Q.}~\bibnamefont {Kong}}, \bibinfo
  {author} {\bibfnamefont {Z.}~\bibnamefont {Chen}}, \bibinfo {author}
  {\bibfnamefont {P.~X.}\ \bibnamefont {Wang}}, \bibinfo {author}
  {\bibfnamefont {J.~J.}\ \bibnamefont {Xu}}, \ and\ \bibinfo {author}
  {\bibfnamefont {Y.~K.}\ \bibnamefont {Ho}},\ }\bibfield  {title} {\enquote
  {\bibinfo {title} {Scaling laws of electron acceleration driven by an intense
  laser pulse},}\ }\href {\doibase 10.1007/s00340-007-2869-2} {\bibfield
  {journal} {\bibinfo  {journal} {Applied Physics B}\ }\textbf {\bibinfo
  {volume} {89}},\ \bibinfo {pages} {549--552} (\bibinfo {year}
  {2007})}\BibitemShut {NoStop}%
\bibitem [{\citenamefont {Wang}\ \emph {et~al.}(2002)\citenamefont {Wang},
  \citenamefont {Ho}, \citenamefont {Yuan}, \citenamefont {Kong}, \citenamefont
  {Cao}, \citenamefont {Shao}, \citenamefont {Sessler}, \citenamefont {Esarey},
  \citenamefont {Moshkovich}, \citenamefont {Nishida}, \citenamefont {Yugami},
  \citenamefont {Ito}, \citenamefont {Wang},\ and\ \citenamefont
  {Scheid}}]{Wang02}%
  \BibitemOpen
  \bibfield  {author} {\bibinfo {author} {\bibfnamefont {P.~X.}\ \bibnamefont
  {Wang}}, \bibinfo {author} {\bibfnamefont {Y.~K.}\ \bibnamefont {Ho}},
  \bibinfo {author} {\bibfnamefont {X.~Q.}\ \bibnamefont {Yuan}}, \bibinfo
  {author} {\bibfnamefont {Q.}~\bibnamefont {Kong}}, \bibinfo {author}
  {\bibfnamefont {N.}~\bibnamefont {Cao}}, \bibinfo {author} {\bibfnamefont
  {L.}~\bibnamefont {Shao}}, \bibinfo {author} {\bibfnamefont {A.~M.}\
  \bibnamefont {Sessler}}, \bibinfo {author} {\bibfnamefont {E.}~\bibnamefont
  {Esarey}}, \bibinfo {author} {\bibfnamefont {E.}~\bibnamefont {Moshkovich}},
  \bibinfo {author} {\bibfnamefont {Y.}~\bibnamefont {Nishida}}, \bibinfo
  {author} {\bibfnamefont {N.}~\bibnamefont {Yugami}}, \bibinfo {author}
  {\bibfnamefont {H.}~\bibnamefont {Ito}}, \bibinfo {author} {\bibfnamefont
  {J.~X.}\ \bibnamefont {Wang}}, \ and\ \bibinfo {author} {\bibfnamefont
  {S.}~\bibnamefont {Scheid}},\ }\bibfield  {title} {\enquote {\bibinfo {title}
  {{Characteristics of laser-driven electron acceleration in vacuum}},}\ }\href
  {\doibase 10.1063/1.1423394} {\bibfield  {journal} {\bibinfo  {journal}
  {Journal of Applied Physics}\ }\textbf {\bibinfo {volume} {91}},\ \bibinfo
  {pages} {856--866} (\bibinfo {year} {2002})}\BibitemShut {NoStop}%
\bibitem [{\citenamefont {He}\ \emph {et~al.}(2003)\citenamefont {He},
  \citenamefont {Yu}, \citenamefont {Lu}, \citenamefont {Xu}, \citenamefont
  {Qian}, \citenamefont {Shen}, \citenamefont {Yuan}, \citenamefont {Li},\ and\
  \citenamefont {Xu}}]{He03}%
  \BibitemOpen
  \bibfield  {author} {\bibinfo {author} {\bibfnamefont {F.}~\bibnamefont
  {He}}, \bibinfo {author} {\bibfnamefont {W.}~\bibnamefont {Yu}}, \bibinfo
  {author} {\bibfnamefont {P.}~\bibnamefont {Lu}}, \bibinfo {author}
  {\bibfnamefont {H.}~\bibnamefont {Xu}}, \bibinfo {author} {\bibfnamefont
  {L.}~\bibnamefont {Qian}}, \bibinfo {author} {\bibfnamefont {B.}~\bibnamefont
  {Shen}}, \bibinfo {author} {\bibfnamefont {X.}~\bibnamefont {Yuan}}, \bibinfo
  {author} {\bibfnamefont {R.}~\bibnamefont {Li}}, \ and\ \bibinfo {author}
  {\bibfnamefont {Z.}~\bibnamefont {Xu}},\ }\bibfield  {title} {\enquote
  {\bibinfo {title} {Ponderomotive acceleration of electrons by a tightly
  focused intense laser beam},}\ }\href {\doibase 10.1103/PhysRevE.68.046407}
  {\bibfield  {journal} {\bibinfo  {journal} {Phys. Rev. E}\ }\textbf {\bibinfo
  {volume} {68}},\ \bibinfo {pages} {046407} (\bibinfo {year}
  {2003})}\BibitemShut {NoStop}%
\bibitem [{\citenamefont {Hartemann}\ \emph {et~al.}(1995)\citenamefont
  {Hartemann}, \citenamefont {Fochs}, \citenamefont {Le~Sage}, \citenamefont
  {Luhmann}, \citenamefont {Woodworth}, \citenamefont {Perry}, \citenamefont
  {Chen},\ and\ \citenamefont {Kerman}}]{Hartemann95}%
  \BibitemOpen
  \bibfield  {author} {\bibinfo {author} {\bibfnamefont {F.~V.}\ \bibnamefont
  {Hartemann}}, \bibinfo {author} {\bibfnamefont {S.~N.}\ \bibnamefont
  {Fochs}}, \bibinfo {author} {\bibfnamefont {G.~P.}\ \bibnamefont {Le~Sage}},
  \bibinfo {author} {\bibfnamefont {N.~C.}\ \bibnamefont {Luhmann}}, \bibinfo
  {author} {\bibfnamefont {J.~G.}\ \bibnamefont {Woodworth}}, \bibinfo {author}
  {\bibfnamefont {M.~D.}\ \bibnamefont {Perry}}, \bibinfo {author}
  {\bibfnamefont {Y.~J.}\ \bibnamefont {Chen}}, \ and\ \bibinfo {author}
  {\bibfnamefont {A.~K.}\ \bibnamefont {Kerman}},\ }\bibfield  {title}
  {\enquote {\bibinfo {title} {Nonlinear ponderomotive scattering of
  relativistic electrons by an intense laser field at focus},}\ }\href
  {\doibase 10.1103/PhysRevE.51.4833} {\bibfield  {journal} {\bibinfo
  {journal} {Phys. Rev. E}\ }\textbf {\bibinfo {volume} {51}},\ \bibinfo
  {pages} {4833--4843} (\bibinfo {year} {1995})}\BibitemShut {NoStop}%
\bibitem [{\citenamefont {Quesnel}\ and\ \citenamefont
  {Mora}(1998)}]{Quesnel98}%
  \BibitemOpen
  \bibfield  {author} {\bibinfo {author} {\bibfnamefont {B.}~\bibnamefont
  {Quesnel}}\ and\ \bibinfo {author} {\bibfnamefont {P.}~\bibnamefont {Mora}},\
  }\bibfield  {title} {\enquote {\bibinfo {title} {Theory and simulation of the
  interaction of ultraintense laser pulses with electrons in vacuum},}\ }\href
  {\doibase 10.1103/PhysRevE.58.3719} {\bibfield  {journal} {\bibinfo
  {journal} {Phys. Rev. E}\ }\textbf {\bibinfo {volume} {58}},\ \bibinfo
  {pages} {3719--3732} (\bibinfo {year} {1998})}\BibitemShut {NoStop}%
\bibitem [{\citenamefont {Yang}, \citenamefont {Craxton},\ and\ \citenamefont
  {Haines}(2011)}]{Yang_2011}%
  \BibitemOpen
  \bibfield  {author} {\bibinfo {author} {\bibfnamefont {J.-H.}\ \bibnamefont
  {Yang}}, \bibinfo {author} {\bibfnamefont {R.~S.}\ \bibnamefont {Craxton}}, \
  and\ \bibinfo {author} {\bibfnamefont {M.~G.}\ \bibnamefont {Haines}},\
  }\bibfield  {title} {\enquote {\bibinfo {title} {Explicit general solutions
  to relativistic electron dynamics in plane-wave electromagnetic fields and
  simulations of ponderomotive acceleration},}\ }\href {\doibase
  10.1088/0741-3335/53/12/125006} {\bibfield  {journal} {\bibinfo  {journal}
  {Plasma Physics and Controlled Fusion}\ }\textbf {\bibinfo {volume} {53}},\
  \bibinfo {pages} {125006} (\bibinfo {year} {2011})}\BibitemShut {NoStop}%
\bibitem [{\citenamefont {Kibble}(1966)}]{Kibble66}%
  \BibitemOpen
  \bibfield  {author} {\bibinfo {author} {\bibfnamefont {T.~W.~B.}\
  \bibnamefont {Kibble}},\ }\bibfield  {title} {\enquote {\bibinfo {title}
  {Mutual refraction of electrons and photons},}\ }\href {\doibase
  10.1103/PhysRev.150.1060} {\bibfield  {journal} {\bibinfo  {journal} {Phys.
  Rev.}\ }\textbf {\bibinfo {volume} {150}},\ \bibinfo {pages} {1060--1069}
  (\bibinfo {year} {1966})}\BibitemShut {NoStop}%
\bibitem [{\citenamefont {Bauer}, \citenamefont {Mulser},\ and\ \citenamefont
  {Steeb}(1995)}]{Bauer95}%
  \BibitemOpen
  \bibfield  {author} {\bibinfo {author} {\bibfnamefont {D.}~\bibnamefont
  {Bauer}}, \bibinfo {author} {\bibfnamefont {P.}~\bibnamefont {Mulser}}, \
  and\ \bibinfo {author} {\bibfnamefont {W.~H.}\ \bibnamefont {Steeb}},\
  }\bibfield  {title} {\enquote {\bibinfo {title} {Relativistic ponderomotive
  force, uphill acceleration, and transition to chaos},}\ }\href {\doibase
  10.1103/PhysRevLett.75.4622} {\bibfield  {journal} {\bibinfo  {journal}
  {Phys. Rev. Lett.}\ }\textbf {\bibinfo {volume} {75}},\ \bibinfo {pages}
  {4622--4625} (\bibinfo {year} {1995})}\BibitemShut {NoStop}%
\bibitem [{\citenamefont {Popov}\ \emph {et~al.}(2008)\citenamefont {Popov},
  \citenamefont {Bychenkov}, \citenamefont {Rozmus},\ and\ \citenamefont
  {Sydora}}]{Popov08}%
  \BibitemOpen
  \bibfield  {author} {\bibinfo {author} {\bibfnamefont {K.~I.}\ \bibnamefont
  {Popov}}, \bibinfo {author} {\bibfnamefont {V.~Y.}\ \bibnamefont
  {Bychenkov}}, \bibinfo {author} {\bibfnamefont {W.}~\bibnamefont {Rozmus}}, \
  and\ \bibinfo {author} {\bibfnamefont {R.~D.}\ \bibnamefont {Sydora}},\
  }\bibfield  {title} {\enquote {\bibinfo {title} {{Electron vacuum
  acceleration by a tightly focused laser pulse}},}\ }\href {\doibase
  10.1063/1.2830651} {\bibfield  {journal} {\bibinfo  {journal} {Physics of
  Plasmas}\ }\textbf {\bibinfo {volume} {15}} (\bibinfo {year} {2008}),\
  10.1063/1.2830651},\ \bibinfo {note} {013108}\BibitemShut {NoStop}%
\bibitem [{\citenamefont {Hegelich}, \citenamefont {Labun},\ and\ \citenamefont
  {Labun}(2023)}]{Hegelich23}%
  \BibitemOpen
  \bibfield  {author} {\bibinfo {author} {\bibfnamefont {B.~M.}\ \bibnamefont
  {Hegelich}}, \bibinfo {author} {\bibfnamefont {L.}~\bibnamefont {Labun}}, \
  and\ \bibinfo {author} {\bibfnamefont {O.~Z.}\ \bibnamefont {Labun}},\
  }\bibfield  {title} {\enquote {\bibinfo {title} {Revisiting experimental
  signatures of the ponderomotive force},}\ }\href {\doibase
  10.3390/photonics10020226} {\bibfield  {journal} {\bibinfo  {journal}
  {Photonics}\ }\textbf {\bibinfo {volume} {10}} (\bibinfo {year} {2023}),\
  10.3390/photonics10020226}\BibitemShut {NoStop}%
\bibitem [{\citenamefont {Sarachik}\ and\ \citenamefont
  {Schappert}(1970)}]{Sarachik70}%
  \BibitemOpen
  \bibfield  {author} {\bibinfo {author} {\bibfnamefont {E.~S.}\ \bibnamefont
  {Sarachik}}\ and\ \bibinfo {author} {\bibfnamefont {G.~T.}\ \bibnamefont
  {Schappert}},\ }\bibfield  {title} {\enquote {\bibinfo {title} {Classical
  theory of the scattering of intense laser radiation by free electrons},}\
  }\href {\doibase 10.1103/PhysRevD.1.2738} {\bibfield  {journal} {\bibinfo
  {journal} {Phys. Rev. D}\ }\textbf {\bibinfo {volume} {1}},\ \bibinfo {pages}
  {2738--2753} (\bibinfo {year} {1970})}\BibitemShut {NoStop}%
\bibitem [{\citenamefont {Gonoskov}\ \emph {et~al.}(2022)\citenamefont
  {Gonoskov}, \citenamefont {Blackburn}, \citenamefont {Marklund},\ and\
  \citenamefont {Bulanov}}]{Gonoskov22}%
  \BibitemOpen
  \bibfield  {author} {\bibinfo {author} {\bibfnamefont {A.}~\bibnamefont
  {Gonoskov}}, \bibinfo {author} {\bibfnamefont {T.~G.}\ \bibnamefont
  {Blackburn}}, \bibinfo {author} {\bibfnamefont {M.}~\bibnamefont {Marklund}},
  \ and\ \bibinfo {author} {\bibfnamefont {S.~S.}\ \bibnamefont {Bulanov}},\
  }\bibfield  {title} {\enquote {\bibinfo {title} {Charged particle motion and
  radiation in strong electromagnetic fields},}\ }\href {\doibase
  10.1103/RevModPhys.94.045001} {\bibfield  {journal} {\bibinfo  {journal}
  {Rev. Mod. Phys.}\ }\textbf {\bibinfo {volume} {94}},\ \bibinfo {pages}
  {045001} (\bibinfo {year} {2022})}\BibitemShut {NoStop}%
\bibitem [{\citenamefont {Jackson}(1999)}]{Jackson:100964}%
  \BibitemOpen
  \bibfield  {author} {\bibinfo {author} {\bibfnamefont {J.~D.}\ \bibnamefont
  {Jackson}},\ }\href@noop {} {\emph {\bibinfo {title} {{Classical
  electrodynamics; 3rd ed.}}}}\ (\bibinfo  {publisher} {Wiley},\ \bibinfo
  {address} {New York, NY},\ \bibinfo {year} {1999})\BibitemShut {NoStop}%
\bibitem [{\citenamefont {Moore}, \citenamefont {Knauer},\ and\ \citenamefont
  {Meyerhofer}(1995)}]{Moore95}%
  \BibitemOpen
  \bibfield  {author} {\bibinfo {author} {\bibfnamefont {C.~I.}\ \bibnamefont
  {Moore}}, \bibinfo {author} {\bibfnamefont {J.~P.}\ \bibnamefont {Knauer}}, \
  and\ \bibinfo {author} {\bibfnamefont {D.~D.}\ \bibnamefont {Meyerhofer}},\
  }\bibfield  {title} {\enquote {\bibinfo {title} {Observation of the
  transition from thomson to compton scattering in multiphoton interactions
  with low-energy electrons},}\ }\href {\doibase 10.1103/PhysRevLett.74.2439}
  {\bibfield  {journal} {\bibinfo  {journal} {Phys. Rev. Lett.}\ }\textbf
  {\bibinfo {volume} {74}},\ \bibinfo {pages} {2439--2442} (\bibinfo {year}
  {1995})}\BibitemShut {NoStop}%
\bibitem [{\citenamefont {Wilks}\ \emph {et~al.}(1992)\citenamefont {Wilks},
  \citenamefont {Kruer}, \citenamefont {Tabak},\ and\ \citenamefont
  {Langdon}}]{Wilks92}%
  \BibitemOpen
  \bibfield  {author} {\bibinfo {author} {\bibfnamefont {S.~C.}\ \bibnamefont
  {Wilks}}, \bibinfo {author} {\bibfnamefont {W.~L.}\ \bibnamefont {Kruer}},
  \bibinfo {author} {\bibfnamefont {M.}~\bibnamefont {Tabak}}, \ and\ \bibinfo
  {author} {\bibfnamefont {A.~B.}\ \bibnamefont {Langdon}},\ }\bibfield
  {title} {\enquote {\bibinfo {title} {Absorption of ultra-intense laser
  pulses},}\ }\href {\doibase 10.1103/PhysRevLett.69.1383} {\bibfield
  {journal} {\bibinfo  {journal} {Phys. Rev. Lett.}\ }\textbf {\bibinfo
  {volume} {69}},\ \bibinfo {pages} {1383--1386} (\bibinfo {year}
  {1992})}\BibitemShut {NoStop}%
\bibitem [{\citenamefont {Peatross}\ \emph {et~al.}(2017)\citenamefont
  {Peatross}, \citenamefont {Berrondo}, \citenamefont {Smith},\ and\
  \citenamefont {Ware}}]{Peatross:17}%
  \BibitemOpen
  \bibfield  {author} {\bibinfo {author} {\bibfnamefont {J.}~\bibnamefont
  {Peatross}}, \bibinfo {author} {\bibfnamefont {M.}~\bibnamefont {Berrondo}},
  \bibinfo {author} {\bibfnamefont {D.}~\bibnamefont {Smith}}, \ and\ \bibinfo
  {author} {\bibfnamefont {M.}~\bibnamefont {Ware}},\ }\bibfield  {title}
  {\enquote {\bibinfo {title} {Vector fields in a tight laser focus: comparison
  of models},}\ }\href@noop {} {\bibfield  {journal} {\bibinfo  {journal} {Opt.
  Express}\ }\textbf {\bibinfo {volume} {25}},\ \bibinfo {pages} {13990--14007}
  (\bibinfo {year} {2017})}\BibitemShut {NoStop}%
\bibitem [{\citenamefont {Erikson}\ and\ \citenamefont
  {Singh}(1994)}]{Erikson94}%
  \BibitemOpen
  \bibfield  {author} {\bibinfo {author} {\bibfnamefont {W.~L.}\ \bibnamefont
  {Erikson}}\ and\ \bibinfo {author} {\bibfnamefont {S.}~\bibnamefont
  {Singh}},\ }\bibfield  {title} {\enquote {\bibinfo {title} {Polarization
  properties of maxwell-gaussian laser beams},}\ }\href@noop {} {\bibfield
  {journal} {\bibinfo  {journal} {Phys. Rev. E}\ }\textbf {\bibinfo {volume}
  {49}},\ \bibinfo {pages} {5778--5786} (\bibinfo {year} {1994})}\BibitemShut
  {NoStop}%
\bibitem [{\citenamefont {Longman}\ and\ \citenamefont
  {Fedosejevs}(2022)}]{Longman22}%
  \BibitemOpen
  \bibfield  {author} {\bibinfo {author} {\bibfnamefont {A.}~\bibnamefont
  {Longman}}\ and\ \bibinfo {author} {\bibfnamefont {R.}~\bibnamefont
  {Fedosejevs}},\ }\bibfield  {title} {\enquote {\bibinfo {title} {{Modeling of
  high intensity orbital angular momentum beams for laser–plasma
  interactions}},}\ }\href {\doibase 10.1063/5.0093067} {\bibfield  {journal}
  {\bibinfo  {journal} {Physics of Plasmas}\ }\textbf {\bibinfo {volume} {29}}
  (\bibinfo {year} {2022}),\ 10.1063/5.0093067},\ \bibinfo {note}
  {063109}\BibitemShut {NoStop}%
\bibitem [{\citenamefont {Augst}\ \emph {et~al.}(1989)\citenamefont {Augst},
  \citenamefont {Strickland}, \citenamefont {Meyerhofer}, \citenamefont
  {Chin},\ and\ \citenamefont {Eberly}}]{Augst89}%
  \BibitemOpen
  \bibfield  {author} {\bibinfo {author} {\bibfnamefont {S.}~\bibnamefont
  {Augst}}, \bibinfo {author} {\bibfnamefont {D.}~\bibnamefont {Strickland}},
  \bibinfo {author} {\bibfnamefont {D.~D.}\ \bibnamefont {Meyerhofer}},
  \bibinfo {author} {\bibfnamefont {S.~L.}\ \bibnamefont {Chin}}, \ and\
  \bibinfo {author} {\bibfnamefont {J.~H.}\ \bibnamefont {Eberly}},\ }\bibfield
   {title} {\enquote {\bibinfo {title} {Tunneling ionization of noble gases in
  a high-intensity laser field},}\ }\href {\doibase
  10.1103/PhysRevLett.63.2212} {\bibfield  {journal} {\bibinfo  {journal}
  {Phys. Rev. Lett.}\ }\textbf {\bibinfo {volume} {63}},\ \bibinfo {pages}
  {2212--2215} (\bibinfo {year} {1989})}\BibitemShut {NoStop}%
\bibitem [{\citenamefont {Augst}\ \emph {et~al.}(1991)\citenamefont {Augst},
  \citenamefont {Meyerhofer}, \citenamefont {Strickland},\ and\ \citenamefont
  {Chin}}]{Augst:91}%
  \BibitemOpen
  \bibfield  {author} {\bibinfo {author} {\bibfnamefont {S.}~\bibnamefont
  {Augst}}, \bibinfo {author} {\bibfnamefont {D.~D.}\ \bibnamefont
  {Meyerhofer}}, \bibinfo {author} {\bibfnamefont {D.}~\bibnamefont
  {Strickland}}, \ and\ \bibinfo {author} {\bibfnamefont {S.~L.}\ \bibnamefont
  {Chin}},\ }\bibfield  {title} {\enquote {\bibinfo {title} {Laser ionization
  of noble gases by coulomb-barrier suppression},}\ }\href {\doibase
  10.1364/JOSAB.8.000858} {\bibfield  {journal} {\bibinfo  {journal} {J. Opt.
  Soc. Am. B}\ }\textbf {\bibinfo {volume} {8}},\ \bibinfo {pages} {858--867}
  (\bibinfo {year} {1991})}\BibitemShut {NoStop}%
\bibitem [{\citenamefont {Tong}, \citenamefont {Zhao},\ and\ \citenamefont
  {Lin}(2002)}]{Tong02}%
  \BibitemOpen
  \bibfield  {author} {\bibinfo {author} {\bibfnamefont {X.~M.}\ \bibnamefont
  {Tong}}, \bibinfo {author} {\bibfnamefont {Z.~X.}\ \bibnamefont {Zhao}}, \
  and\ \bibinfo {author} {\bibfnamefont {C.~D.}\ \bibnamefont {Lin}},\
  }\bibfield  {title} {\enquote {\bibinfo {title} {Theory of molecular
  tunneling ionization},}\ }\href {\doibase 10.1103/PhysRevA.66.033402}
  {\bibfield  {journal} {\bibinfo  {journal} {Phys. Rev. A}\ }\textbf {\bibinfo
  {volume} {66}},\ \bibinfo {pages} {033402} (\bibinfo {year}
  {2002})}\BibitemShut {NoStop}%
\bibitem [{\citenamefont {Kramida}\ \emph {et~al.}(2022)\citenamefont
  {Kramida}, \citenamefont {{Yu.~Ralchenko}}, \citenamefont {Reader},\ and\
  \citenamefont {Team}}]{NIST_ASD}%
  \BibitemOpen
  \bibfield  {author} {\bibinfo {author} {\bibfnamefont {A.}~\bibnamefont
  {Kramida}}, \bibinfo {author} {\bibnamefont {{Yu.~Ralchenko}}}, \bibinfo
  {author} {\bibfnamefont {J.}~\bibnamefont {Reader}}, \ and\ \bibinfo {author}
  {\bibfnamefont {N.~A.}\ \bibnamefont {Team}},\ }\href@noop {} {}\bibinfo
  {howpublished} {{NIST Atomic Spectra Database (ver. 5.10), [Online].
  Available: {\tt{https://physics.nist.gov/asd}} [2023, May 15]. National
  Institute of Standards and Technology, Gaithersburg, MD.}} (\bibinfo {year}
  {2022})\BibitemShut {NoStop}%
\bibitem [{\citenamefont {Boutoux}\ \emph {et~al.}(2015)\citenamefont
  {Boutoux}, \citenamefont {Rabhi}, \citenamefont {Batani}, \citenamefont
  {Binet}, \citenamefont {Ducret}, \citenamefont {Jakubowska}, \citenamefont
  {Nègre}, \citenamefont {Reverdin},\ and\ \citenamefont
  {Thfoin}}]{Boutoux15}%
  \BibitemOpen
  \bibfield  {author} {\bibinfo {author} {\bibfnamefont {G.}~\bibnamefont
  {Boutoux}}, \bibinfo {author} {\bibfnamefont {N.}~\bibnamefont {Rabhi}},
  \bibinfo {author} {\bibfnamefont {D.}~\bibnamefont {Batani}}, \bibinfo
  {author} {\bibfnamefont {A.}~\bibnamefont {Binet}}, \bibinfo {author}
  {\bibfnamefont {J.-E.}\ \bibnamefont {Ducret}}, \bibinfo {author}
  {\bibfnamefont {K.}~\bibnamefont {Jakubowska}}, \bibinfo {author}
  {\bibfnamefont {J.-P.}\ \bibnamefont {Nègre}}, \bibinfo {author}
  {\bibfnamefont {C.}~\bibnamefont {Reverdin}}, \ and\ \bibinfo {author}
  {\bibfnamefont {I.}~\bibnamefont {Thfoin}},\ }\bibfield  {title} {\enquote
  {\bibinfo {title} {{Study of imaging plate detector sensitivity to 5-18 MeV
  electrons}},}\ }\href {\doibase 10.1063/1.4936141} {\bibfield  {journal}
  {\bibinfo  {journal} {Review of Scientific Instruments}\ }\textbf {\bibinfo
  {volume} {86}} (\bibinfo {year} {2015}),\ 10.1063/1.4936141},\ \bibinfo
  {note} {113304}\BibitemShut {NoStop}%
\bibitem [{\citenamefont {Berger}\ \emph {et~al.}(2005)\citenamefont {Berger},
  \citenamefont {Coursey}, \citenamefont {Zucker},\ and\ \citenamefont
  {Chang}}]{NIST_PML}%
  \BibitemOpen
  \bibfield  {author} {\bibinfo {author} {\bibfnamefont {M.}~\bibnamefont
  {Berger}}, \bibinfo {author} {\bibfnamefont {J.}~\bibnamefont {Coursey}},
  \bibinfo {author} {\bibfnamefont {M.}~\bibnamefont {Zucker}}, \ and\ \bibinfo
  {author} {\bibfnamefont {J.}~\bibnamefont {Chang}},\ }\href@noop {}
  {}\bibinfo {howpublished} {{ESTAR, PSTAR, and ASTAR: Computer Programs for
  Calculating Stopping-Power and Range Tables for Electrons, Protons, and
  Helium Ions (ver.1.2.3), [Online]. Available:
  {\tt{https://physics.nist.gov/Star}} [2023, May 15]. National Institute of
  Standards and Technology, Gaithersburg, MD.}} (\bibinfo {year}
  {2005})\BibitemShut {NoStop}%
\bibitem [{\citenamefont {Fienup}(1982)}]{Fienup:82}%
  \BibitemOpen
  \bibfield  {author} {\bibinfo {author} {\bibfnamefont {J.~R.}\ \bibnamefont
  {Fienup}},\ }\bibfield  {title} {\enquote {\bibinfo {title} {Phase retrieval
  algorithms: a comparison},}\ }\href {\doibase 10.1364/AO.21.002758}
  {\bibfield  {journal} {\bibinfo  {journal} {Appl. Opt.}\ }\textbf {\bibinfo
  {volume} {21}},\ \bibinfo {pages} {2758--2769} (\bibinfo {year}
  {1982})}\BibitemShut {NoStop}%
\bibitem [{\citenamefont {Trebino}(2000)}]{Trebino}%
  \BibitemOpen
  \bibfield  {author} {\bibinfo {author} {\bibfnamefont {R.}~\bibnamefont
  {Trebino}},\ }\href@noop {} {\emph {\bibinfo {title} {{Frequency-Resolved
  Optical Gating: The Measurement of Ultrashort Laser Pulses}}}},\ \bibinfo
  {edition} {1st}\ ed.\ (\bibinfo  {publisher} {Springer},\ \bibinfo {year}
  {2000})\BibitemShut {NoStop}%
\bibitem [{\citenamefont {Williams}\ \emph {et~al.}(2014)\citenamefont
  {Williams}, \citenamefont {Maddox}, \citenamefont {Chen}, \citenamefont
  {Kojima},\ and\ \citenamefont {Millecchia}}]{Williams14}%
  \BibitemOpen
  \bibfield  {author} {\bibinfo {author} {\bibfnamefont {G.~J.}\ \bibnamefont
  {Williams}}, \bibinfo {author} {\bibfnamefont {B.~R.}\ \bibnamefont
  {Maddox}}, \bibinfo {author} {\bibfnamefont {H.}~\bibnamefont {Chen}},
  \bibinfo {author} {\bibfnamefont {S.}~\bibnamefont {Kojima}}, \ and\ \bibinfo
  {author} {\bibfnamefont {M.}~\bibnamefont {Millecchia}},\ }\bibfield  {title}
  {\enquote {\bibinfo {title} {{Calibration and equivalency analysis of image
  plate scanners}},}\ }\href {\doibase 10.1063/1.4886390} {\bibfield  {journal}
  {\bibinfo  {journal} {Review of Scientific Instruments}\ }\textbf {\bibinfo
  {volume} {85}} (\bibinfo {year} {2014}),\ 10.1063/1.4886390}\BibitemShut
  {NoStop}%
\bibitem [{\citenamefont {Volpe}\ \emph {et~al.}(2019)\citenamefont {Volpe},
  \citenamefont {Fedosejevs}, \citenamefont {Gatti}, \citenamefont
  {Perez-Hernandez}, \citenamefont {Mendez}, \citenamefont {Apinaniz},
  \citenamefont {Vaisseau}, \citenamefont {Salgado}, \citenamefont {Huault},
  \citenamefont {Malko}, \citenamefont {Zeraouli}, \citenamefont {Ospina},
  \citenamefont {Longman}, \citenamefont {De~Luis}, \citenamefont {Li},
  \citenamefont {Varela}, \citenamefont {Garcia}, \citenamefont {Hernandez},
  \citenamefont {Pisonero}, \citenamefont {Garcia~Ajates}, \citenamefont
  {Alvarez}, \citenamefont {Garcia}, \citenamefont {Rico}, \citenamefont
  {Arana}, \citenamefont {Hernandez-Toro},\ and\ \citenamefont
  {Roso}}]{volpe_2019}%
  \BibitemOpen
  \bibfield  {author} {\bibinfo {author} {\bibfnamefont {L.}~\bibnamefont
  {Volpe}}, \bibinfo {author} {\bibfnamefont {R.}~\bibnamefont {Fedosejevs}},
  \bibinfo {author} {\bibfnamefont {G.}~\bibnamefont {Gatti}}, \bibinfo
  {author} {\bibfnamefont {J.~A.}\ \bibnamefont {Perez-Hernandez}}, \bibinfo
  {author} {\bibfnamefont {C.}~\bibnamefont {Mendez}}, \bibinfo {author}
  {\bibfnamefont {J.}~\bibnamefont {Apinaniz}}, \bibinfo {author}
  {\bibfnamefont {X.}~\bibnamefont {Vaisseau}}, \bibinfo {author}
  {\bibfnamefont {C.}~\bibnamefont {Salgado}}, \bibinfo {author} {\bibfnamefont
  {M.}~\bibnamefont {Huault}}, \bibinfo {author} {\bibfnamefont
  {S.}~\bibnamefont {Malko}}, \bibinfo {author} {\bibfnamefont
  {G.}~\bibnamefont {Zeraouli}}, \bibinfo {author} {\bibfnamefont
  {V.}~\bibnamefont {Ospina}}, \bibinfo {author} {\bibfnamefont
  {A.}~\bibnamefont {Longman}}, \bibinfo {author} {\bibfnamefont
  {D.}~\bibnamefont {De~Luis}}, \bibinfo {author} {\bibfnamefont
  {K.}~\bibnamefont {Li}}, \bibinfo {author} {\bibfnamefont {O.}~\bibnamefont
  {Varela}}, \bibinfo {author} {\bibfnamefont {E.}~\bibnamefont {Garcia}},
  \bibinfo {author} {\bibfnamefont {I.}~\bibnamefont {Hernandez}}, \bibinfo
  {author} {\bibfnamefont {J.}~\bibnamefont {Pisonero}}, \bibinfo {author}
  {\bibfnamefont {J.}~\bibnamefont {Garcia~Ajates}}, \bibinfo {author}
  {\bibfnamefont {J.}~\bibnamefont {Alvarez}}, \bibinfo {author} {\bibfnamefont
  {C.}~\bibnamefont {Garcia}}, \bibinfo {author} {\bibfnamefont
  {M.}~\bibnamefont {Rico}}, \bibinfo {author} {\bibfnamefont {D.}~\bibnamefont
  {Arana}}, \bibinfo {author} {\bibfnamefont {J.}~\bibnamefont
  {Hernandez-Toro}}, \ and\ \bibinfo {author} {\bibfnamefont {L.}~\bibnamefont
  {Roso}},\ }\bibfield  {title} {\enquote {\bibinfo {title} {Generation of high
  energy laser-driven electron and proton sources with the 200 tw system vega 2
  at the centro de laseres pulsados},}\ }\href@noop {} {\bibfield  {journal}
  {\bibinfo  {journal} {High Power Laser Science and Engineering}\ }\textbf
  {\bibinfo {volume} {7}},\ \bibinfo {pages} {e25} (\bibinfo {year}
  {2019})}\BibitemShut {NoStop}%
\bibitem [{\citenamefont {Cummings}\ and\ \citenamefont
  {Thomas}(2011)}]{Cummings11}%
  \BibitemOpen
  \bibfield  {author} {\bibinfo {author} {\bibfnamefont {P.}~\bibnamefont
  {Cummings}}\ and\ \bibinfo {author} {\bibfnamefont {A.~G.~R.}\ \bibnamefont
  {Thomas}},\ }\bibfield  {title} {\enquote {\bibinfo {title} {{A computational
  investigation of the impact of aberrated Gaussian laser pulses on electron
  beam properties in laser-wakefield acceleration experiments}},}\ }\href
  {\doibase 10.1063/1.3587111} {\bibfield  {journal} {\bibinfo  {journal}
  {Physics of Plasmas}\ }\textbf {\bibinfo {volume} {18}} (\bibinfo {year}
  {2011}),\ 10.1063/1.3587111},\ \bibinfo {note} {053110}\BibitemShut {NoStop}%
\bibitem [{\citenamefont {Born}\ and\ \citenamefont {Wolf}(1999)}]{BornWolf99}%
  \BibitemOpen
  \bibfield  {author} {\bibinfo {author} {\bibfnamefont {M.}~\bibnamefont
  {Born}}\ and\ \bibinfo {author} {\bibfnamefont {E.}~\bibnamefont {Wolf}},\
  }\href@noop {} {\emph {\bibinfo {title} {Principles of Optics:
  Electromagnetic Theory of Propagation, Interference and Diffraction of Light
  (7th Edition)}}},\ \bibinfo {edition} {7th}\ ed.\ (\bibinfo  {publisher}
  {Cambridge University Press},\ \bibinfo {year} {1999})\BibitemShut {NoStop}%
\bibitem [{\citenamefont {Palastro}\ \emph {et~al.}(2020)\citenamefont
  {Palastro}, \citenamefont {Shaw}, \citenamefont {Franke}, \citenamefont
  {Ramsey}, \citenamefont {Simpson},\ and\ \citenamefont
  {Froula}}]{Palastro20}%
  \BibitemOpen
  \bibfield  {author} {\bibinfo {author} {\bibfnamefont {J.~P.}\ \bibnamefont
  {Palastro}}, \bibinfo {author} {\bibfnamefont {J.~L.}\ \bibnamefont {Shaw}},
  \bibinfo {author} {\bibfnamefont {P.}~\bibnamefont {Franke}}, \bibinfo
  {author} {\bibfnamefont {D.}~\bibnamefont {Ramsey}}, \bibinfo {author}
  {\bibfnamefont {T.~T.}\ \bibnamefont {Simpson}}, \ and\ \bibinfo {author}
  {\bibfnamefont {D.~H.}\ \bibnamefont {Froula}},\ }\bibfield  {title}
  {\enquote {\bibinfo {title} {Dephasingless laser wakefield acceleration},}\
  }\href {\doibase 10.1103/PhysRevLett.124.134802} {\bibfield  {journal}
  {\bibinfo  {journal} {Phys. Rev. Lett.}\ }\textbf {\bibinfo {volume} {124}},\
  \bibinfo {pages} {134802} (\bibinfo {year} {2020})}\BibitemShut {NoStop}%
\bibitem [{\citenamefont {Goodman}(2005)}]{goodman2005}%
  \BibitemOpen
  \bibfield  {author} {\bibinfo {author} {\bibfnamefont {J.~W.}\ \bibnamefont
  {Goodman}},\ }\href@noop {} {\emph {\bibinfo {title} {Introduction to Fourier
  optics}}},\ \bibinfo {edition} {3rd}\ ed.,\ Vol.~\bibinfo {volume} {1}\
  (\bibinfo  {publisher} {Roberts \& Co. Publishers},\ \bibinfo {year}
  {2005})\BibitemShut {NoStop}%
\bibitem [{\citenamefont {Longman}\ and\ \citenamefont
  {Fedosejevs}(2017)}]{Longman:17}%
  \BibitemOpen
  \bibfield  {author} {\bibinfo {author} {\bibfnamefont {A.}~\bibnamefont
  {Longman}}\ and\ \bibinfo {author} {\bibfnamefont {R.}~\bibnamefont
  {Fedosejevs}},\ }\bibfield  {title} {\enquote {\bibinfo {title} {Mode
  conversion efficiency to laguerre-gaussian oam modes using spiral phase
  optics},}\ }\href@noop {} {\bibfield  {journal} {\bibinfo  {journal} {Opt.
  Express}\ }\textbf {\bibinfo {volume} {25}},\ \bibinfo {pages} {17382--17392}
  (\bibinfo {year} {2017})}\BibitemShut {NoStop}%
\end{thebibliography}%

\end{document}